\documentclass[showpacs,superscriptaddress]{revtex4-2}
\usepackage{booktabs}
\usepackage{hyperref}
\hypersetup{
  colorlinks=true,
  linkcolor=blue,
  citecolor=red,
}
\usepackage{graphicx}
\usepackage{dcolumn}
\usepackage{bm}
\usepackage{amsmath}
\usepackage{amssymb}
\usepackage{color}
\usepackage{enumitem}
\usepackage{mathtools}
\usepackage{orcidlink}

\begin{document}

\title{Analytical Solutions of One-Dimensional ($1\mathcal{D}$) Potentials for Spin-0 Particles via the Feshbach-Villars Formalism}

\author{Abdelmalek Boumali\orcidlink{0000-0003-2552-0427}}
\email{abdelmalek.boumali@gmail.com}
\affiliation{Echahid Cheikh Larbi Tebessi University 12001, Tebessa, Algeria}

\author{Abdelmalek Bouzenada\orcidlink{0000-0002-3363-980X}}
\email{abdelmalekbouzenada@gmail.com}
\affiliation{Echahid Cheikh Larbi Tebessi University 12001, Tebessa, Algeria}
\affiliation{Research Center of Astrophysics and Cosmology, Khazar University, Baku, AZ1096, 41 Mehseti Street, Azerbaijan}

\author{Edilberto O.\ Silva\orcidlink{0000-0002-0297-5747}}
\email{edilberto.silva@ufma.br (Corresp. author)}
\affiliation{Departamento de F\'{\i}sica, Universidade Federal do Maranh\~{a}o,
65085-580, S\~{a}o Lu\'{\i}s, Maranh\~{a}o, Brazil}

\begin{abstract}
We present a unified analytical and numerical study of the one-dimensional Feshbach--Villars (FV) equation for spin-0 particles in the presence of several representative external potentials. Starting from the FV formulation of the Klein--Gordon equation, we derive the corresponding one-dimensional master equation and analyse its solutions for Coulomb, power-exponential, Cornell, P\"oschl--Teller, and Woods--Saxon interactions. For the singular Coulomb and Cornell cases, a Loudon-type cutoff regularisation is implemented on the full line, allowing a mathematically controlled treatment of the origin and an explicit classification of the states by parity. The Coulomb problem exhibits the expected near-degenerate even--odd structure in the cutoff limit, while the Cornell potential combines short-distance Coulomb behaviour with long-distance confinement and produces a finite set of bound states for fixed parameters. The power-exponential potential with $p=1$ is reduced to a Whittaker-type equation and yields an intrinsically relativistic spectrum with no standard Schr\"odinger bound-state limit in the parameter regime considered. For the smooth short-range P\"oschl--Teller and Woods-Saxon potentials, the FV formalism reveals, respectively, the effects of definite parity and spatial asymmetry on the spectrum, wave functions, and particle--antiparticle mixing. In all cases, we reconstruct the full FV spinor, analyse the associated charge density, and compare the relativistic behaviour with the corresponding non-relativistic expectations whenever such a limit exists. The results provide a coherent set of analytical and numerical benchmarks for relativistic scalar bound states in one dimension.
\end{abstract}

\maketitle

\section{Introduction}\label{s1}

Quantum mechanics provides the fundamental description of nature at the smallest scales~\cite{B1}, introducing phenomena such as superposition and entanglement that have no classical counterpart. The Schr\"odinger equation~\cite{B2} governs the non-relativistic evolution of the wave function, yielding a probabilistic description of observables such as position and momentum and encapsulating the uncertainty inherent in microscopic systems. This framework plays a decisive role in the understanding of atomic and molecular structure, and its implications extend to modern quantum technologies, including quantum computing~\cite{B3} and quantum cryptography~\cite{B4}.

When particles attain relativistic energies or when particle--antiparticle effects become relevant, a relativistic extension of quantum mechanics is indispensable. The Klein--Gordon equation~\cite{k1} is the natural relativistic wave equation for scalar (spin-0) particles, describing bosons whose quantum state is represented by a Lorentz-scalar field. Spin-0 particles obey Bose--Einstein statistics and, unlike spin-$1/2$ fermions~\cite{k2}, are not subject to the Pauli exclusion principle. Their most celebrated representative is the Higgs boson~\cite{k3,k3-1,k3-2,k3-3}, discovered in 2012 at the Large Hadron Collider, which plays a central role in the Standard Model by endowing other elementary particles with mass through the Higgs mechanism. Beyond collider physics, scalar fields also arise naturally in models of dark matter~\cite{k4} and in effective descriptions of nuclear and hadronic interactions~\cite{k5}.

A recurring theme in both non-relativistic and relativistic quantum mechanics is the study of exactly solvable or analytically tractable models, which serve as benchmarks for numerical methods and clarify the roles of symmetry, boundary conditions, and asymptotic behaviour. The form of the external potential strongly influences the spectrum, the spatial profile of the wave function, and the structure of the corresponding bound states. Coulomb potentials, characterised by an inverse-distance dependence, are fundamental to atomic physics and arise naturally in systems governed by electromagnetic interactions~\cite{c1,c2}. The Cornell potential, which combines a short-range Coulomb term with long-range linear confinement, is widely used in phenomenological descriptions of quark--antiquark bound states in quantum chromodynamics~\cite{c3,c4,c5,c6,c7}. The P\"oschl--Teller potential provides a smooth short-range well with definite parity on the full line, making it particularly useful in the study of bound states and exactly solvable models. The Woods--Saxon potential, central to nuclear physics, describes a diffuse finite-depth interaction with an asymmetric profile and realistic surface behaviour.

Power-exponential potentials of the form $V(x)\propto e^{-(x/x_0)^p}$ constitute a flexible and physically motivated family that interpolates between the Gaussian ($p=2$) and simple exponential ($p=1$) cases. Such potentials arise in cosmology, where they appear in models of inflation and scalar-field dynamics~\cite{p2,p3}, and in condensed-matter physics, where they can describe confinement in quantum dots and effective interactions in low-dimensional systems~\cite{p1,p4}. Their analytical tractability makes them attractive in relativistic quantum mechanics, especially when one wishes to compare smooth short-range interactions with singular or confining potentials.

The Feshbach--Villars (FV) formalism~\cite{F1} provides an elegant reformulation of the Klein--Gordon equation. By introducing a two-component wave function, the FV approach recasts the original second-order equation into a system of coupled first-order equations of Schr\"odinger type. This representation facilitates the inclusion of external potentials and boundary conditions, makes the particle--antiparticle structure explicit, and clarifies the interpretation of the conserved charge density. The FV framework has been applied to a wide range of problems, including the harmonic oscillator~\cite{F2}, Coulomb- and Cornell-type interactions~\cite{F3,F4}, curved space-time backgrounds~\cite{F5,F6}, magnetic-field configurations~\cite{F7}, and systems involving topological defects and Aharonov--Bohm-type effects~\cite{F8,Hamla2024}. More recent work has examined the FV equation in Lorentz-violating settings~\cite{F9,F10} and explored its connection with the Foldy--Wouthuysen (FW) transformation~\cite{F11,Silenko2013}.

The Foldy--Wouthuysen transformation is a canonical unitary transformation that block-diagonalises a relativistic Hamiltonian, thereby separating positive-energy (particle) states from negative-energy (antiparticle) states and making the non-relativistic limit more transparent. Although originally developed for the Dirac equation, it is also applicable to the FV equation for scalar particles, where it helps clarify the physical meaning of relativistic corrections and provides a bridge between the fully relativistic dynamics and its non-relativistic approximation~\cite{Silenko2013,F11}.

Despite this extensive literature, a unified treatment of the one-dimensional FV equation across several qualitatively distinct external potentials remains useful both pedagogically and physically. In particular, it is valuable to compare, within a single framework, singular interactions requiring regularisation, smooth short-range wells, confining potentials, and asymmetric finite-range profiles. The present work addresses this goal by analysing the FV equation for five representative cases: the Coulomb, power-exponential, Cornell, P\"oschl--Teller, and Woods--Saxon potentials. Depending on the potential, the analysis combines exact reductions to special-function equations with cutoff regularisation, matching conditions, or direct numerical shooting.

Our main objectives are threefold. First, we derive and solve the one-dimensional FV master equation for a representative set of potentials spanning singular, short-range, confining, and asymmetric regimes. Second, we reconstruct the full FV spinor in each case and analyse the corresponding particle--antiparticle mixing and conserved charge density. Third, whenever appropriate, we compare the relativistic results with their non-relativistic counterparts and identify the features that are genuinely relativistic.

The paper is organised as follows. Section~\ref{s2} reviews the FV formalism and derives the one-dimensional master equation. Section~\ref{s3_intro} introduces the class of applications considered in this work. Section~\ref{s3} presents the regularised Coulomb problem. Section~\ref{s4} analyses the power-exponential potential with $p=1$. Section~\ref{s5} discusses the non-relativistic limit. Section~\ref{s6} is devoted to the regularised Cornell potential. Section~\ref{s7} studies the P\"oschl--Teller potential, and Section~\ref{s8} treats the Woods--Saxon case. Finally, Section~\ref{s_conclusion} summarises the main results and outlines possible extensions.

\section{Feshbach-Villars Formalism}\label{s2}

The Klein-Gordon equation is a cornerstone of relativistic quantum mechanics for spin-0 particles, yet its second-order time derivative leads to conceptual and practical difficulties. In particular, the wave function does not admit an immediate probabilistic interpretation analogous to that of the Schr\"odinger theory, and the separation between particle and antiparticle components is not manifest. The Feshbach-Villars (FV) formalism~\cite{Feshbach1958} addresses these issues by introducing a two-component representation that recasts the Klein-Gordon equation into a Schr\"odinger-like first-order form. This formulation makes the conserved charge density and current more transparent, facilitates the coupling to external electromagnetic fields, and provides a natural route to the non-relativistic limit.

In this section, we derive the FV equations from the Klein-Gordon equation, present the associated Hamiltonian structure and conserved quantities, and obtain the one-dimensional stationary equation that serves as the starting point for the exact solutions discussed in the following sections. Throughout this work, natural units $\hbar=c=1$ are used.

\subsection{From the Klein-Gordon equation to the FV representation}

We begin with the Klein-Gordon equation for a scalar particle of mass $m$ in the presence of an external electrostatic potential $V(x,t)$,
\begin{equation}
  \left( i\frac{\partial}{\partial t} - eV \right)^2 \psi
  = \left( -\nabla^2 + m^2 \right) \psi.
  \label{eq:KG_original}
\end{equation}
Following Ref.~\cite{Feshbach1958}, we introduce the two-component wave function
\begin{equation}
  \Psi =
  \begin{pmatrix}
    \psi_1 \\
    \psi_2
  \end{pmatrix},
  \label{eq:KGF}
\end{equation}
together with the definitions
\begin{equation}
  \psi = \psi_1 + \psi_2,
  \label{eq:psi_sum}
\end{equation}
and
\begin{equation}
  \left( i\frac{\partial}{\partial t} - eV \right)\psi
  = m(\psi_1 - \psi_2).
  \label{eq:kg_hamiltonian}
\end{equation}
The quantities $\psi_1$ and $\psi_2$ may be interpreted as the particle and antiparticle components of the FV wave function, respectively. For stationary states, $\psi(x,t)=e^{-iEt}\psi(x)$, Eqs.~\eqref{eq:psi_sum} and~\eqref{eq:kg_hamiltonian} immediately yield
\begin{equation}
  \psi_1(x)
  = \frac{1}{2}\left[1+\frac{E-eV(x)}{m}\right]\psi(x),
  \label{eq:psi1}
\end{equation}
\begin{equation}
  \psi_2(x)
  = \frac{1}{2}\left[1-\frac{E-eV(x)}{m}\right]\psi(x).
  \label{eq:psi2}
\end{equation}
These relations make explicit how the external potential mixes the two components.

\subsection{Derivation of the Feshbach-Villars equations}

Substituting the decomposition~\eqref{eq:psi_sum} and the auxiliary relation~\eqref{eq:kg_hamiltonian} into the Klein-Gordon equation~\eqref{eq:KG_original}, one obtains a coupled system for the sum and difference of the FV components:
\begin{equation}
  \left( i\frac{\partial}{\partial t} - eV \right)(\psi_1 + \psi_2)
  = m(\psi_1 - \psi_2),
  \label{eq:kg_hamiltonian_modified}
\end{equation}
\begin{equation}
  \left( i\frac{\partial}{\partial t} - eV \right)(\psi_1 - \psi_2)
  = \left( \frac{\hat{p}^2}{m} + m \right)(\psi_1 + \psi_2),
  \label{eq:kg_hamiltonian_modified2}
\end{equation}
where $\hat{p}=-i\nabla$ is the momentum operator. By taking the sum and the difference of Eqs.~\eqref{eq:kg_hamiltonian_modified} and~\eqref{eq:kg_hamiltonian_modified2}, one arrives at the Feshbach-Villars equations for a spin-0 particle:
\begin{equation}
  i\frac{\partial \psi_1}{\partial t}
  = \frac{\hat{p}^2}{2m}(\psi_1+\psi_2) + (m+eV)\psi_1,
  \label{eq:feshbach1}
\end{equation}
\begin{equation}
  i\frac{\partial \psi_2}{\partial t}
  = -\frac{\hat{p}^2}{2m}(\psi_1+\psi_2) - (m-eV)\psi_2.
  \label{eq:feshbach2}
\end{equation}
These two coupled first-order equations are Schr\"odinger-like in structure and constitute the basis of the FV representation~\cite{Feshbach1958}.

\subsection{Hamiltonian form and Pauli-matrix structure}

Equations~\eqref{eq:feshbach1} and~\eqref{eq:feshbach2} can be written compactly as
\begin{equation}
  i\frac{\partial\Psi}{\partial t} = \mathcal{H}_{\rm FV}\,\Psi,
  \label{eq:HFV0_tdep}
\end{equation}
with Hamiltonian
\begin{equation}
  \mathcal{H}_{\rm FV}
  = (\tau_3+i\tau_2)\frac{\hat{p}^2}{2m}
    + \tau_3 m + eV,
  \label{eq:hfv0_components}
\end{equation}
where $\tau_i$ ($i=1,2,3$) are the Pauli matrices,
\begin{equation}
  \tau_1 =
  \begin{pmatrix}
    0 & 1\\
    1 & 0
  \end{pmatrix},\quad
  \tau_2 =
  \begin{pmatrix}
    0 & -i\\
    i & 0
  \end{pmatrix},\quad
  \tau_3 =
  \begin{pmatrix}
    1 & 0\\
    0 & -1
  \end{pmatrix}.
  \label{eq:pauli}
\end{equation}
The matrix combination $(\tau_3+i\tau_2)$ takes the explicit form
\begin{equation}
  \tau_3+i\tau_2 =
  \begin{pmatrix}
    1 & 1\\
    1 & 1
  \end{pmatrix},
  \label{eq:tau3_itau2}
\end{equation}
showing that the kinetic term couples symmetrically to both FV components, as expected from Eqs.~\eqref{eq:feshbach1} and~\eqref{eq:feshbach2}.

An important property of the FV Hamiltonian is its pseudo-Hermiticity~\cite{Silenko2013},
\begin{equation}
  \mathcal{H}_{\rm FV}
  = \tau_3\,\mathcal{H}_{\rm FV}^\dagger\,\tau_3,
  \qquad
  \mathcal{H}_{\rm FV}^\dagger
  = \tau_3\,\mathcal{H}_{\rm FV}\,\tau_3,
  \label{eq:hermitian}
\end{equation}
which guarantees the conservation of the $\tau_3$-weighted inner product and is compatible with a real energy spectrum under the appropriate physical boundary conditions.

\subsection{Electromagnetic interaction and conserved current}

In the presence of a general external electromagnetic field described by the four-potential $(A_0,\mathbf{A})$, the minimal-coupling prescription $\hat{p}\to \hat{p}-e\mathbf{A}$ leads to the FV Hamiltonian~\cite{Feshbach1958}
\begin{equation}
  \mathcal{H}_{\rm FV}
  = (\tau_3+i\tau_2)\frac{(\hat{p}-e\mathbf{A})^2}{2m}
    + m\tau_3 + eA_0.
  \label{eq:hfv_interaction}
\end{equation}
The associated conserved charge density and current are~\cite{Hamla2024}
\begin{equation}
  \rho(\mathbf{r},t)
  = \Psi^\dagger \tau_3 \Psi
  = |\psi_1|^2 - |\psi_2|^2,
  \label{eq:probability_density}
\end{equation}
and
\begin{equation}
  \mathbf{J}(\mathbf{r},t)
  = \frac{i}{2m}
    \left[
      \Psi^\dagger \tau_3(\tau_3+i\tau_2)\nabla\Psi
      - (\nabla\Psi^\dagger)\tau_3(\tau_3+i\tau_2)\Psi
    \right]
    - \frac{e}{m}\,\mathbf{A}\,
    \Psi^\dagger\tau_3(\tau_3+i\tau_2)\Psi,
  \label{eq:current_density}
\end{equation}
which satisfy the continuity equation
\begin{equation}
  \frac{\partial \rho}{\partial t} + \nabla\cdot\mathbf{J} = 0.
  \label{eq:continuity}
\end{equation}
Accordingly, $\rho$ is interpreted not as a probability density in the non-relativistic sense, but as a charge density. It is positive for positive-energy solutions and negative for negative-energy solutions, thereby distinguishing particle and antiparticle sectors.

\subsection{One-dimensional stationary equation}

For the one-dimensional problem considered in this work, we restrict ourselves to a purely electrostatic potential $V(x)$ and seek stationary solutions of the form
\begin{equation}
  \Psi(x,t)
  = e^{-iEt}
  \begin{pmatrix}
    \psi_1(x)\\
    \psi_2(x)
  \end{pmatrix},
  \label{eq:stationary_ansatz}
\end{equation}
where $E$ is the energy eigenvalue. Substituting Eq.~\eqref{eq:stationary_ansatz} into Eqs.~\eqref{eq:feshbach1} and~\eqref{eq:feshbach2}, we obtain the coupled stationary equations~\cite{F12}
\begin{equation}
  E\psi_1
  = -\frac{1}{2m}\frac{d^2}{dx^2}(\psi_1+\psi_2)
    + m\psi_1 + eV(x)\psi_1,
  \label{eq:stationary_psi1}
\end{equation}
\begin{equation}
  E\psi_2
  = +\frac{1}{2m}\frac{d^2}{dx^2}(\psi_1+\psi_2)
    - m\psi_2 + eV(x)\psi_2.
  \label{eq:stationary_psi2}
\end{equation}
To decouple the system, we define the sum and difference functions
\begin{equation}
  \psi_s = \psi_1 + \psi_2,
  \qquad
  \psi_d = \psi_1 - \psi_2.
\end{equation}
Adding Eqs.~\eqref{eq:stationary_psi1} and~\eqref{eq:stationary_psi2}, one obtains
\begin{equation}
  E\psi_s = m\psi_d + eV(x)\psi_s,
  \label{eq:sum_eq}
\end{equation}
whereas subtracting Eq.~\eqref{eq:stationary_psi2} from Eq.~\eqref{eq:stationary_psi1} yields
\begin{equation}
  E\psi_d
  = -\frac{1}{m}\frac{d^2\psi_s}{dx^2} + m\psi_d.
  \label{eq:diff_eq}
\end{equation}
Equation~\eqref{eq:sum_eq} gives
\begin{equation}
  \psi_d = \frac{E-eV(x)}{m}\,\psi_s.
  \label{eq:psi_d_relation}
\end{equation}
Substituting Eq.~\eqref{eq:psi_d_relation} into Eq.~\eqref{eq:diff_eq}, we arrive at the master equation for $\psi_s$,
\begin{equation}
  \frac{d^2\psi_s(x)}{dx^2}
  + \left[(E-eV(x))^2 - m^2\right]\psi_s(x) = 0.
  \label{eq:KG_sumfun}
\end{equation}
This equation has exactly the form of the one-dimensional Klein-Gordon equation with electrostatic potential $eV(x)$, thus confirming the self-consistency of the FV formulation. Once $\psi_s(x)$ is known, the individual FV components are reconstructed from
\begin{equation}
  \psi_1(x)
  = \frac{1}{2}\left[1+\frac{E-eV(x)}{m}\right]\psi_s(x),
  \label{eq:psi1_from_psid}
\end{equation}
\begin{equation}
  \psi_2(x)
  = \frac{1}{2}\left[1-\frac{E-eV(x)}{m}\right]\psi_s(x).
  \label{eq:psi2_from_psid}
\end{equation}
These expressions make explicit the role of the external field in mixing the particle and antiparticle sectors: for $V=0$ and positive energy, $\psi_1$ dominates over $\psi_2$, whereas a sufficiently strong attractive potential may enhance the antiparticle component significantly. Equation~\eqref{eq:KG_sumfun} is the fundamental equation to be solved analytically for the external potentials considered in the following sections.

\section{Applications: Motivation and Overview}\label{s3_intro}

The exact analytical study of the FV equation for specific external potentials is important for several reasons. From a theoretical perspective, closed-form solutions provide direct insight into the interplay between relativistic kinematics and the spatial profile of the external field, an interplay that is absent in the non-relativistic Schr\"odinger theory. Within the FV framework, this interplay is encoded in the term $(E-eV(x))^2$ in Eq.~\eqref{eq:KG_sumfun}: even when the external potential has a simple functional form, the resulting effective relativistic dynamics is highly non-trivial.

From a practical perspective, exact or quasi-exact solutions provide valuable benchmarks for numerical, variational, and semiclassical methods. They also allow a transparent discussion of quantum numbers, level ordering, degeneracy patterns, component mixing, and the non-relativistic limit. In addition, comparing several analytically tractable potentials within the same relativistic formalism helps clarify which spectral and wave-function features are universal and which depend sensitively on the detailed shape of the interaction.

The potentials considered in this work are both physically relevant and mathematically complementary. The Coulomb interaction is the prototype of a singular long-range potential and, in the relativistic spin-0 context, plays a role analogous to that of the hydrogen atom in elementary quantum mechanics. Confining and screened interactions, such as the Cornell, power-exponential, P\"oschl-Teller, and Woods-Saxon potentials, probe different combinations of short-range structure, finite-range smoothness, and asymptotic behaviour. Taken together, these examples illustrate how qualitatively distinct external fields generate markedly different spectral structures within the same Feshbach-Villars framework.

\section{Coulomb-Type Potential: cutoff regularization \`a la Loudon}\label{s3}

The one-dimensional Coulomb problem requires special care because the potential is singular at the origin. In the full-line formulation, $V(x)\propto 1/|x|$ is not a regular potential in the ordinary Sturm--Liouville sense, and a direct treatment of the singular equation may obscure the status of even-parity states, the origin of the odd--even degeneracy, and the meaning of the deeply localised state near $x=0$. For this reason, rather than solving the singular problem \emph{ab initio}, we follow the regularisation strategy introduced by Loudon for the one-dimensional hydrogen atom~\cite{Loudon1959,Loudon2016}: the singularity is first smoothed by a short-distance cutoff, the bound-state problem is solved for the regularised potential, and only afterwards is the limit in which the cutoff tends to zero considered. This procedure provides a mathematically controlled route to the singular problem and makes transparent the physical origin of the spectral features that survive in the hard-core limit.

We therefore replace the singular Coulomb interaction by the regularised even potential
\begin{equation}
V_{\delta}(x)=
\begin{cases}
\dfrac{a}{\delta}, & |x|<\delta,\\[6pt]
\dfrac{a}{|x|}, & |x|\ge \delta,
\end{cases}
\qquad a>0,
\label{eq:coulomb_cutoff_potential}
\end{equation}
where $\delta>0$ is a small cutoff length, and we define, as before, $\alpha \equiv ea$. In the Feshbach--Villars formalism, the sum component $\psi_s$ satisfies
\begin{equation}
\frac{d^2\psi_s}{dx^2}
+\left[(E-eV_\delta(x))^2-m^2\right]\psi_s=0.
\label{eq:fv_cutoff_master}
\end{equation}
Because $V_\delta(x)$ is even, the eigenstates can be classified by parity, in direct analogy with Loudon's regularised treatment of the one-dimensional hydrogen atom~\cite{Loudon1959,Loudon2016},
\begin{equation}
\psi_s(-x)=\pm \psi_s(x),
\label{eq:parity_condition}
\end{equation}
with the plus (minus) sign corresponding to even (odd) states.

For the exterior region $|x|\ge\delta$, Eq.~\eqref{eq:fv_cutoff_master} reduces to
\begin{equation}
\frac{d^2\psi_s}{dx^2}
+\left[\frac{\alpha^2}{x^2}-\frac{2E\alpha}{|x|}-k^2\right]\psi_s=0,
\qquad
k^2\equiv m^2-E^2>0,
\label{eq:exterior_cutoff_eq}
\end{equation}
which is the same Coulomb equation obtained from the singular problem, but now posed only outside the regularised core. Introducing $z=2k|x|$, the decaying solution can be written as
\begin{equation}
\psi_s^{\rm(out)}(x)=
\mathcal{C}\,
W_{\lambda,\mu}(2k|x|),
\label{eq:outer_whittaker}
\end{equation}
where $W_{\lambda,\mu}$ is the Whittaker function and
\begin{equation}
\lambda=\frac{E\alpha}{k},
\qquad
\mu=\frac{1}{2}\sqrt{1-4\alpha^2}.
\label{eq:lambda_mu_cutoff}
\end{equation}
The restriction $|\alpha|\le 1/2$ is again required in order for $\mu$ to remain real and for the exterior solution to display the standard bound-state behaviour.

Inside the core, $|x|<\delta$, the potential is constant and Eq.~\eqref{eq:fv_cutoff_master} becomes
\begin{equation}
\frac{d^2\psi_s}{dx^2}+p_\delta^2\,\psi_s=0,
\qquad
p_\delta^2=\left(E-\frac{\alpha}{\delta}\right)^2-m^2.
\label{eq:inner_cutoff_eq}
\end{equation}
For the parameter range of interest and sufficiently small $\delta$, one has $p_\delta^2>0$, and the regular interior solutions are
\begin{equation}
\psi_s^{\rm(in,even)}(x)=A\cos(p_\delta x),
\qquad
\psi_s^{\rm(in,odd)}(x)=B\sin(p_\delta x).
\label{eq:inner_even_odd}
\end{equation}
The spectrum is determined by matching $\psi_s$ and $d\psi_s/dx$ at $x=\delta$. Defining the logarithmic derivative of the exterior solution by
\begin{equation}
\mathcal{L}(E,\delta)
\equiv
\left.
\frac{d}{dx}\ln \psi_s^{\rm(out)}(x)
\right|_{x=\delta^+},
\label{eq:log_derivative}
\end{equation}
the matching conditions become
\begin{equation}
\mathcal{L}(E,\delta)=-p_\delta\tan(p_\delta\delta)
\qquad
\text{(even states)},
\label{eq:even_matching}
\end{equation}
\begin{equation}
\mathcal{L}(E,\delta)=p_\delta\cot(p_\delta\delta)
\qquad
\text{(odd states)}.
\label{eq:odd_matching}
\end{equation}
Equations~\eqref{eq:even_matching} and~\eqref{eq:odd_matching} therefore provide the regularised quantisation conditions for the full one-dimensional problem.

\begin{figure}[htbp]
  \centering
  \includegraphics[width=\linewidth]{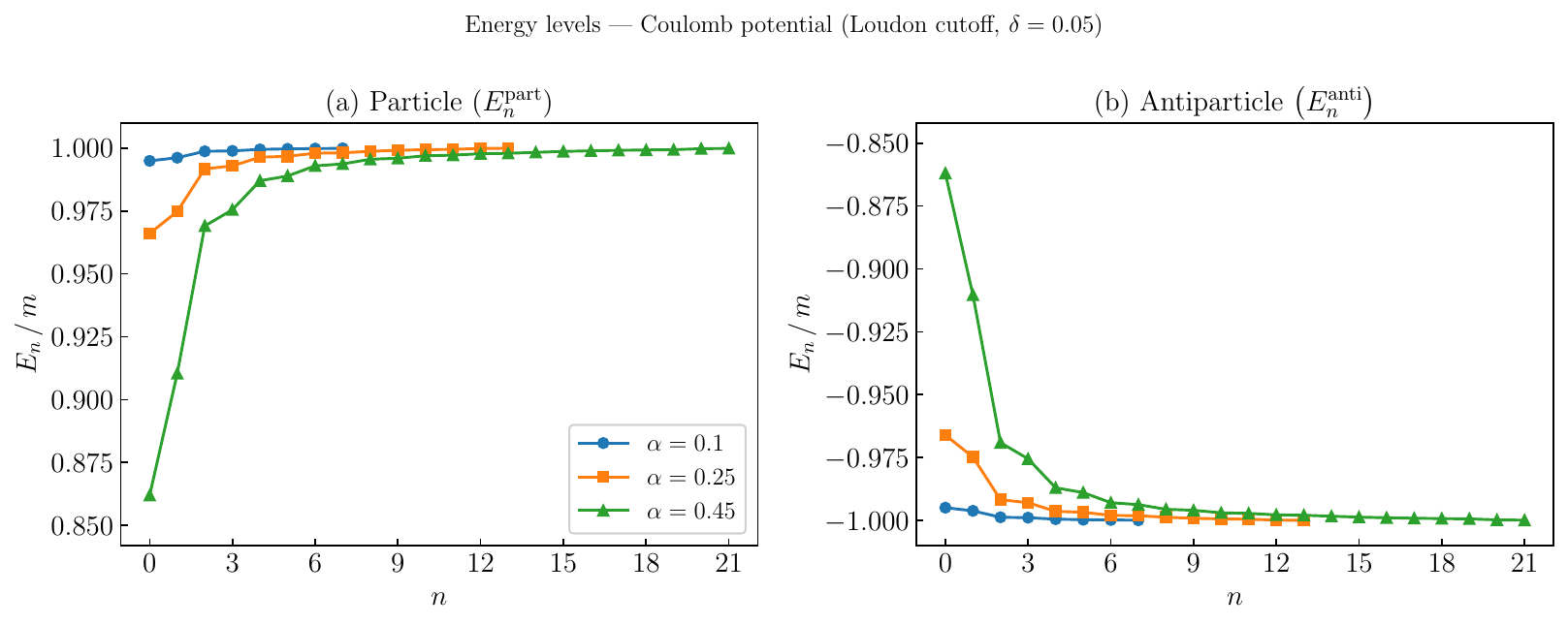}
  \caption{Bound-state energy spectrum of the regularised one-dimensional Coulomb potential in the Feshbach--Villars formalism, for $\delta = 0.05\,m^{-1}$ and three values of the coupling constant $\alpha$.
    \emph{Left panel}: particle eigenvalues $E_n^{\rm part}/m$ as a function of the principal quantum number $n$.
    \emph{Right panel}: antiparticle eigenvalues $E_n^{\rm anti}/m = -E_n^{\rm part}$, obtained from the charge-conjugation symmetry $E \to -E$, $\alpha \to -\alpha$ of Eq.~\protect\eqref{eq:fv_cutoff_master}.
    All levels lie inside the mass gap, $|E_n|<m$, and accumulate toward $|E|=m$ as $n$ increases. Deeply localised states (Loudon core states) are excluded from both panels, since they are artefacts of the finite cutoff and disappear in the limit $\delta\to 0^+$ (see text).}
  \label{fig:energy_levels}
\end{figure}

The structure of the bound-state spectrum that emerges from these quantisation conditions is illustrated in Fig.~\ref{fig:energy_levels}. The left panel shows the particle eigenvalues $E_n^{\rm part}/m$ as a function of the quantum number $n$ for three values of the coupling constant, $\alpha=0.10$, $0.25$, and $0.45$, with the cutoff fixed at $\delta = 0.05\,m^{-1}$. All levels lie inside the mass gap, $0 < E_n^{\rm part} < m$, and accumulate monotonically toward the continuum threshold $E = m$ as $n$ increases, reproducing the Rydberg-like compression of the spectrum at large $n$. The stronger the coupling, the more deeply the ground state is pulled away from the threshold. The right panel displays the antiparticle counterpart $E_n^{\rm anti}/m$, obtained from the particle spectrum through the charge-conjugation symmetry of the FV equation: the simultaneous substitutions $E \to -E$ and $\alpha \to -\alpha$ leave Eq.~\eqref{eq:fv_cutoff_master} invariant, so that $E_n^{\rm anti} = -E_n^{\rm part}$, with all levels in the range $-m < E_n^{\rm anti} < 0$.

\begin{figure}[htbp]
  \centering
  \includegraphics[width=0.7\linewidth]{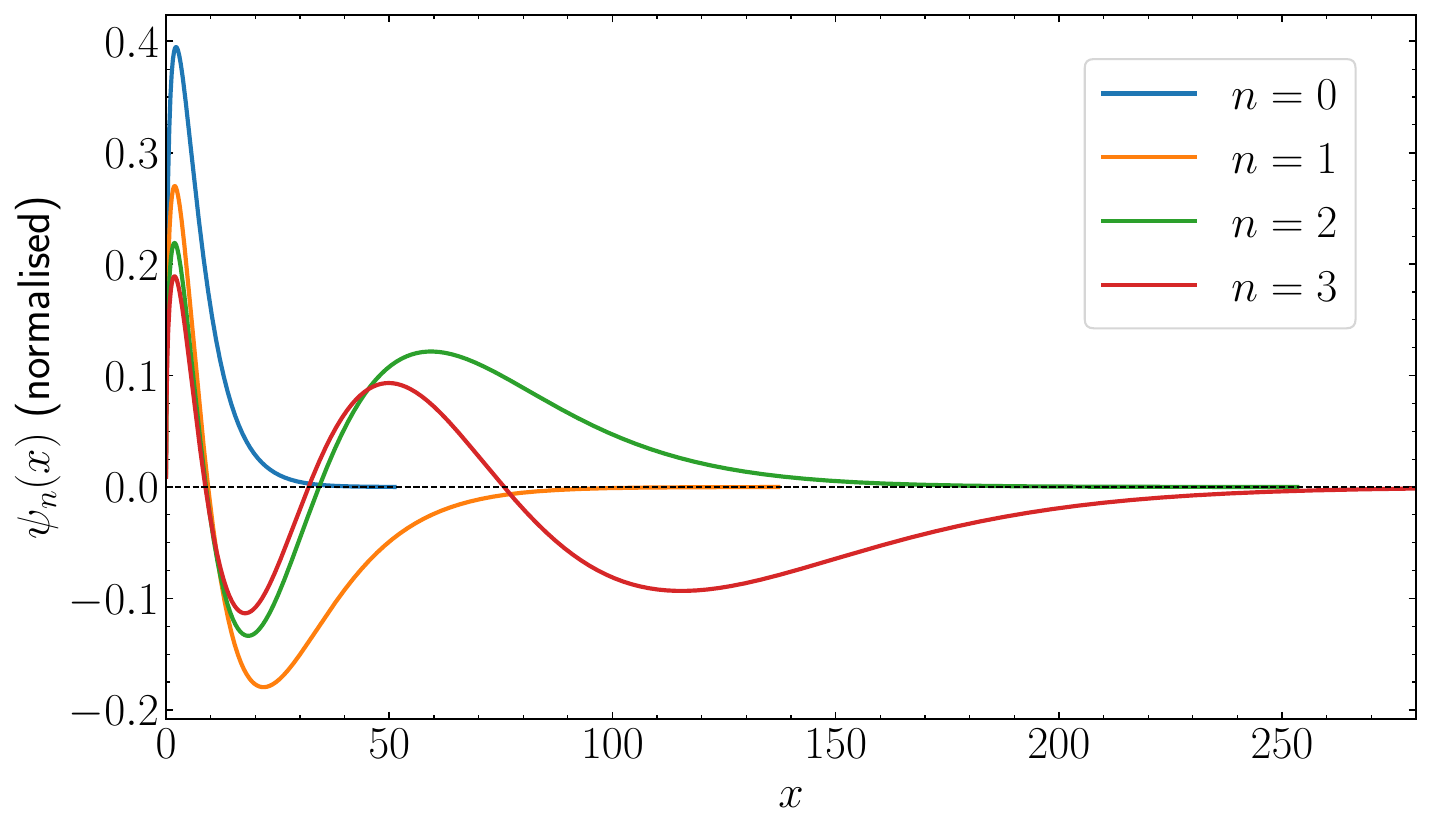}
  \caption{Normalised Coulomb wave functions $\psi_n(x)$ for the one-dimensional problem, with $\alpha = 0.20$ and $m=1$, shown for the four lowest states $n=0,1,2,3$. Each curve is plotted over a sufficiently large interval to capture both its oscillatory structure and asymptotic decay, and all of them are displayed together on a common axis for direct comparison. As $n$ increases, the wave functions extend over progressively larger spatial regions and develop the expected increase in the number of nodes, while their amplitudes become more spread out over the classically allowed region.}
  \label{fig:coulomb_wavefunctions}
\end{figure}

To complement the spectral information, it is useful to examine the spatial structure of the corresponding Coulomb eigenfunctions. This is illustrated in Fig.~\ref{fig:coulomb_wavefunctions}, which displays the normalised wave functions $\psi_n(x)$ for the four lowest states, $n=0,1,2,3$, at fixed coupling $\alpha = 0.20$. As expected, the ground state is nodeless and strongly concentrated near the origin, while the excited states develop an increasing number of nodes and extend over progressively larger spatial regions. This behaviour reflects the usual hierarchy of bound states in an attractive Coulomb-like interaction: higher values of $n$ correspond to less tightly bound states, with broader spatial support and a more pronounced oscillatory structure before the asymptotic decay sets in. When all profiles are shown on the same axis, one clearly sees the combined effect of the increasing radial extent and the growing number of oscillations as the excitation level rises.

\begin{figure}[htbp]
  \centering
  \includegraphics[width=\linewidth]{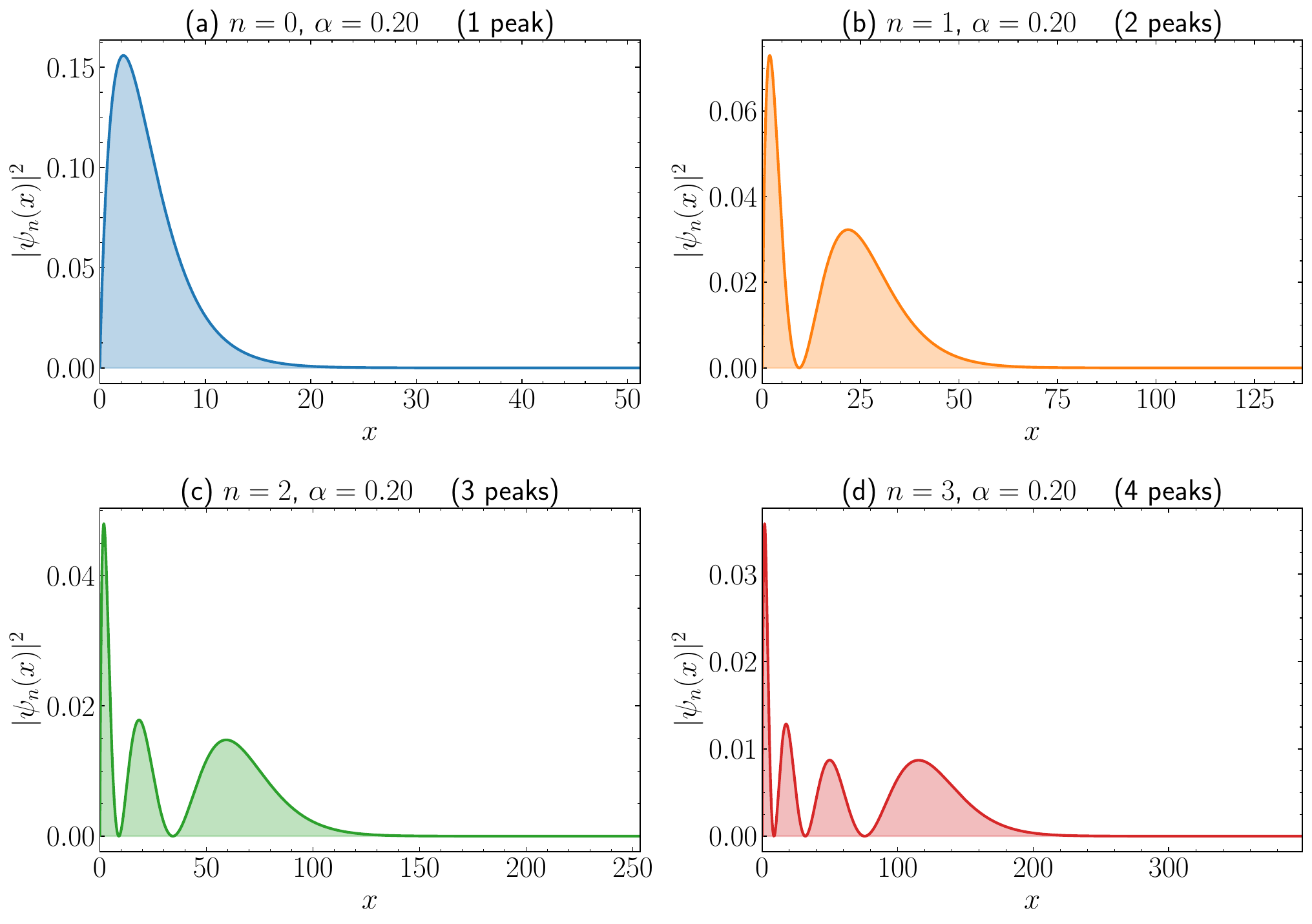}
  \caption{Probability densities $|\psi_n(x)|^2$ for the one-dimensional Coulomb problem with $\alpha = 0.20$ and $m=1$, for the four lowest states: (a) $n=0$, (b) $n=1$, (c) $n=2$, and (d) $n=3$. Each panel is displayed over an adaptive spatial interval chosen to make the full profile visible. The densities become progressively broader as $n$ increases and exhibit the corresponding increase in radial structure, with the number of local maxima growing from the ground state to the excited states. This behaviour reflects the increasingly extended character of the higher Coulomb eigenstates.}
  \label{fig:coulomb_probability_density}
\end{figure}

The corresponding probability densities are shown in Fig.~\ref{fig:coulomb_probability_density} for the same value $\alpha = 0.20$. The ground-state density is sharply localised near the origin and decreases monotonically with $x$, whereas the excited-state densities become progressively broader and develop the expected internal structure associated with the nodes of the wave functions. In particular, the number of local maxima increases with $n$, and the dominant support of the distribution is displaced toward larger values of $x$, indicating that the higher states are more spatially extended. Taken together, these panels provide a complementary visualisation of the bound-state hierarchy: while the wave functions exhibit alternating signs and nodes, the densities emphasise the outward spreading and increasing spatial complexity of the excited Coulomb states.

This regularised formulation clarifies several points that remain hidden in the strictly singular treatment. First, odd states are unproblematic: because $\psi_s(0)=0$, they are naturally protected from the core singularity and, in the limit $\delta\to 0^+$, they approach the usual Coulombic Balmer-type sequence. Second, even states are also perfectly well defined at finite $\delta$, but their existence is controlled by the matching condition~\eqref{eq:even_matching}; when the cutoff is sent to zero, the regularised even levels collapse onto the odd ones, thereby explaining the well-known pairwise odd--even degeneracy of the one-dimensional Coulomb problem in the cutoff limit, exactly as in Loudon's construction~\cite{Loudon1959,Loudon2016}. Third, the regularised problem contains an additional deeply localised even state whose probability density becomes increasingly concentrated inside the core as $\delta$ decreases. In Loudon's non-relativistic analysis, its binding energy diverges as $\delta\to 0^+$~\cite{Loudon1959,Loudon2016}; in the present relativistic FV setting, the same tendency indicates that the state is driven beyond the regime in which a single-particle interpretation can be trusted. For this reason, in what follows we focus on the regular bound-state branch continuously connected to the finite-gap sector $|E|<m$.

The cutoff construction thus provides the appropriate mathematical foundation for the Coulomb problem in one dimension. The singular equation is not taken as the starting point, but rather as the limiting form of a family of regular problems~\cite{Loudon1959,Loudon2016}. This has two important consequences for the present work. On the one hand, it justifies the use of parity as a good quantum number in the full-line problem and explains the emergence of odd--even degeneracy in the hard-core limit. On the other hand, it shows that the near-origin behaviour of the FV spinor components must be interpreted through a limiting procedure rather than by assigning an independent physical meaning to the formally singular point $x=0$.

Once the regularised eigenvalue $E_{n,\pm}(\delta)$ is obtained from Eqs.~\eqref{eq:even_matching} and~\eqref{eq:odd_matching}, the FV spinor components are reconstructed from Eqs.~\eqref{eq:psi1_from_psid} and~\eqref{eq:psi2_from_psid}:
\begin{equation}
\psi_1(x)=\frac{1}{2}
\left[1+\frac{E-eV_\delta(x)}{m}\right]\psi_s(x),
\qquad
\psi_2(x)=\frac{1}{2}
\left[1-\frac{E-eV_\delta(x)}{m}\right]\psi_s(x).
\label{eq:spinor_cutoff}
\end{equation}
Physical densities and observables are then computed at finite $\delta$ and only afterwards examined in the limit $\delta\to 0^+$.

\begin{figure}[htbp]
  \centering
  \includegraphics[width=\linewidth]{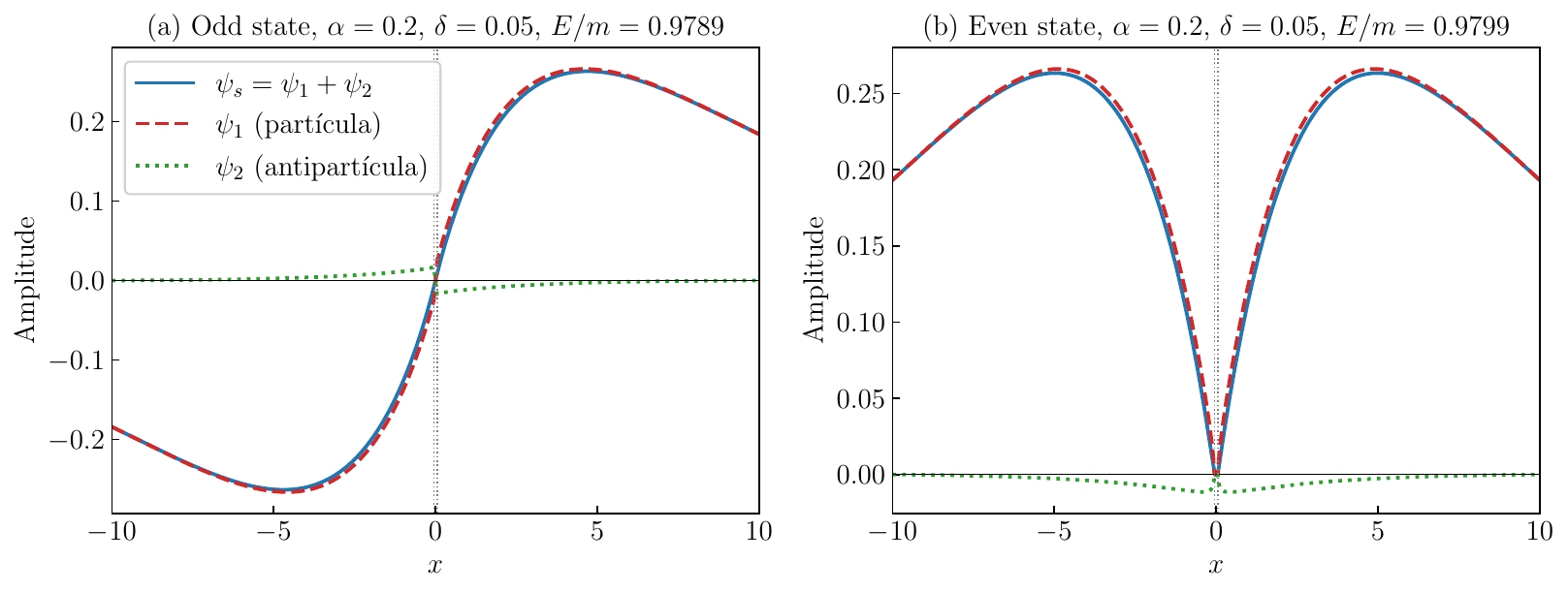}
  \caption{Feshbach--Villars spinor components for the regularised Coulomb potential with $\alpha = 0.20$ and $\delta = 0.05\,m^{-1}$.
    (a) lowest odd-parity state ($n=0$, odd). (b) lowest even-parity state ($n=0$, even).
    Solid blue: sum component $\psi_s = \psi_1 + \psi_2$.
    Dashed red: particle component $\psi_1$.
    Dotted green: antiparticle component $\psi_2$.
    The vertical dotted grey lines mark the cutoff boundary $x = \pm\delta$, across which the interior oscillatory solution is matched to the exterior Whittaker function.
    The particle component dominates throughout, while $\psi_2$ acquires a visible enhancement inside the core owing to the large constant potential $eV_\delta = \alpha/\delta$.}
  \label{fig:fv_components}
\end{figure}

The internal structure of the FV spinor for the regularised Coulomb potential is shown in Fig.~\ref{fig:fv_components}, for $\alpha = 0.20$ and $\delta = 0.05\,m^{-1}$. The left (right) panel corresponds to the lowest odd (even) state. In both cases the sum component $\psi_s$ (solid blue) is a smooth function that joins the oscillatory interior solution across $|x| = \delta$ (marked by vertical dashed lines) to the monotonically decaying Whittaker exterior. The particle component $\psi_1$ (red dashed) dominates in magnitude throughout the domain, while the antiparticle component $\psi_2$ (green dotted) remains smaller but non-zero and encodes the relativistic correction to the non-relativistic limit. Its spatial profile mirrors that of $\psi_1$, modulated by the local value of the potential through the factor $[1 - (E-eV_\delta)/m]$; the enhancement near the origin visible inside the core reflects the large value of $eV_\delta = \alpha/\delta$ in that region.

\begin{figure}[htbp]
  \centering
  \includegraphics[width=0.6\linewidth]{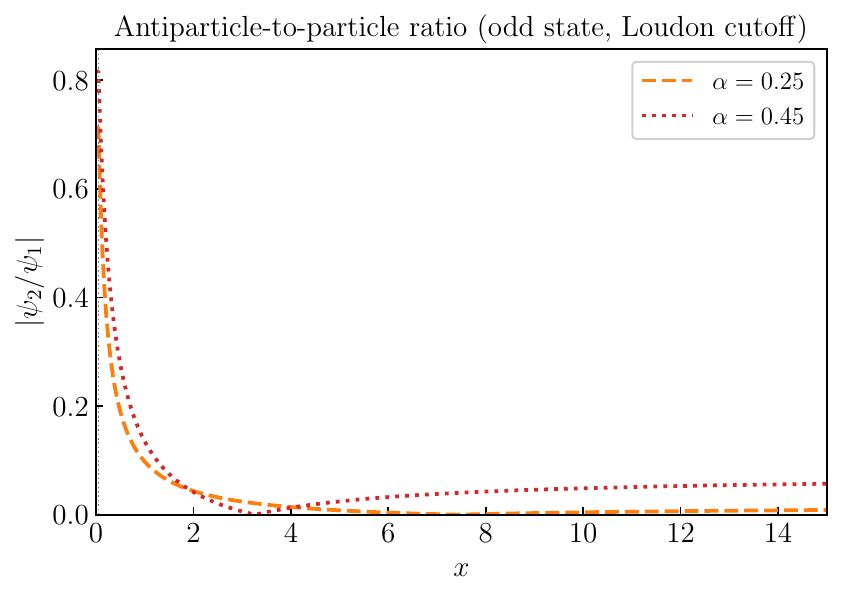}
  \caption{Ratio $|\psi_2/\psi_1|$ of the antiparticle to particle components as a function of $x > 0$, for the lowest odd state with $\delta = 0.05\,m^{-1}$ and three values of the coupling constant $\alpha$. The vertical dotted line marks the cutoff radius $x = \delta$. For $x > \delta$ the ratio is given analytically by Eq.~\protect\eqref{eq:psi2_psi1_ratio} and decreases monotonically from a maximum at the cutoff boundary toward the asymptotic value $(m-E_0)/(m+E_0)$ in magnitude at large $x$. Larger coupling produces a uniformly higher ratio, reflecting the increasing relativistic character of the more deeply bound states.}
  \label{fig:fv_ratio}
\end{figure}

A quantitative measure of the relativistic content of each state is provided by the ratio $|\psi_2/\psi_1|$, plotted in Fig.~\ref{fig:fv_ratio} for the lowest odd state and three values of $\alpha$. In the exterior region $x > \delta$, where $eV_\delta(x) = \alpha/x$, this ratio takes the explicit form
\begin{equation}
\left|\frac{\psi_2}{\psi_1}\right|
= \left|
\frac{1 - (E-\alpha/x)/m}{1 + (E-\alpha/x)/m}
\right|,
\label{eq:psi2_psi1_ratio}
\end{equation}
which is a monotonically decreasing function of $x$ for fixed $E$ and $\alpha$ in the parameter range considered here. At large distances, where $\alpha/x \ll E$, it approaches the constant non-relativistic limit $(m-E)/(m+E)$ in magnitude, so the ratio is suppressed whenever the binding energy $m-E$ is small. Near the cutoff, by contrast, the growing potential amplifies $\psi_2$ and the ratio increases. The figure confirms all three trends: for larger $\alpha$ the ratio is higher at every $x$ (stronger coupling implies stronger relativistic effects), it increases noticeably as $x \to \delta^+$, and it settles to a nearly constant value at large $x$.

\begin{figure}[htbp]
  \centering
  \includegraphics[width=\linewidth]{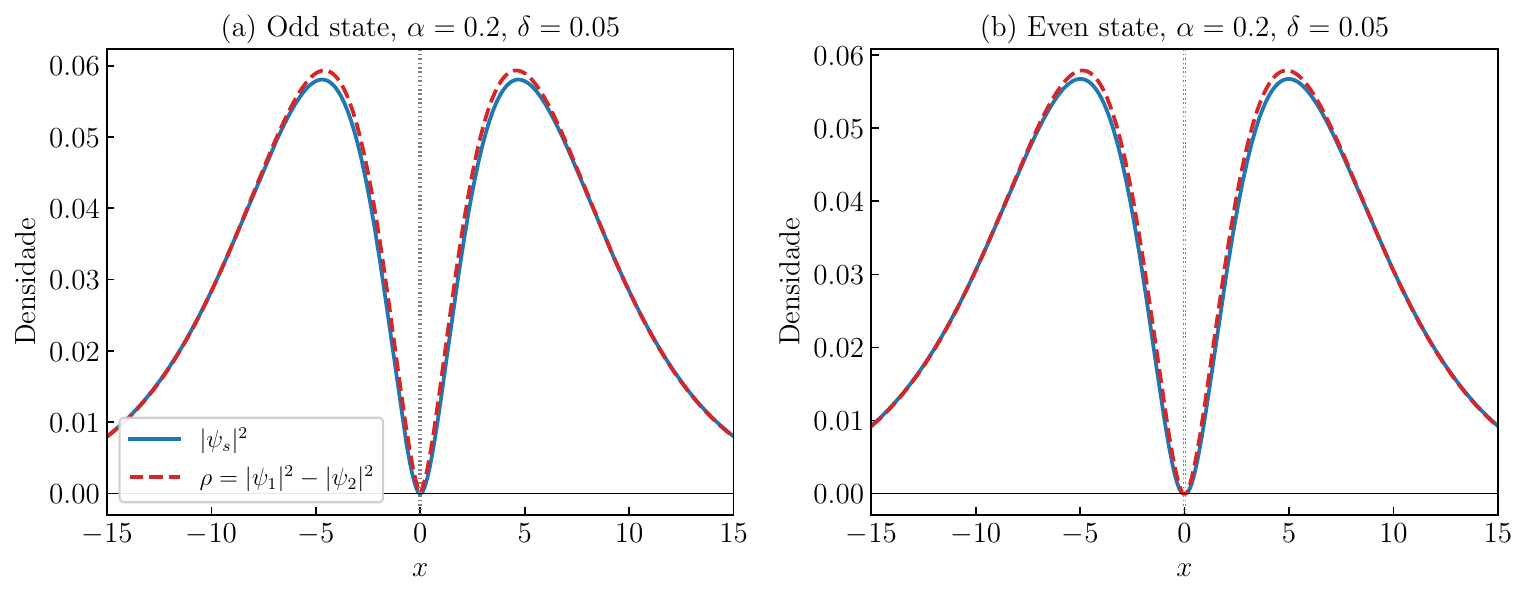}
  \caption{Comparison of the FV charge density $\rho = |\psi_1|^2 - |\psi_2|^2$ (dashed red) and the squared sum component $|\psi_s|^2$ (solid blue) for $\alpha = 0.20$ and $\delta = 0.05\,m^{-1}$.
    (a) lowest odd-parity state. (b) lowest even-parity state.
    Vertical dotted grey lines indicate the cutoff boundary $x = \pm\delta$. In the exterior region, the two curves nearly coincide, since the antiparticle amplitude is small there. Inside the core the constant potential $eV_\delta = \alpha/\delta$ produces a visible separation between $\rho$ and $|\psi_s|^2$. The density $\rho$ is non-negative throughout and satisfies the FV normalisation $\int\rho\,dx = 1$, confirming the consistency of the one-particle interpretation within the mass gap.}
  \label{fig:fv_density}
\end{figure}

Finally, Fig.~\ref{fig:fv_density} compares the FV charge density $\rho = |\psi_1|^2 - |\psi_2|^2$ with the squared sum component $|\psi_s|^2$ for the same parameters as Fig.~\ref{fig:fv_components}. The two quantities agree closely in the exterior region, where the relativistic correction encoded in $\psi_2$ is small, but differ visibly inside the core: the large constant value $eV_\delta = \alpha/\delta$ enhances the local component mixing and produces a noticeable separation between $\rho$ and $|\psi_s|^2$ for $|x|<\delta$. Crucially, $\rho$ remains non-negative for the positive-energy gap states considered here and satisfies the FV normalisation condition $\int_{-\infty}^{+\infty}\rho\,dx = 1$, confirming the consistency of the single-particle charge-density interpretation within the gap sector.

In numerical practice, one may solve the regularised problem for a sequence of decreasing cutoffs $\delta$, verify the convergence of the odd and even excited levels toward common limiting values, and monitor separately the collapse of the deeply localised even branch. This strategy removes the mathematical ambiguity at the origin and yields a clearer physical interpretation of the relativistic one-dimensional Coulomb problem within the Feshbach--Villars formalism.

\section{Power-Exponential Potential ($p=1$)}\label{s4}

The power-exponential potential belongs to a physically distinct class from the Coulomb interaction: whereas the Coulomb potential is long-ranged and singular at the origin, the exponential potential is short-ranged and bounded everywhere. This complementary character makes it a useful counterpart to the Coulomb case. We consider the general power-exponential form~\cite{p1,p2,p3,p4}
\begin{equation}
  V_{p}(x) = -V_0\exp\!\left[-\left(\frac{x}{x_0}\right)^{p}\right],
  \qquad V_0 > 0,\quad p \geq 1,
  \label{eq:exp_potential_general}
\end{equation}
and focus here on $p = 1$, which yields a simple decaying exponential and admits an exact reduction to Whittaker functions. With the shorthands $b \equiv eV_0$ and $q \equiv x_0$, the potential becomes
\begin{equation}
eV(x) = -b\,e^{-x/q}.
\end{equation}
For definiteness, we work on the half-line $x\ge 0$. The master equation~\eqref{eq:KG_sumfun} then becomes
\begin{equation}
  \frac{d^{2}\psi_s}{dx^{2}}
  + \left[b^{2}\,e^{-2x/q} + 2bE\,e^{-x/q}
    + \kappa^{2}\right]\psi_s = 0,
  \label{eq:exp_ode}
\end{equation}
where
\begin{equation}
  \kappa^{2} \equiv E^{2} - m^{2}.
\end{equation}

To reduce Eq.~\eqref{eq:exp_ode} to standard form, we first set
\begin{equation}
  \psi_s(x) = e^{x/(2q)}\,\mathcal{D}(x),
\end{equation}
and then introduce the change of variable
\begin{equation}
  z = 2ibq\,e^{-x/q},
  \label{eq:z_substitution}
\end{equation}
which maps $x\in(0,\infty)$ onto a curve in the complex $z$-plane. Under these substitutions, Eq.~\eqref{eq:exp_ode} transforms into the Whittaker equation~\cite{Abramowitz}
\begin{equation}
  \frac{d^{2}\mathcal{D}}{dz^{2}}
  + \left(-\frac{1}{4} + \frac{\mu_{W}}{z}
    + \frac{1-4\nu_{W}^{2}}{4z^{2}}\right)\mathcal{D} = 0,
  \label{eq:whittaker}
\end{equation}
with parameters
\begin{equation}
  \mu_{W} = -iEq,
  \qquad
  \nu_{W} = iq\kappa.
\end{equation}
The two independent solutions are the Whittaker functions $M_{\mu_{W},\nu_{W}}(z)$ and $W_{\mu_{W},\nu_{W}}(z)$. As $x \to \infty$, one has $z \to 0$, and the behaviour of $W_{\mu_{W},\nu_{W}}(z)$ near $z=0$ yields a divergent contribution. Requiring the physical solution to remain finite as $x\to\infty$ therefore eliminates that branch. Using the connection between the Whittaker $M$ function and the confluent hypergeometric function ${}_1F_1(a,b,z)$~\cite{Abramowitz}, the resulting solution may be written as
\begin{align}
  \psi_s(x) &= \mathcal{N}_{n}\,e^{x/(2q)}
    \!\left(2ibq\,e^{-x/q}\right)^{\!\frac{1}{2}+iq\kappa}
    \exp\!\left(-ibq\,e^{-x/q}\right)  {}_1F_1\!\left(\tfrac{1}{2}+iEq+iq\kappa,\;
    1+2iq\kappa,\; 2ibq\,e^{-x/q}\right),
  \label{eq:exp_explicit}
\end{align}
where $\mathcal{N}_n$ is a normalisation constant to be interpreted in a box-normalisation sense, or else over a finite plotting interval, since the states discussed below are not square-integrable in the usual sense.

For large $|z|$, the confluent hypergeometric function ${}_1F_1(a,b,z)$ grows exponentially unless the series terminates. Termination occurs when
\begin{equation}
  \tfrac{1}{2} + iEq + iq\kappa = -n,
  \qquad n = 0,1,2,\ldots,
  \label{eq:exp_quantisation}
\end{equation}
which yields the quantisation condition. Using $\kappa^{2}=E^{2}-m^{2}$, one obtains the corresponding spectrum
\begin{equation}
  E_{n} = \pm\,\frac{m^{2}q^{2}
    + \left|n + \tfrac{1}{2}\right|^{2}}
    {2q\,\left|n + \tfrac{1}{2}\right|}.
  \label{eq:exp_energy}
\end{equation}
The $\pm$ sign corresponds to the particle and antiparticle branches. The characteristic half-integer shift $\left|n+\tfrac{1}{2}\right|$ is a direct consequence of the Whittaker parameters and has no counterpart in the Coulomb spectrum. For large $n$, the energy grows approximately linearly,
\begin{equation}
E_n \approx \frac{|n+\tfrac{1}{2}|}{2q},
\end{equation}
whereas for fixed $n$ the levels decrease as $q$ increases. As written, Eq.~\eqref{eq:exp_energy} depends on $m$, $q$, and $n$, while the parameter $b$ affects the local spatial structure of the wave function through the argument of the hypergeometric function and the FV component mixing, rather than the discrete energies themselves.

\begin{figure}[h!]
\centering
\includegraphics[width=0.50\textwidth]{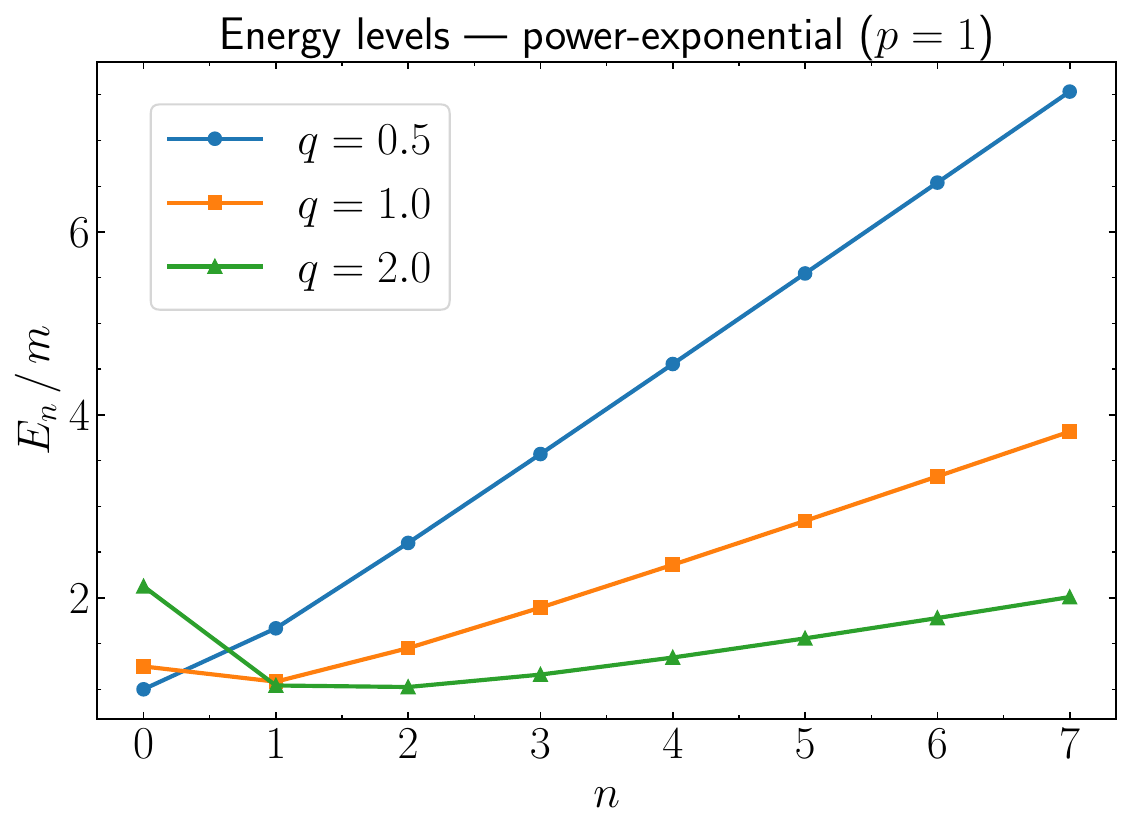}
\caption{Energy levels $E_{n}$ (particle branch) as a function of the quantum number $n$ for the power-exponential potential ($p=1$, $m=1$) and $q = 0.5$ (circles), $q = 1.0$ (squares), and $q = 2.0$ (triangles). The levels grow approximately linearly with $n$ for large $n$, in contrast to the Coulomb case. Smaller $q$ corresponds to a narrower interaction region and larger level spacings.}
\label{fig:energy_exp}
\end{figure}

The energy levels $E_{n}$ are displayed in Fig.~\ref{fig:energy_exp} for three values of $q$. For small $q$ (a narrower effective interaction region) the levels are more widely spaced; as $q$ increases, the potential broadens and the levels decrease. In all cases the levels grow with $n$, and the growth is faster for smaller $q$.

The complete two-component FV wave function is reconstructed by inserting $\psi_s(x)$ into Eqs.~\eqref{eq:psi1_from_psid} and~\eqref{eq:psi2_from_psid} with $eV(x)=-b\,e^{-x/q}$:
\begin{align}
  \Psi_{\mathrm{total}}(x) &= \frac{\mathcal{N}_{n}}{2}
  \begin{pmatrix}
    1 + \dfrac{1}{m}\bigl(E + b\,e^{-x/q}\bigr)\\[8pt]
    1 - \dfrac{1}{m}\bigl(E + b\,e^{-x/q}\bigr)
  \end{pmatrix}
  e^{x/(2q)}
  \!\left(2ibq\,e^{-x/q}\right)^{\!\frac{1}{2}+iq\kappa}
  e^{-ibq\,e^{-x/q}} \notag \\
  &\quad \times {}_1F_1\!\left(\tfrac{1}{2}+iEq+iq\kappa,\;
    1+2iq\kappa,\; 2ibq\,e^{-x/q}\right),
  \label{eq:exp_total_wf}
\end{align}
where $\mathcal{N}_{n}$ is fixed by the chosen box normalisation or by numerical normalisation over the displayed interval. The two-component spinor structure reflects the interplay between the energy $E$, the rest mass $m$, and the local value of the potential $b\,e^{-x/q}$: where the potential is stronger, the two components are more strongly mixed, while far from the source the mixing approaches the constant value set by $E/m$.

The two-component structure of the FV spinor for the power-exponential potential differs qualitatively from the Coulomb case and therefore deserves separate discussion. A key point follows immediately from Eq.~\eqref{eq:exp_energy}: for the particle branch one has $E_n>m$ for all $n$, so
\begin{equation}
\kappa = \sqrt{E^2-m^2}
\end{equation}
is real and positive. Accordingly, $\psi_s$ oscillates asymptotically rather than decaying exponentially. The corresponding stationary states are therefore delta-normalisable rather than $L^2$-normalisable, and the profiles shown below should be understood in a box-normalisation sense or as numerically normalised over a finite plotting window.

\begin{figure}[h!]
\centering
\includegraphics[width=0.90\textwidth]{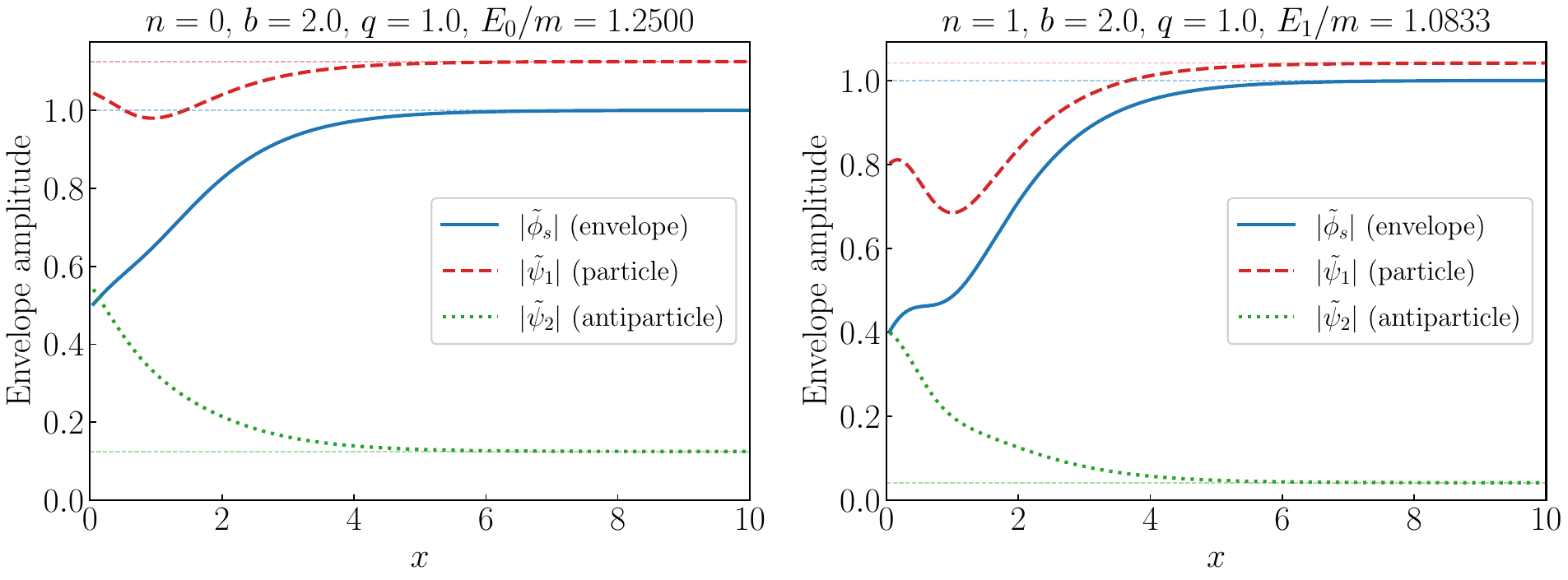}
\caption{Real parts of the FV spinor components $\psi_1$ (particle, red dashed), $\psi_2$ (antiparticle, green dotted), and the sum function $\psi_s=\psi_1+\psi_2$ (blue solid) for the power-exponential potential ($p=1$) with $b = eV_0 = 2$, $q = 1$, and $m = 1$. Left panel: $n = 0$, $E_0/m = 1.2500$. Right panel: $n = 1$, $E_1/m = 1.0833$. The wave functions are oscillatory throughout the domain because $E_n > m$ for all $n$, so $\kappa = \sqrt{E^2 - m^2}$ is real. The displayed profiles are normalised numerically over the plotted interval. The antiparticle component $\psi_2$ is largest near the origin, where the potential is deepest, and approaches a finite non-zero asymptotic amplitude ratio relative to $\psi_1$ as $x \to \infty$.}
\label{fig:pe_components}
\end{figure}

The individual components $\psi_1$ and $\psi_2$, reconstructed from Eqs.~\eqref{eq:psi1_from_psid} and~\eqref{eq:psi2_from_psid} with $eV(x)=-b\,e^{-x/q}$, are displayed in Fig.~\ref{fig:pe_components} for $n=0$ and $n=1$ with $b=2$ and $q=1$. The particle component $\psi_1$ dominates over $\psi_2$ throughout the domain, as expected for a positive-energy state, but the antiparticle component is appreciably larger near the origin than at large $x$. This spatial dependence is controlled by the local mixing factor $(E+b\,e^{-x/q})/m$: at the origin it takes the value $(E+b)/m$, reflecting the full depth of the potential well, and it decreases monotonically to $E/m$ as $x\to\infty$, where the potential vanishes. Unlike the Coulomb case, where the mixing diverges at the origin because of the $1/x$ singularity, here the mixing remains bounded everywhere and approaches a finite asymptotic value determined by $E/m$.

\begin{figure}[h!]
\centering
\includegraphics[scale=0.5]{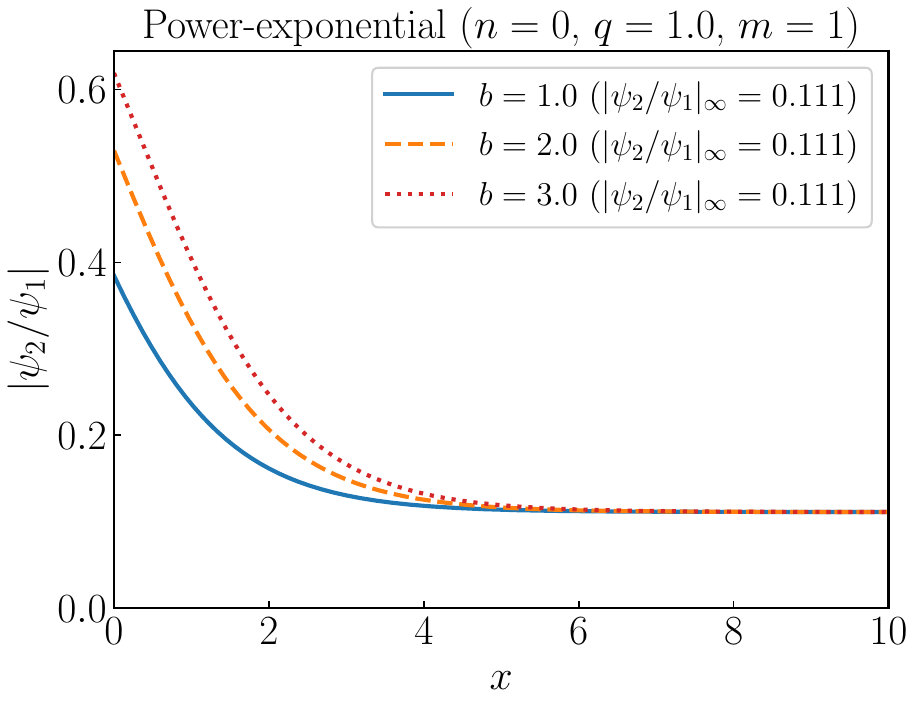}
\caption{Spatial profile of the antiparticle-to-particle ratio $|\psi_2/\psi_1|$ for the ground state ($n=0$, $q=1$) of the power-exponential potential and three values of the potential depth: $b = 1.0$ (blue solid), $b = 2.0$ (orange dashed), and $b = 3.0$ (red dotted). In all cases the ratio decreases monotonically from a maximum at $x=0$, where the potential reaches its full depth, to the common asymptotic plateau $\left|(1-E/m)/(1+E/m)\right|$ at large $x$. Increasing $b$ enhances the near-origin mixing, while leaving the asymptotic plateau unchanged for fixed $q$ and $n$. This behaviour contrasts sharply with the Coulomb case, where the ratio diverges near the origin and vanishes asymptotically.}
\label{fig:pe_ratio}
\end{figure}

The dependence of the particle--antiparticle mixing on the potential depth is shown in Fig.~\ref{fig:pe_ratio}, which displays the ratio $|\psi_2/\psi_1|$ for the ground state and three values of $b$. In all cases the ratio decreases monotonically from a maximum at $x=0$ to the asymptotic value
\begin{equation}
\left|\frac{1-E/m}{1+E/m}\right|
\end{equation}
as $x\to\infty$, reflecting the approach of the local mixing factor to its free-field limit. The spatial profile is smooth and monotonic, without the sharp near-origin divergence characteristic of the Coulomb case. Since Eq.~\eqref{eq:exp_energy} is independent of $b$, the asymptotic plateau is the same for fixed $q$ and $n$; changing $b$ affects only the near-origin part of the curve, where the local potential still contributes appreciably to the FV mixing.

\begin{figure}[h!]
\centering
\includegraphics[width=0.90\textwidth]{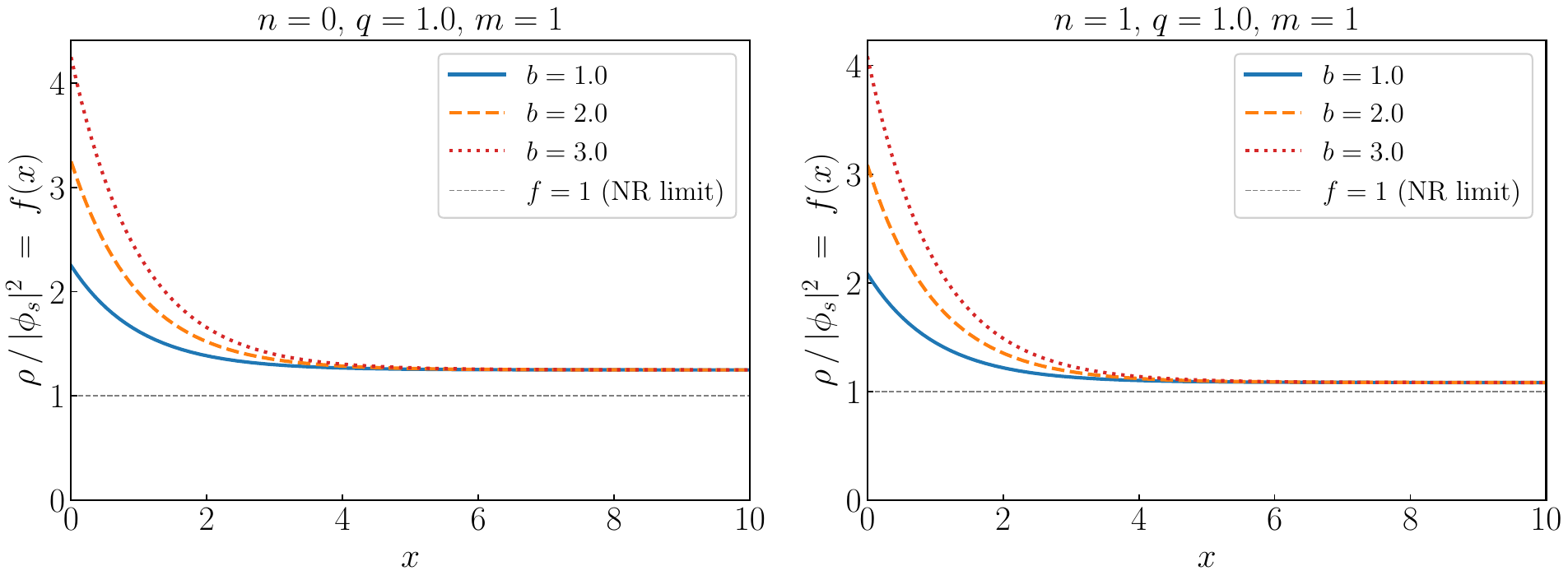}
\caption{Comparison between the squared sum component $|\psi_s|^2$ (blue solid) and the conserved FV charge density $\rho = |\psi_1|^2 - |\psi_2|^2$ (red dashed) for the power-exponential potential with $b = 2$, $q = 1$, and $m = 1$. Left panel: $n=0$. Right panel: $n=1$. Both quantities are computed from the numerically normalised profiles over the displayed interval and remain oscillatory at large $x$, consistent with the delta-normalisable character of the states. For the parameters shown, the conserved density $\rho$ lies above $|\psi_s|^2$ throughout the domain because the particle component remains dominant at all $x$.}
\label{fig:pe_density}
\end{figure}

A further consequence of the non-vanishing asymptotic mixing is visible in Fig.~\ref{fig:pe_density}, which compares $|\psi_s|^2$ with the conserved FV charge density $\rho = |\psi_1|^2 - |\psi_2|^2$ for $n=0$ and $n=1$. Both quantities display oscillatory behaviour and approach non-zero asymptotic envelopes over the plotted interval, consistent with the delta-normalisable character of the states. For the parameter range shown, $\rho$ lies systematically above $|\psi_s|^2$, reflecting the dominance of the particle component over the antiparticle component throughout the domain. The difference between the two densities is largest where the potential is deepest and the local mixing is strongest.

\section{Non-Relativistic Limit}\label{s5}

A fundamental consistency requirement for any relativistic quantum-mechanical framework is that it should reduce to the corresponding non-relativistic theory in the appropriate limit. For the FV equation, it is convenient to parameterise the energy as
\begin{equation}
E = m + \frac{\varepsilon}{m},
\qquad
|\varepsilon|\ll m^2,
\end{equation}
where $\varepsilon$ plays the role of the non-relativistic binding energy. One then expands the relevant spectral formulas in inverse powers of $m$.

For the Coulomb case,
\begin{equation}
k = \sqrt{m^{2}-E^{2}} \approx \sqrt{-2\varepsilon}
\end{equation}
to leading order. Substituting $E = m + \varepsilon/m$ into the Coulomb quantisation condition and expanding for large $m$, the binding energy satisfies
\begin{equation}
  \varepsilon_{n} \approx
  -\frac{\alpha^{2}m}{2\!\left(n + \dfrac{1}{2}
    + \dfrac{1}{2}\sqrt{1-4\alpha^{2}}\right)^{2}}.
  \label{eq:NR_Coulomb_general}
\end{equation}
For weak coupling, $\alpha \ll 1$, one has $\sqrt{1-4\alpha^{2}} \approx 1 - 2\alpha^{2}$, and Eq.~\eqref{eq:NR_Coulomb_general} reduces to
\begin{equation}
  \varepsilon_{n} \approx
  -\frac{\alpha^{2}m}{2(n+1)^{2}}
  \left[1 + \mathcal{O}(\alpha^{2})\right],
  \label{eq:NR_Coulomb_weak}
\end{equation}
which is precisely the hydrogen-like Bohr formula with principal quantum number $N=n+1$. This provides a non-trivial consistency check of the FV formalism: the relativistic Klein--Gordon equation, solved through the FV representation, reproduces the correct non-relativistic limit without any additional ad hoc assumption. It is also worth noting that the effective principal quantum number in the relativistic treatment,
\begin{equation}
n + \frac{1}{2} + \frac{1}{2}\sqrt{1-4\alpha^2},
\end{equation}
is not, in general, an integer. This fractional shift is a purely relativistic effect and disappears as $\alpha\to 0$, where the standard Bohr sequence is recovered.

\begin{figure}[h!]
\centering
\includegraphics[width=0.50\textwidth]{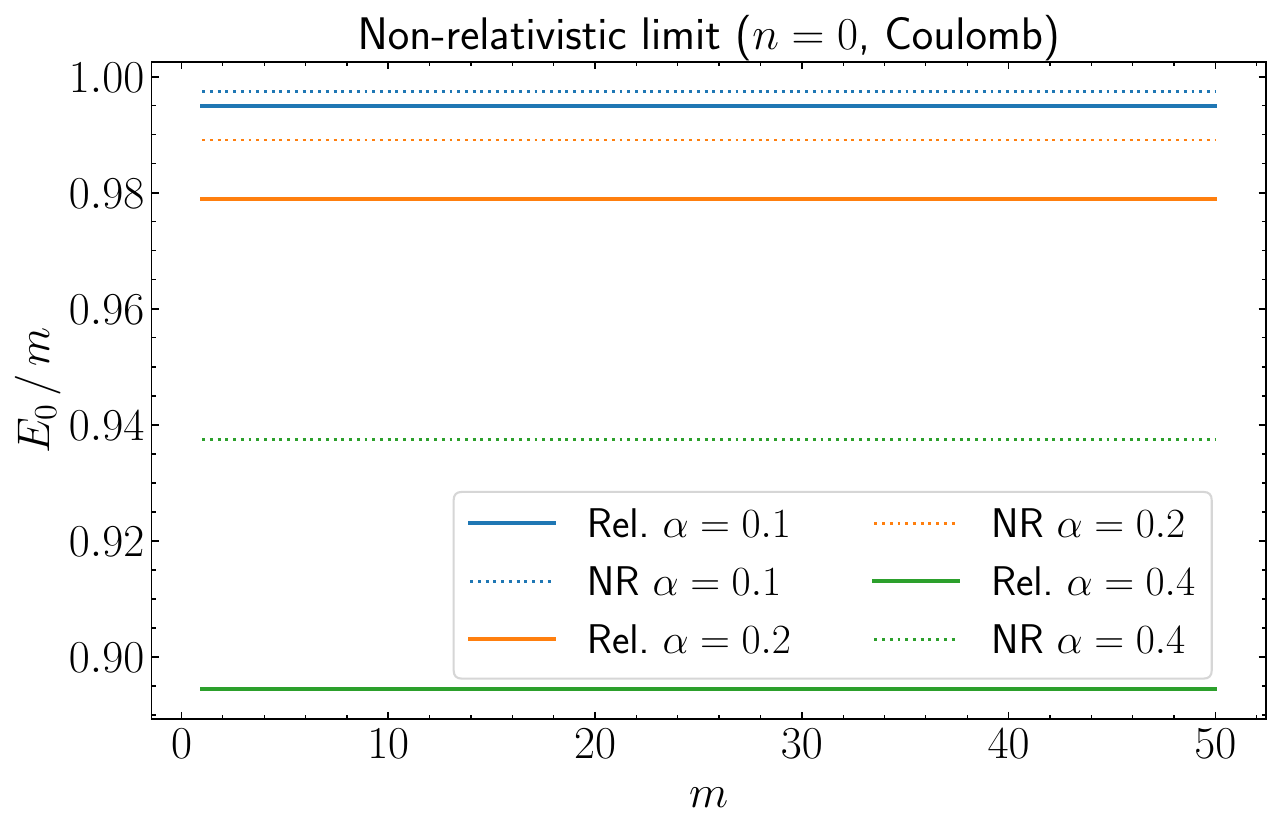}
\caption{Non-relativistic limit of the ground-state ($n=0$) Coulomb energy as a function of the mass $m$, for $\alpha = 0.10$ (blue), $0.20$ (orange), and $0.40$ (green). Solid lines: exact relativistic FV result $E_{0}/m$. Dotted lines: non-relativistic approximation $E_{\mathrm{NR}}/m = 1 - \alpha^{2}/(2m(n+1)^{2})$. The two curves converge for large $m$; convergence is faster for smaller $\alpha$.}
\label{fig:nr_limit}
\end{figure}

The convergence of the relativistic result to its non-relativistic limit is illustrated in Fig.~\ref{fig:nr_limit}, which shows the ground-state energy $E_{0}/m$ as a function of the mass $m$ for three values of $\alpha$. For small $m$ (strongly relativistic regime), the two results differ appreciably, but they converge rapidly as $m$ increases, and for $m \gtrsim 10$ they are essentially indistinguishable. The convergence is faster for smaller $\alpha$ because the relativistic correction to the binding energy is controlled by the ratio $\alpha^{2}/m^{2}$: for small $\alpha$ this ratio is already negligible at moderate $m$, whereas for $\alpha = 0.40$ the corrections remain perceptible over a wider range.

For the power-exponential case, the non-relativistic limit is more subtle. Using Eq.~\eqref{eq:exp_energy} and expanding for large $m$, one finds that the corresponding energies remain above the mass threshold, so that the stationary states discussed in Sec.~\ref{s4} do not reduce to conventional Schr\"odinger bound states. Equivalently, because $\kappa^2 = E^2 - m^2$ remains positive, the asymptotic behaviour is oscillatory rather than exponentially decaying, and the solutions are delta-normalisable rather than square-integrable. In this sense, the $p=1$ family treated here is intrinsically relativistic: a standard Schr\"odinger bound-state limit is not recovered unless a different scaling of the potential parameters is introduced.

The contrast with the Coulomb case is therefore instructive. Whereas the Coulomb spectrum connects smoothly to the hydrogenic Schr\"odinger result in the weak-coupling and large-mass regime, the power-exponential model considered here does not possess a standard non-relativistic bound-state counterpart in the same parameter scaling. The FV treatment thus reveals an important structural difference between the two cases: the Coulomb branch admits a conventional non-relativistic interpretation, while the exponential branch, in the form analysed here, remains essentially relativistic.

\section{Cornell Potential}\label{s6}

The Cornell potential combines a short-range Coulomb term with a long-range linear confining term and plays a standard phenomenological role in the description of quark--antiquark bound states. In the full-line formulation, however, the one-dimensional Coulombic part of the interaction, $V(x)\propto 1/|x|$, is singular at the origin. To treat this difficulty in a mathematically controlled way and to make the parity structure of the spectrum explicit, we follow the same Loudon-type regularisation strategy adopted in Section~\ref{s3}~\cite{Loudon1959,Loudon2016}. We therefore replace the singular Cornell interaction by the regularised even potential
\begin{equation}
V_{\mathrm{C},\delta}(x)=
\begin{cases}
\dfrac{a}{\delta}+b\delta, & |x|<\delta,\\[6pt]
\dfrac{a}{|x|}+b|x|, & |x|\ge \delta,
\end{cases}
\qquad a>0,\quad b>0,
\label{eq:cornell_cutoff_potential}
\end{equation}
where $a$ is the Coulomb coupling, $b$ is the string tension, and $\delta>0$ is a small cutoff length. Defining, as before, $\alpha \equiv ea$ and $\beta \equiv eb$, the FV master equation~\eqref{eq:KG_sumfun} becomes
\begin{equation}
\frac{d^2\psi_s}{dx^2}
+\left[\bigl(E-eV_{\mathrm{C},\delta}(x)\bigr)^2-m^2\right]\psi_s=0.
\label{eq:fv_cornell_cutoff_master}
\end{equation}
Because $V_{\mathrm{C},\delta}(x)$ is even, the eigenstates can be classified by parity,
\begin{equation}
\psi_s(-x)=\pm\psi_s(x),
\label{eq:cornell_parity_condition}
\end{equation}
with the plus (minus) sign corresponding to even (odd) states.

\subsection{Interior Region ($|x|<\delta$)}

Inside the regularised core the potential is constant and Eq.~\eqref{eq:fv_cornell_cutoff_master} reduces to
\begin{equation}
\frac{d^2\psi_s}{dx^2}+p_\delta^2\,\psi_s=0,
\qquad
p_\delta^2=
\left(E-\frac{\alpha}{\delta}-\beta\delta\right)^2-m^2.
\label{eq:cornell_inner_eq}
\end{equation}
For sufficiently small $\delta$, the Coulomb term $\alpha/\delta$ dominates, so that $p_\delta^2>0$, and the regular interior solutions classified by parity are
\begin{equation}
\psi_s^{\rm(in,even)}(x)=A\cos(p_\delta x),
\qquad
\psi_s^{\rm(in,odd)}(x)=B\sin(p_\delta x),
\label{eq:cornell_inner_even_odd}
\end{equation}
where $A$ and $B$ are constants fixed by continuity at $x=\delta$. When $p_\delta^2<0$, the trigonometric functions are replaced by their hyperbolic counterparts; the matching conditions below accommodate both cases in a unified way.

\subsection{Exterior Region ($|x|\ge\delta$)}

In the exterior region, the master equation takes the form
\begin{equation}
\frac{d^2\psi_s}{dx^2}
+\left[\left(E-\frac{\alpha}{|x|}-\beta|x|\right)^2-m^2\right]\psi_s=0.
\label{eq:cornell_outer_eq}
\end{equation}
We factor out the asymptotic behaviour by writing
\begin{equation}
\psi_s(x)=|x|^{\tilde{\mu}+1/2}e^{-\beta x^2/2}\chi(x),
\qquad
\tilde{\mu}=\frac{1}{2}\sqrt{1-4\alpha^2},
\label{eq:cornell_outer_ansatz}
\end{equation}
and define
\begin{equation}
\mu\equiv \tfrac{1}{2}+\tilde{\mu}.
\label{eq:cornell_mu_definition}
\end{equation}
Introducing the variable $\xi=\sqrt{2\beta}\,|x|$, the equation for $\chi$ is transformed into Kummer's confluent hypergeometric equation. Since the singular point $x=0$ is excluded from the exterior region, neither independent solution needs to be discarded on regularity grounds. The general exterior solution therefore reads
\begin{equation}
\psi_s^{\rm(out)}(x)=
|x|^{\tilde{\mu}+1/2}e^{-\beta x^2/2}
\left[
C_1\,M\!\left(a_{\mathrm{K}},b_{\mathrm{K}},\beta x^2\right)
+
C_2\,U\!\left(a_{\mathrm{K}},b_{\mathrm{K}},\beta x^2\right)
\right],
\label{eq:cornell_outer_general}
\end{equation}
where $M$ and $U$ are the confluent hypergeometric functions of the first and second kind, respectively, and
\begin{equation}
a_{\mathrm{K}}
=
\frac{\mu+1}{2}
+\frac{m^2-E^2}{4\beta}
-\frac{E\alpha}{\sqrt{2\beta}},
\qquad
b_{\mathrm{K}}
=
\tilde{\mu}+\frac{1}{2}.
\label{eq:cornell_kummer_params_cutoff}
\end{equation}

\paragraph{Normalisability.}
For large $z=\beta x^2$, the Kummer function satisfies $M(a,b,z)\sim e^z z^{a-b}$ and therefore grows exponentially, so square-integrability requires $C_1=0$. By contrast, the Tricomi function behaves as $U(a,b,z)\sim z^{-a}$ for large $z$; any resulting polynomial factor is absorbed by the Gaussian envelope $e^{-\beta x^2/2}$, leaving a normalisable bound-state solution throughout the parameter range of interest. The physical exterior solution is therefore
\begin{equation}
\psi_s^{\rm(out)}(x)=
\mathcal{C}\,
|x|^{\tilde{\mu}+1/2}e^{-\beta x^2/2}
U\!\left(a_{\mathrm{K}},\,b_{\mathrm{K}},\,\beta x^2\right),
\label{eq:cornell_outer_physical}
\end{equation}
with $\mathcal{C}$ a normalisation constant. We emphasise that, in the present cutoff formulation, the quantisation condition is provided by the matching procedure derived below, not by the truncation condition $a_{\mathrm{K}}=-n$, which arises only in the singular half-line problem where regularity at the origin enforces a different boundary condition.

\subsection{Matching Conditions}

The spectrum is determined by matching $\psi_s$ and its first derivative at $x=\delta$. Defining the logarithmic derivative of the exterior solution by
\begin{equation}
\mathcal{L}_{\mathrm{C}}(E,\delta)
\equiv
\left.
\frac{d}{dx}\ln \psi_s^{\rm(out)}(x)
\right|_{x=\delta^+}
=
\frac{\tilde{\mu}+\tfrac{1}{2}}{\delta}
-\beta\delta
-2\beta\delta\,
\frac{a_{\mathrm{K}}\,
U\!\left(a_{\mathrm{K}}+1,\,b_{\mathrm{K}}+1,\,\beta\delta^2\right)}
{U\!\left(a_{\mathrm{K}},\,b_{\mathrm{K}},\,\beta\delta^2\right)},
\label{eq:cornell_log_derivative}
\end{equation}
where we used $dU/dz = -a_{\mathrm{K}}\,U(a_{\mathrm{K}}+1,b_{\mathrm{K}}+1,z)$, the matching conditions become
\begin{equation}
\mathcal{L}_{\mathrm{C}}(E,\delta)
=
-p_\delta\tan(p_\delta\delta)
\qquad
\text{(even states)},
\label{eq:cornell_even_matching}
\end{equation}
and
\begin{equation}
\mathcal{L}_{\mathrm{C}}(E,\delta)
=
p_\delta\cot(p_\delta\delta)
\qquad
\text{(odd states)}.
\label{eq:cornell_odd_matching}
\end{equation}
When $p_\delta^2<0$, the trigonometric functions are replaced by their hyperbolic counterparts. Equations~\eqref{eq:cornell_even_matching} and~\eqref{eq:cornell_odd_matching} thus define the regularised bound-state problem for the full one-dimensional Cornell potential.

A numerical subtlety arises in the root-finding step: the logarithmic derivative $\mathcal{L}_{\mathrm{C}}$ has poles whenever
\begin{equation}
U\!\left(a_{\mathrm{K}}(E),b_{\mathrm{K}},\beta\delta^2\right)=0,
\end{equation}
and a naive sign-change search may mistake these poles for roots of the matching equations. In the numerical implementation we therefore require
\begin{equation}
\left|U\!\left(a_{\mathrm{K}},b_{\mathrm{K}},\beta\delta^2\right)\right|>\varepsilon,
\qquad
\varepsilon=10^{-3},
\end{equation}
at the endpoints of every candidate bracket, and we additionally verify
\begin{equation}
\left|\mathcal{L}_{\mathrm{C}}-\mathcal{L}_{\rm in}\right|<0.1
\end{equation}
at the refined root before accepting an eigenvalue. Without this pole-rejection step, spurious states may appear in the numerical scan. For the parameter range examined here, no bound states are found in the negative-energy sector: because $eV(x)>0$ everywhere, the effective potential $(E-eV)^2-m^2$ does not generate a confining exterior region for $E<0$.

\subsection{Spectrum and Parity Pairing}

\begin{figure}[t]
  \centering
  \includegraphics[width=0.86\textwidth]{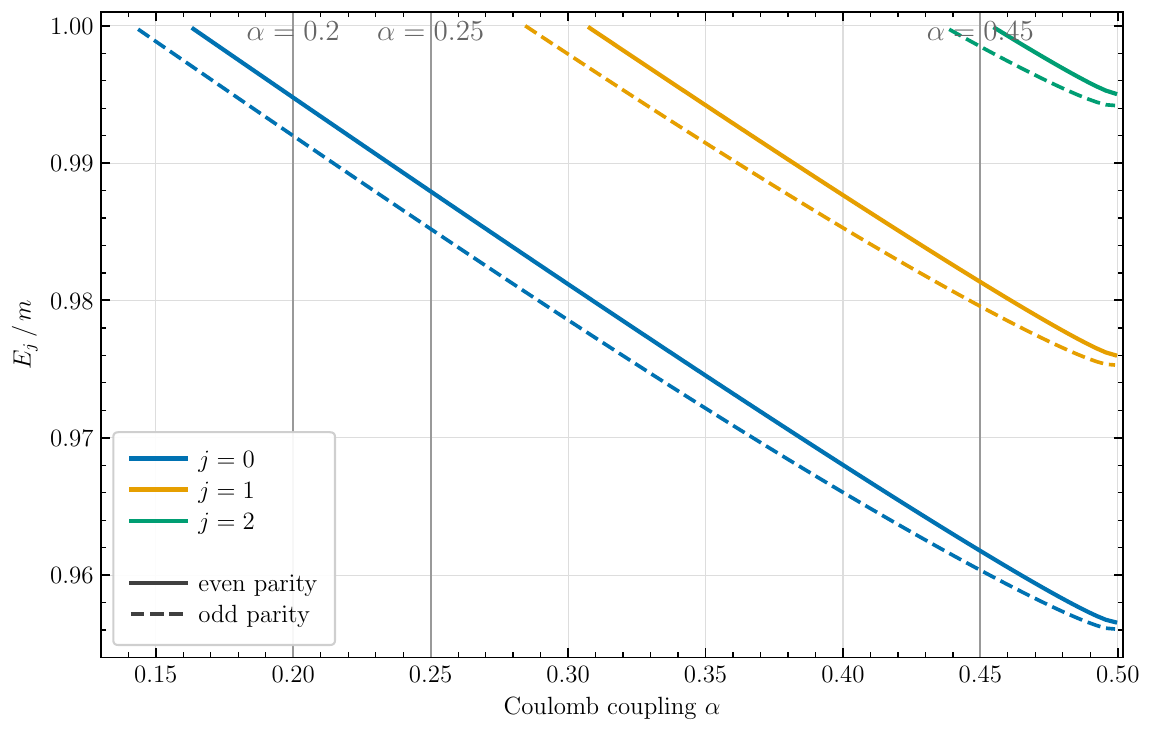}
  \caption{Positive-energy eigenvalues $E_j/m$ of the regularised Cornell potential~\eqref{eq:cornell_cutoff_potential} as a function of the Coulomb coupling $\alpha$, for string tension $\beta=0.01$, cutoff $\delta=0.10$, and mass $m=1$. Solid (dashed) curves denote even (odd) parity states; colour distinguishes the level index $j=0$ (blue), $j=1$ (orange), and $j=2$ (green). Vertical grey lines mark the reference couplings $\alpha\in\{0.20,0.25,0.45\}$. Energies are obtained from the matching conditions~\eqref{eq:cornell_even_matching} and~\eqref{eq:cornell_odd_matching} with pole rejection applied to the root finder (see text). The near-degeneracy between same-colour solid and dashed curves reflects the even--odd pairing inherited from the half-line limit $\delta\to0^+$. Each level enters the spectrum from above ($E\to m^-$) at a threshold coupling indicated by the left endpoint of the corresponding curve.}
  \label{fig:cornell_levels}
\end{figure}

Figure~\ref{fig:cornell_levels} shows how the positive-energy eigenvalues $E_j/m$ evolve with the Coulomb coupling $\alpha$ for $\beta=0.01$ and $\delta=0.10$. Solid curves correspond to even-parity states and dashed curves to odd-parity states, while colour distinguishes the level index $j$. The three vertical reference lines mark the coupling values $\alpha\in\{0.20,0.25,0.45\}$ discussed explicitly below.

Several features are immediately apparent. All bound-state energies cluster just below the continuum threshold $E=m$, reflecting the combined effect of the weak string tension $\beta=0.01$ and the Coulomb attraction. New levels enter the spectrum from above ($E\to m^-$) as $\alpha$ increases: the lowest odd-parity state first appears at $\alpha\approx0.14$, the lowest even-parity state at $\alpha\approx0.16$, and the first excited pair enters around $\alpha\approx0.29$--$0.31$. At $\alpha=0.20$ and $\alpha=0.25$ only one state per parity sector is present; at $\alpha=0.45$ two even and three odd roots are found, of which the four lowest distinct states are used in the wave-function figures below.

A key structural feature is the \emph{near-degeneracy} of even and odd levels in pairs. Each solid curve lies within $\Delta E\sim10^{-3}\,m$ of a dashed curve of the same colour throughout the displayed range. For $\alpha=0.45$, the two lowest pairs are
\begin{align}
E^{(\mathrm{o})}_0 &= 0.960356\,m, &
E^{(\mathrm{e})}_0 &= 0.961761\,m, &
\Delta E_0 &= 1.41\times10^{-3}\,m,
\label{eq:cornell_pair1}
\\
E^{(\mathrm{o})}_1 &= 0.979567\,m, &
E^{(\mathrm{e})}_1 &= 0.981343\,m, &
\Delta E_1 &= 1.78\times10^{-3}\,m.
\label{eq:cornell_pair2}
\end{align}
This even--odd pairing is a remnant of the exact two-fold degeneracy of the half-line problem in the limit $\delta\to0^+$, where the even and odd full-line states become degenerate copies of the same half-line solution. At finite $\delta$, the regularised core breaks this symmetry and lifts the degeneracy by a small amount that vanishes as $\delta\to0^+$.

\subsection{Wave Functions}\label{s6.4}

Once an eigenvalue has been obtained from the matching conditions, the full-line wave function is reconstructed piecewise. In the interior ($|x|<\delta$), the solution is the even or odd trigonometric (or hyperbolic) function anchored to the exterior value at $x=\delta$:
\begin{equation}
\psi_s^{\rm(in)}(x)
=
\begin{cases}
\dfrac{\psi_s^{\rm(out)}(\delta)}{\cos(p_\delta\delta)}\,
  \cos(p_\delta x), & \text{even, } p_\delta^2>0,\\[6pt]
\dfrac{\psi_s^{\rm(out)}(\delta)}{\sin(p_\delta\delta)}\,
  \sin(p_\delta x), & \text{odd, } p_\delta^2>0,\\[6pt]
\dfrac{\psi_s^{\rm(out)}(\delta)}{\cosh(q_\delta\delta)}\,
  \cosh(q_\delta x), & \text{even, } p_\delta^2<0,\\[6pt]
\dfrac{\psi_s^{\rm(out)}(\delta)}{\sinh(q_\delta\delta)}\,
  \sinh(q_\delta x), & \text{odd, } p_\delta^2<0,
\end{cases}
\label{eq:cornell_interior_wf}
\end{equation}
where $q_\delta=\sqrt{-p_\delta^2}$. In the exterior ($|x|\ge\delta$), the solution is given by Eq.~\eqref{eq:cornell_outer_physical}, with the appropriate parity sign for negative $x$:
\begin{equation}
\psi_s^{\rm(out)}(x)
=
\mathcal{C}\,|x|^{\tilde{\mu}+1/2}e^{-\beta x^2/2}
U\!\left(a_{\mathrm{K}},b_{\mathrm{K}},\beta x^2\right)
\times
\begin{cases}
+1, & \text{even},\\
\operatorname{sgn}(x), & \text{odd}.
\end{cases}
\label{eq:cornell_exterior_wf_signed}
\end{equation}
The constant $\mathcal{C}$ is fixed by $L^2$ normalisation over the full line.

For $\alpha=0.45$, $\beta=0.01$, $\delta=0.10$, the corrected numerical analysis yields the four lowest distinct bound states shown below: odd $j=0$, even $j=0$, odd $j=1$, and even $j=1$. Two practical points are worth noting. First, for excited states with sufficiently negative $a_{\mathrm{K}}$, the Tricomi function may develop a zero in the exterior region; the corresponding sign change of $\psi_s^{\rm(out)}$ is physical and reflects the excited-state character of the solution. Second, the probability density $|\psi_s|^2$ reaches its maxima around $|x|\simeq 11$--$21$, depending on the state, so a domain extending to $|x|\le 60$ is required to capture more than $99.99\%$ of the norm for all states shown.

\begin{figure}[t]
  \centering
  \includegraphics[width=0.88\textwidth]{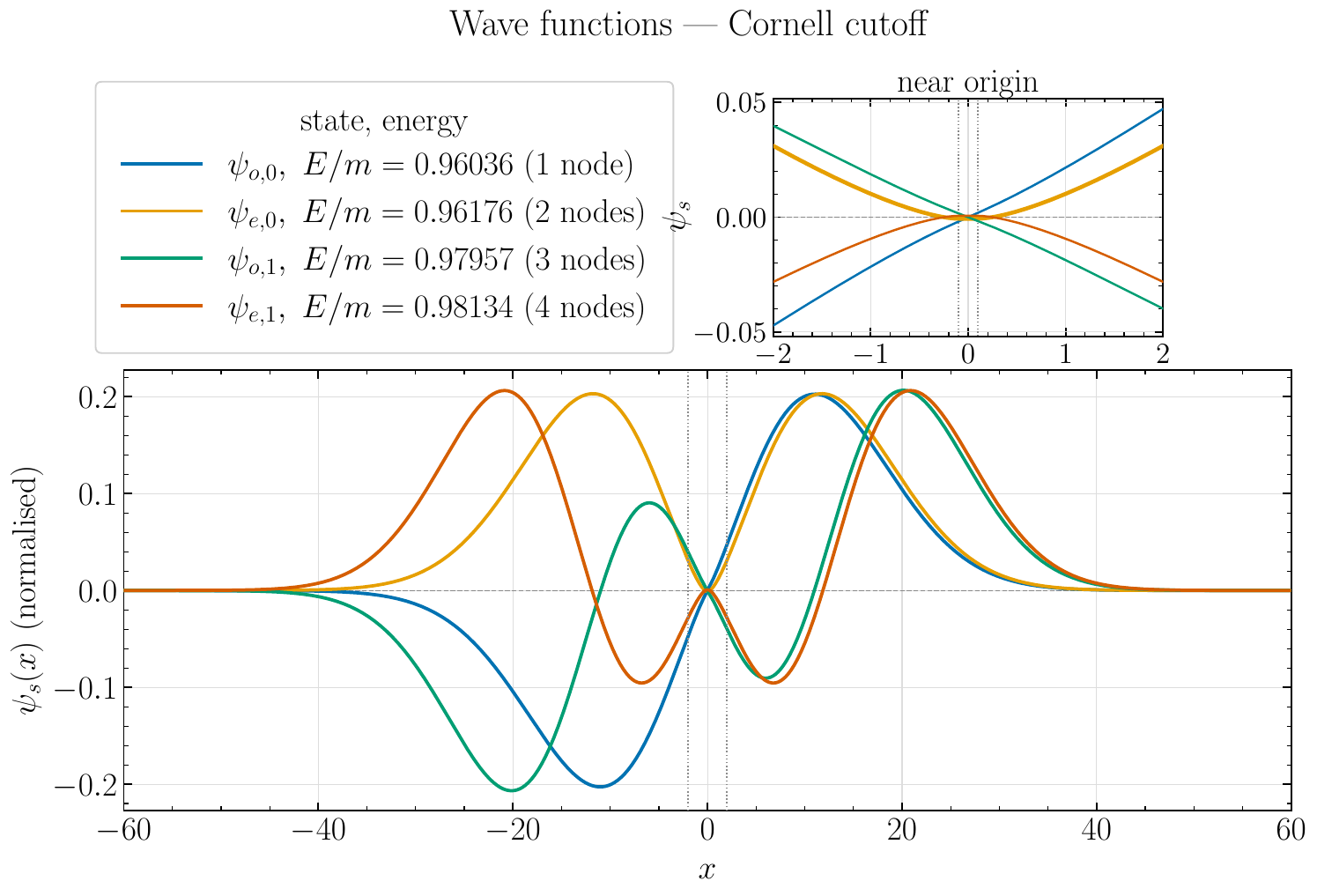}
  \caption{Normalised full-line wave functions $\psi_s(x)$ for the four lowest distinct states of the regularised Cornell potential, with $\alpha=0.45$, $\beta=0.01$, $\delta=0.10$, and $m=1$. The number of nodes of $\psi_s$ is indicated in the legend. Inset: region $|x|\le2$, showing the unresolved near-origin nodes of the even-parity states at $|x|\approx0.25$ (arrow). Dotted vertical lines in the inset mark the regularisation boundary $x=\pm\delta$. The dotted vertical lines on the main panel at $x=\pm2$ delimit the inset domain. The states, in order of increasing energy, are odd $j=0$ ($E/m=0.960356$, $a_{\mathrm{K}}=-0.254$, 1 node), even $j=0$ ($E/m=0.961761$, $a_{\mathrm{K}}=-0.326$, 2 nodes), odd $j=1$ ($E/m=0.979567$, $a_{\mathrm{K}}=-1.247$, 3 nodes), and even $j=1$ ($E/m=0.981343$, $a_{\mathrm{K}}=-1.340$, 4 nodes). The domain $|x|\le60$ captures more than $99.99\%$ of the norm for each state.}
  \label{fig:cornell_wf}
\end{figure}

Figure~\ref{fig:cornell_wf} displays the four normalised wave functions on the full line, together with a near-origin inset ($|x|\le2$). The two lowest states (odd $j=0$, even $j=0$) form a near-degenerate pair with splitting $\Delta E=1.41\times10^{-3}\,m$; their amplitudes are nearly identical in magnitude but differ in sign for $x<0$. The odd state has an exact zero at $x=0$ (parity node), whereas the even state has two nodes at $|x|\approx0.25$, visible only in the inset. The odd $j=1$ state (3 nodes) and the even $j=1$ state (4 nodes) both display one exterior zero crossing at $|x|\approx11$--$12$, which gives rise to the visible inner lobes in the probability density.

\subsection{Probability Densities}

\begin{figure}[t]
  \centering
  \includegraphics[width=\textwidth]{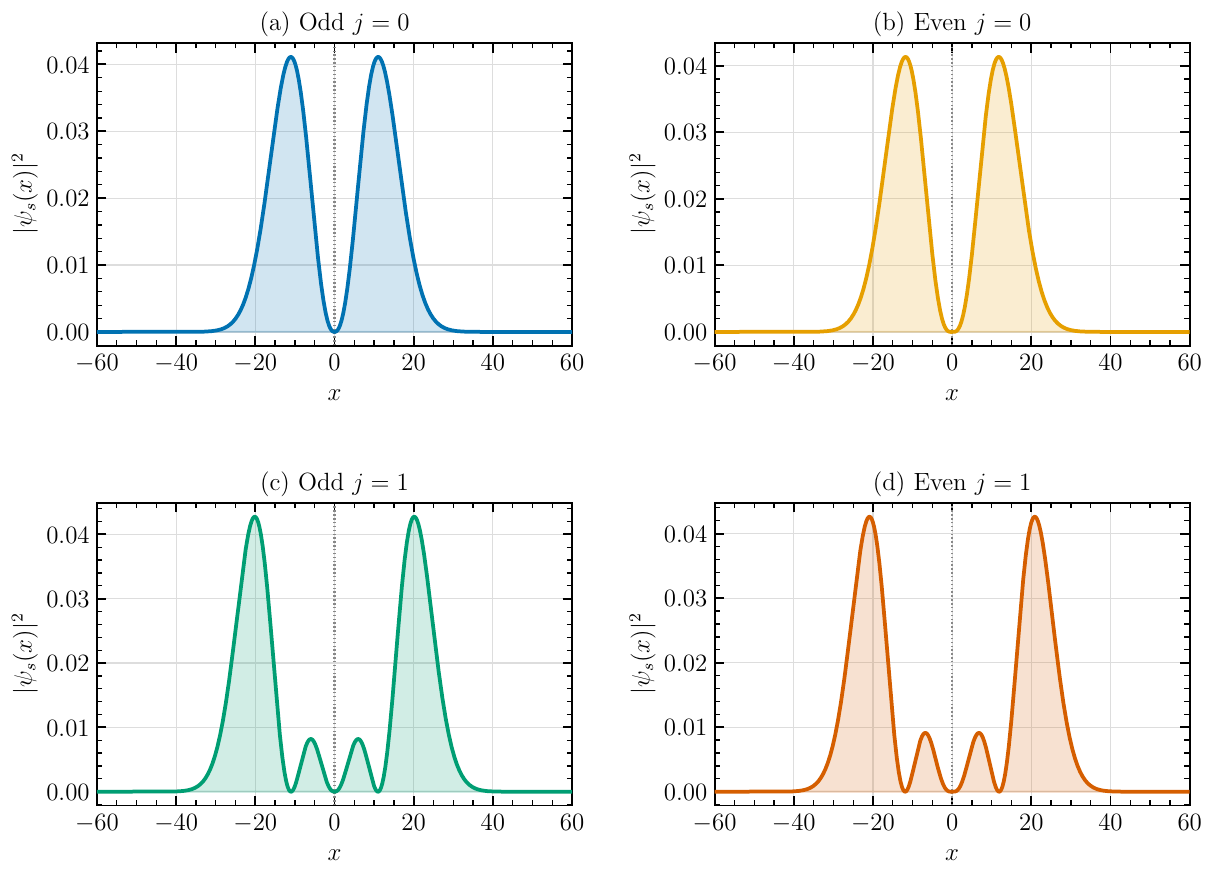}
  \caption{Probability densities $|\psi_s(x)|^2$ for the four lowest distinct states shown in Fig.~\ref{fig:cornell_wf}. Panel~(a): odd $j=0$, $E/m=0.960356$ (2 visible lobes). Panel~(b): even $j=0$, $E/m=0.961761$ (2 visible lobes). Panel~(c): odd $j=1$, $E/m=0.979567$ (4 visible lobes). Panel~(d): even $j=1$, $E/m=0.981343$ (4 visible lobes). Parameters: $\alpha=0.45$, $\beta=0.01$, $\delta=0.10$, $m=1$. Lobe counts refer to the visible peaks of $|\psi_s|^2$; the node count of $\psi_s$ itself, including the unresolved near-origin nodes of the even states, is given in Fig.~\ref{fig:cornell_wf}. The near-identical profiles of panels~(a) and~(b) reflect the even--odd pairing of the two lowest levels, while the inner lobes in panels~(c) and~(d) arise from the exterior nodes at $|x|\approx11$--$12$.}
  \label{fig:cornell_prob}
\end{figure}

Figure~\ref{fig:cornell_prob} shows the probability density $|\psi_s(x)|^2$ for each of the four states on separate panels. The panel titles indicate the number of \emph{visible} lobes of $|\psi_s|^2$, which differs from the node count of $\psi_s$ reported in Fig.~\ref{fig:cornell_wf}: the even-parity states contain an additional near-origin pair of nodes at $|x|\approx0.25$ that is not resolved at the scale of Fig.~\ref{fig:cornell_prob}.

\emph{Panels~(a) and~(b).}
The two lowest states (odd $j=0$ and even $j=0$) each display two visible lobes concentrated in the range $5\lesssim|x|\lesssim25$, with maxima near $|x|\approx11$. Their near-degeneracy ($\Delta E/m\approx1.4\times10^{-3}$) makes the two density profiles almost identical in shape and scale. The physically meaningful distinction between them is obscured in $|\psi_s|^2$: the odd state has $|\psi_s(0)|^2=0$ exactly, whereas the even state has a very small but non-zero central value because its true zeros lie slightly away from the origin, at $|x|\approx0.25$. The inset in Fig.~\ref{fig:cornell_wf} resolves this difference directly at the level of $\psi_s$.

\emph{Panel~(c).}
The odd $j=1$ state has three full-line nodes: one exact zero at $x=0$ and two exterior zeros at $|x|\approx11$. This produces four visible lobes in $|\psi_s|^2$: two outer lobes centred near $|x|\approx20$ and two smaller inner lobes near $|x|\approx6$. The inner lobes have significantly smaller amplitude than the outer ones, reflecting the stronger localisation in the confining outer region.

\emph{Panel~(d).}
The even $j=1$ state has four full-line nodes: a near-origin pair at $|x|\approx0.25$ and an exterior pair at $|x|\approx12$. At the scale of the figure, only the exterior pair produces visible structure in $|\psi_s|^2$, yielding four lobes arranged symmetrically about the origin. Compared with panel~(c), the outer lobes are shifted slightly outward and the inner lobes slightly inward, consistent with the small energy shift $\Delta E_1=1.78\times10^{-3}\,m$ between the two members of the pair. This state provides the clearest illustration of how the string tension $\beta$ controls the confinement scale: decreasing $\beta$ would move all lobes farther from the origin, whereas increasing $\beta$ would compress the density toward the core.

\subsection{FV Spinor Reconstruction}\label{s6.5}

Once a regularised eigenvalue has been obtained from Eqs.~\eqref{eq:cornell_even_matching} and~\eqref{eq:cornell_odd_matching}, and the scalar wave function $\psi_s$ has been normalised as described above, the FV spinor components follow from the general reconstruction formulas~\eqref{eq:psi1_from_psid} and~\eqref{eq:psi2_from_psid}:
\begin{equation}
\psi_1(x)=\frac{1}{2}
\left[1+\frac{E-eV_{\mathrm{C},\delta}(x)}{m}\right]\psi_s(x),
\qquad
\psi_2(x)=\frac{1}{2}
\left[1-\frac{E-eV_{\mathrm{C},\delta}(x)}{m}\right]\psi_s(x).
\label{eq:cornell_spinor_cutoff}
\end{equation}
The piecewise definition of $V_{\mathrm{C},\delta}$ propagates directly into the spinor components: inside the core ($|x|<\delta$) the potential is constant, so $\psi_{1,2}$ are proportional to $\psi_s$ with energy-dependent but spatially uniform coefficients; in the exterior ($|x|\ge\delta$) the coefficients inherit the $1/|x|$ and $|x|$ dependences of the Cornell potential.

The local mixing factor governing the particle--antiparticle decomposition is
\begin{equation}
f(x)\equiv\frac{E-eV_{\mathrm{C},\delta}(x)}{m},
\label{eq:cornell_mixing_factor}
\end{equation}
so that
\begin{equation}
\psi_1=\frac{1}{2}(1+f)\psi_s,
\qquad
\psi_2=\frac{1}{2}(1-f)\psi_s.
\end{equation}
In the exterior, where $eV(x)=\alpha/|x|+\beta|x|$, the function $f(x)$ varies non-trivially with position. Near the cutoff boundary, the Coulomb term makes $eV$ large, so $f$ may become negative and can even drop below $-1$ in a narrow near-core region; in that case the antiparticle component $|\psi_2|$ locally exceeds the particle component $|\psi_1|$. Moving outward, $eV$ decreases, $f$ crosses zero at the classical turning point $x_{\rm tp}$ defined by $eV(x_{\rm tp})=E$, and for $x>x_{\rm tp}$ the particle component again dominates. At large distances one has $f\to E/m<1$, and the asymptotic ratio becomes
\begin{equation}
\left|\frac{\psi_2}{\psi_1}\right|_{x\to\infty}
=
\frac{1-E/m}{1+E/m},
\label{eq:cornell_asymptotic_ratio}
\end{equation}
which equals $0.020$ for the lowest odd state ($E/m=0.960356$) and decreases as the binding energy becomes smaller.

\begin{figure}[t]
  \centering
  \includegraphics[width=0.8\textwidth]{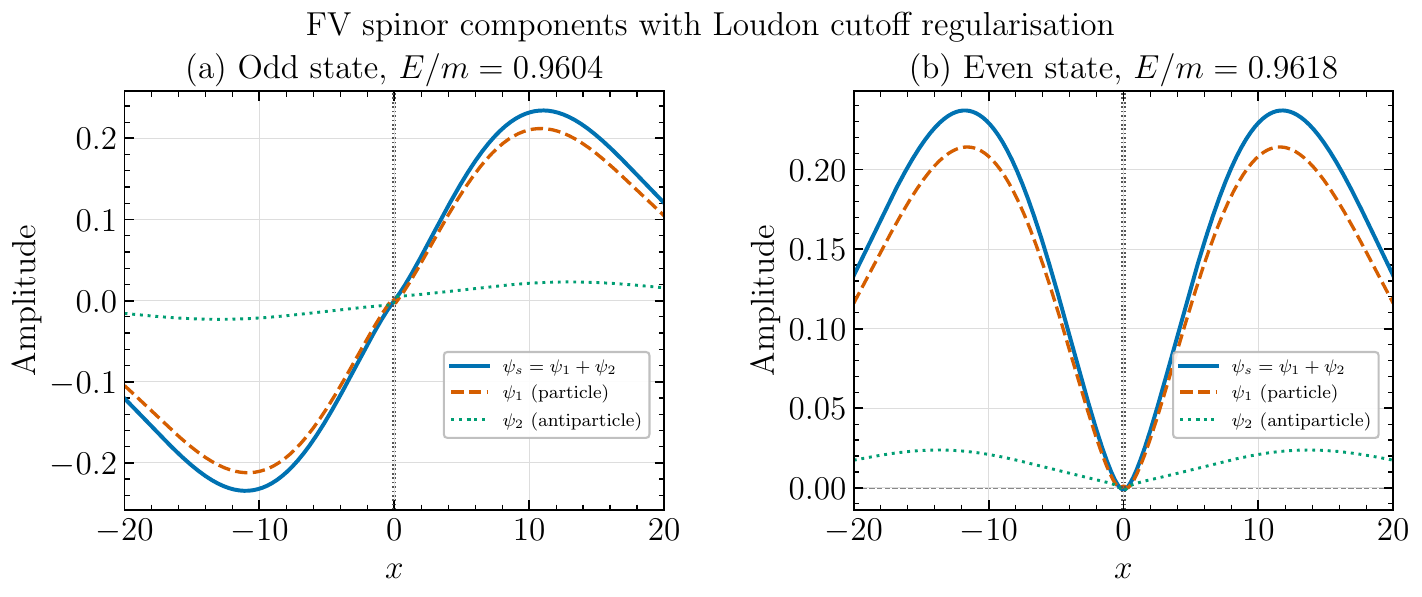}
  \caption{Feshbach--Villars spinor components for the regularised Cornell potential with $\alpha=0.45$, $\beta=0.01$, $\delta=0.10$, and $m=1$, normalised so that $\int\rho\,dx=1$. (a) Lowest odd-parity state ($j=0$, $E/m=0.960356$). (b) Lowest even-parity state ($j=0$, $E/m=0.961761$). Solid blue: sum component $\psi_s=\psi_1+\psi_2$. Dashed red: particle component $\psi_1$. Dotted green: antiparticle component $\psi_2$. Dotted grey vertical lines mark the cutoff boundary $x=\pm\delta$, across which the interior oscillatory solution is matched to the exterior Tricomi-function solution. The particle component dominates over most of the domain, while $\psi_2$ is visibly enhanced near the core, where the large regularised potential increases the local particle--antiparticle mixing.}
  \label{fig:cornell_fv_components}
\end{figure}

The internal structure of the FV spinor for the regularised Cornell potential is shown in Fig.~\ref{fig:cornell_fv_components} for the two lowest states with $\alpha=0.45$. In both panels the sum component $\psi_s$ (solid blue) smoothly joins the oscillatory interior solution across $|x|=\delta$ (marked by dotted vertical lines) to the decaying exterior solution. The particle component $\psi_1$ (dashed red) dominates over most of the domain, while the antiparticle component $\psi_2$ (dotted green) remains smaller except for a narrow near-core region, where the strong potential enhances the local particle--antiparticle mixing. For the odd state [panel~(a)] the antisymmetry under $x\to-x$ is manifest in all three components, whereas for the even state [panel~(b)] all components are symmetric about the origin.

\begin{figure}[t]
  \centering
  \includegraphics[width=0.6\textwidth]{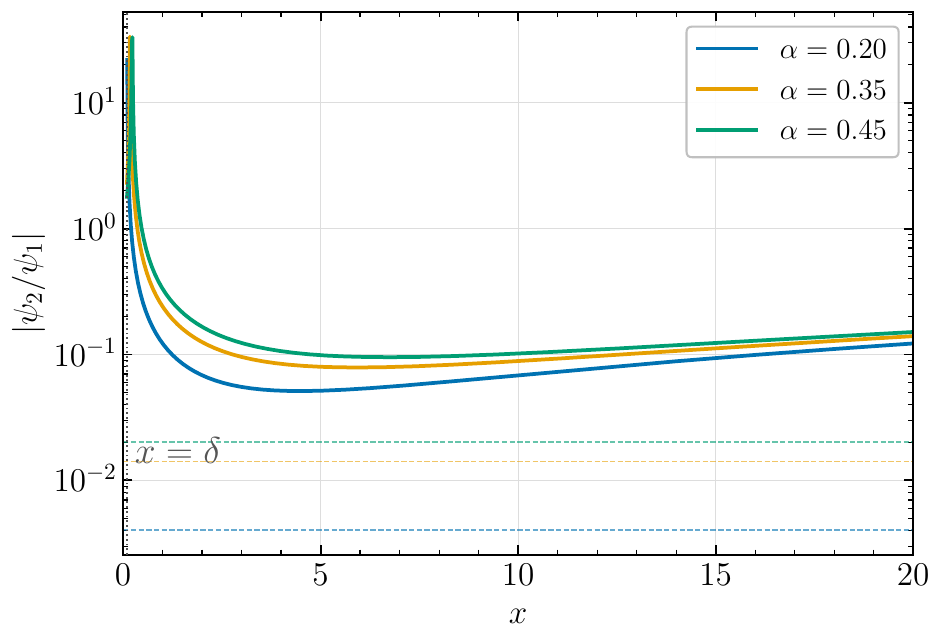}
  \caption{Antiparticle-to-particle ratio $|\psi_2/\psi_1|$ as a function of $x>0$ for the lowest odd-parity state ($j=0$) of the regularised Cornell potential with $\beta=0.01$, $\delta=0.10$, $m=1$, and three Coulomb couplings: $\alpha=0.20$ (blue), $\alpha=0.35$ (orange), and $\alpha=0.45$ (green). The ratio is plotted on a logarithmic scale starting just above the classical turning point $x_{\rm tp}$, where $eV=E$ and $|\psi_1|=|\psi_2|$. Horizontal dashed lines indicate the asymptotic values $(1-E/m)/(1+E/m)$, equal to $0.004$, $0.014$, and $0.020$ for the three couplings. Larger coupling produces a uniformly higher ratio, reflecting the increasing relativistic character of the more deeply bound states.}
  \label{fig:cornell_ratio}
\end{figure}

A quantitative measure of the relativistic content of each state is provided by the antiparticle-to-particle ratio $|\psi_2/\psi_1|$, shown in Fig.~\ref{fig:cornell_ratio} for the lowest odd state and three values of the coupling $\alpha\in\{0.20,0.35,0.45\}$. In the exterior ($x>\delta$) the ratio is given analytically by $|(1-f)/(1+f)|$, with $f(x)=(E-\alpha/x-\beta x)/m$. Three features are noteworthy. \emph{First}, for $x\lesssim x_{\rm tp}$ the denominator $1+f$ approaches zero and the ratio rises sharply on the logarithmic scale, reflecting the region where the strong Coulomb field drives the antiparticle component above the particle component; for visual clarity, the curves are therefore shown only for $x>x_{\rm tp}$. \emph{Second}, for $x>x_{\rm tp}$ the ratio decreases monotonically and approaches the asymptotic plateau~\eqref{eq:cornell_asymptotic_ratio}, equal to $0.020$, $0.014$, and $0.004$ for $\alpha=0.45$, $0.35$, and $0.20$, respectively. \emph{Third}, the ratio is uniformly larger for stronger coupling because increasing $\alpha$ lowers $E/m$ and thus enhances the asymptotic particle--antiparticle mixing.

\begin{figure}[t]
\centering
\includegraphics[width=0.8\textwidth]{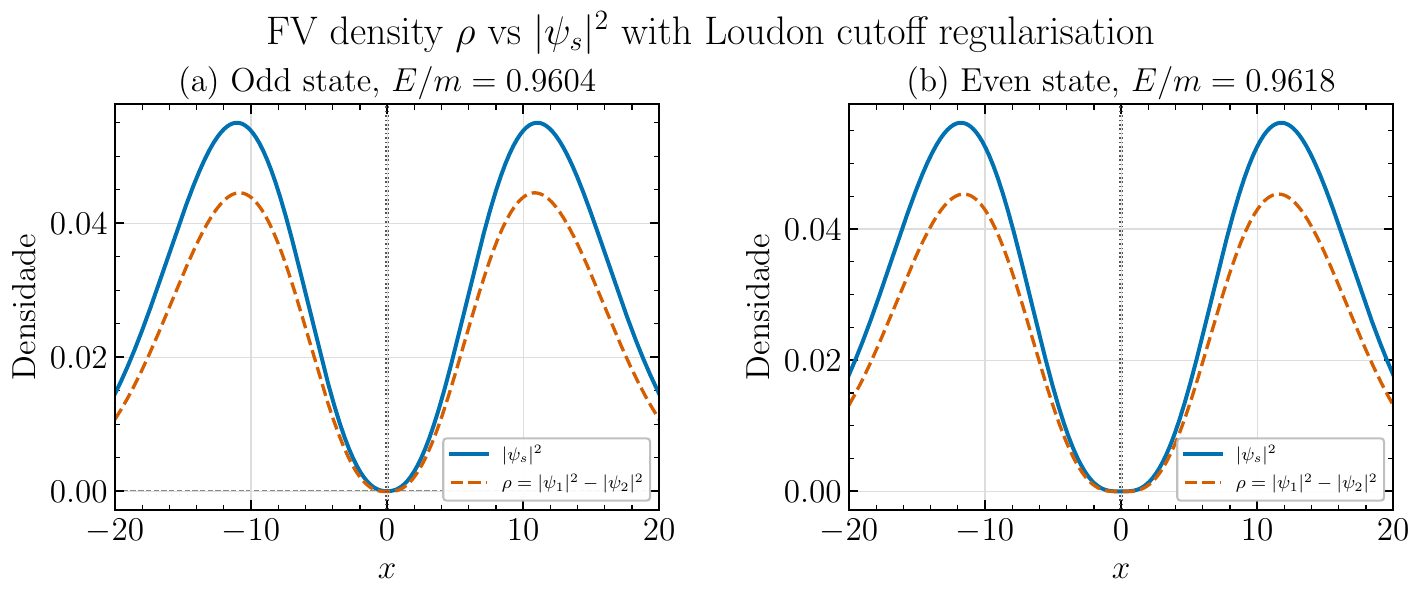}
\caption{Comparison of the FV charge density $\rho=|\psi_1|^2-|\psi_2|^2$ (dashed red) and the squared sum component $|\psi_s|^2$ (solid blue) for $\alpha=0.45$, $\beta=0.01$, $\delta=0.10$, and $m=1$, normalised so that $\int\rho\,dx=1$. (a) Lowest odd-parity state ($j=0$, $E/m=0.960356$). (b) Lowest even-parity state ($j=0$, $E/m=0.961761$). Dotted grey vertical lines mark the cutoff boundary $x=\pm\delta$. In the outer region the two curves nearly coincide; in the narrow shell where $eV_{\mathrm{C},\delta}>E$, the charge density dips slightly below zero, signalling a local enhancement of the antiparticle component. The integrated charge remains positive and normalised.}
\label{fig:cornell_density}
\end{figure}

Figure~\ref{fig:cornell_density} compares the FV charge density
\begin{equation}
\rho=|\psi_1|^2-|\psi_2|^2
\end{equation}
with the squared sum component $|\psi_s|^2$ for the two lowest states, normalised so that $\int\rho\,dx=1$. In the bulk of the exterior region ($|x|\gtrsim0.5$), the two quantities nearly coincide because the antiparticle correction is small. They differ visibly in the narrow shell $|x|\lesssim x_{\rm tp}\approx0.47$, where $eV>E$ and hence $f<0$: in that region the charge density becomes slightly negative, while $|\psi_s|^2$ remains positive. This local sign reversal indicates that the antiparticle component momentarily dominates in the near-core shell. Nevertheless, the integrated charge remains positive and normalised, so the one-particle charge-density interpretation within the mass gap is preserved.

The physically meaningful densities and observables are computed at finite $\delta$ and only afterwards analysed in the limit $\delta\to0^+$. The reconstruction formula~\eqref{eq:cornell_spinor_cutoff} is well defined for all $x\neq0$ at any finite $\delta$, and the limit is smooth because the scalar wave function vanishes at the origin for odd states and remains finite there for even states.

\section{P\"{o}schl--Teller Potential}\label{s7}

The P\"{o}schl--Teller (PT) potential~\cite{Poschl1933} is defined by
\begin{equation}
  V_{\mathrm{PT}}(x) = -\frac{V_0}{\cosh^{2}(x/d)},
  \qquad V_0 > 0,\quad d > 0,
  \label{eq:pt_potential}
\end{equation}
where $V_0$ is the well depth and $d$ is the range parameter. The
potential is attractive, even in $x$, and bounded everywhere: it
reaches its maximum depth $-V_0$ at the origin and vanishes as
$|x|\to\infty$. These properties make the PT potential a prototype of
a short-range, regular interaction without singularities, in clear
contrast with the Coulomb and Cornell potentials discussed in the
preceding sections. Since the problem is defined on the full real line
and the potential is even, the eigenstates may be classified by
definite parity. The PT potential arises naturally in several branches
of physics: in molecular physics it models smooth finite-depth
interactions~\cite{Poschl1933}; in nuclear and condensed-matter
contexts it provides a useful short-range model; and in mathematical
physics it is one of the classic exactly solvable potentials of
supersymmetric quantum mechanics~\cite{Cooper1995}.

Substituting $V=V_{\mathrm{PT}}(x)$ into the master
equation~\eqref{eq:KG_sumfun}, one obtains
\begin{equation}
  \frac{d^{2}\psi_{s}}{dx^{2}}
  + \left[
      \frac{2eEV_0}{\cosh^{2}(x/d)}
      + \frac{(eV_0)^{2}}{\cosh^{4}(x/d)}
      - \kappa^{2}
    \right]\psi_{s}=0,
  \qquad
  \kappa^{2}\equiv m^{2}-E^{2}>0.
  \label{eq:pt_master}
\end{equation}
Equivalently, one may write
\begin{equation}
  \psi_{s}'' + [V_{\mathrm{eff}}(x)-\kappa^{2}]\,\psi_{s}=0,
  \label{eq:pt_schrodinger_like}
\end{equation}
with the effective potential
\begin{equation}
  V_{\mathrm{eff}}(x)
  = \frac{2eEV_0}{\cosh^{2}(x/d)}
    + \frac{(eV_0)^{2}}{\cosh^{4}(x/d)}.
  \label{eq:pt_Veff}
\end{equation}
The first term is the familiar non-relativistic P\"{o}schl--Teller
profile, whereas the second term is a genuinely relativistic
contribution arising from the square of the potential in
$(E-eV)^2$. For weak coupling, $eV_0 \ll m$, the $\cosh^{-4}$ term is
subleading and the equation approaches the textbook Schr\"odinger PT
problem. In the full relativistic FV setting, however, both terms must
be retained.

To analyse the structure of the equation, it is convenient to
introduce
\begin{equation}
  \xi = \tanh(x/d),
  \qquad \xi\in(-1,1),
  \label{eq:pt_xi_def}
\end{equation}
so that
\begin{equation}
  \frac{1}{\cosh^{2}(x/d)} = 1-\xi^{2}.
\end{equation}
Equation~\eqref{eq:pt_master} then becomes
\begin{equation}
  (1-\xi^{2})\frac{d^{2}\psi_s}{d\xi^{2}}
  - 2\xi\frac{d\psi_s}{d\xi}
  + \left[
      d^{2}(eV_0)^{2}(1-\xi^{2})
      + 2eEV_0d^{2}
      - \frac{d^{2}\kappa^{2}}{1-\xi^{2}}
    \right]\psi_s = 0.
  \label{eq:pt_xi_eq}
\end{equation}
Unlike the non-relativistic PT equation, the full relativistic problem
contains the additional term $d^{2}(eV_0)^2(1-\xi^{2})$, which
prevents a direct reduction to the associated Legendre equation. In
other words, the standard closed-form Schr\"odinger solution is
recovered only in the weak-coupling limit where the
$\cosh^{-4}(x/d)$ contribution may be neglected. In the parameter
regime studied here, the relativistic correction is not negligible, so
the bound-state spectrum is determined numerically from the master
equation itself.

Because the potential is even, the stationary states have definite
parity:
\begin{equation}
  \psi_s(-x)=\pm\psi_s(x),
\end{equation}
with the plus (minus) sign corresponding to even (odd) states. This
symmetry makes a half-line shooting formulation especially convenient.
For even states we impose
\begin{equation}
  \psi_s(0)=1,\qquad \psi_s'(0)=0,
  \label{eq:pt_even_bc}
\end{equation}
whereas for odd states we impose
\begin{equation}
  \psi_s(0)=0,\qquad \psi_s'(0)=1.
  \label{eq:pt_odd_bc}
\end{equation}
The equation is then integrated from $x=0$ to a large cutoff $L$,
chosen here as $L=8d$, for which the potential is already negligible.
For a trial energy $E\in(0,m)$, square-integrability requires the
solution to match the asymptotic decay
\begin{equation}
  \psi_s(x)\propto e^{-\kappa x},
  \qquad x\to+\infty.
\end{equation}
The bound-state energies are therefore obtained by searching for zeros
of the mismatch function at $x=L$, with each root refined by Brent's
method to a tolerance of $10^{-10}$.

For the representative parameter set $eV_0=1.0$, $d=3$, and $m=1$,
the numerical procedure yields five positive-energy bound states,
\begin{equation}
  \frac{E_n}{m}\approx
  0.196,\quad 0.533,\quad 0.770,\quad 0.925,\quad 0.997,
  \qquad n=0,1,2,3,4.
  \label{eq:pt_numerics}
\end{equation}
The antiparticle branch follows by charge conjugation,
$E_n^{(\mathrm{anti})}=-E_n^{(\mathrm{part})}$. The finiteness of the
spectrum is a hallmark of short-range potentials and contrasts
strongly with the infinite tower of Coulomb states.

\begin{figure}[h!]
\centering
\includegraphics[width=0.90\linewidth]{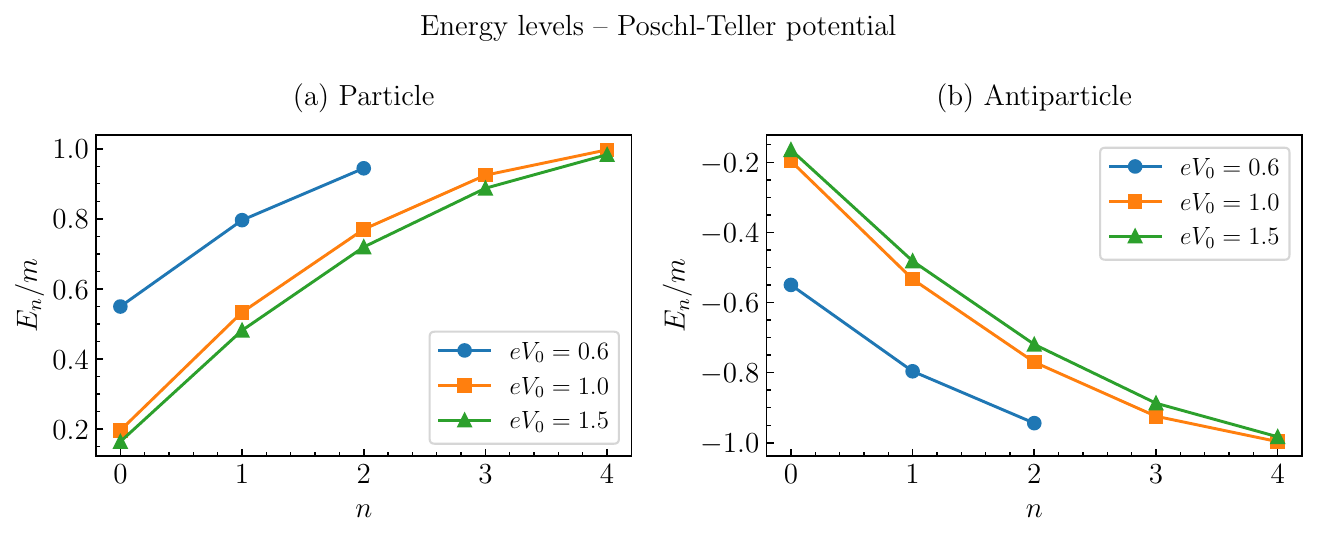}
\caption{Energy levels $E_{n}/m$ as a function of the quantum number
$n$ for the P\"{o}schl--Teller potential ($m = 1$, $d = 3$) and
three coupling values: $eV_0 = 0.6$ (circles), $1.0$ (squares),
$1.5$ (triangles). (a)~Particle branch
$E_{n}^{(\mathrm{part})} > 0$. (b)~Antiparticle branch
$E_{n}^{(\mathrm{anti})} < 0$. Only a finite number of bound states
exists for each coupling, as expected for a short-range well.}
\label{fig:pt_levels}
\end{figure}

The bound-state energies are displayed in Fig.~\ref{fig:pt_levels} for
$m=1$, $d=3$, and three coupling values: $eV_0=0.6$ (circles), $1.0$
(squares), and $1.5$ (triangles). Panel~(a) shows the particle branch
and panel~(b) the antiparticle branch. As $eV_0$ increases,
additional bound states appear and the existing levels move deeper
into the gap, reflecting stronger binding. For each coupling only a
finite number of levels exists, in contrast with the Coulomb case and
with the near-threshold accumulation characteristic of the Cornell
spectrum.

\begin{figure}[h!]
\centering
\includegraphics[width=0.75\linewidth]{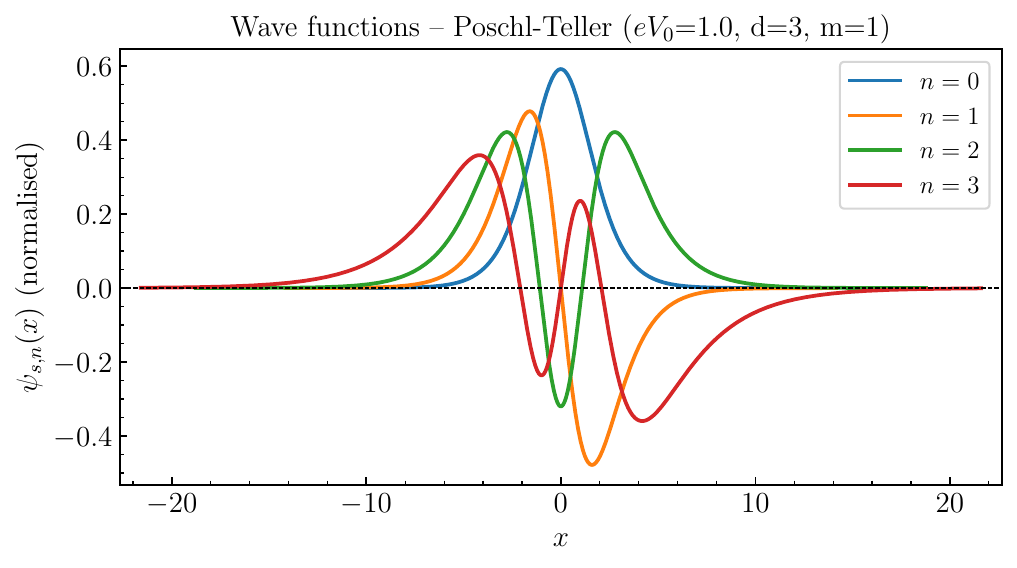}
\caption{Normalised wave functions $\psi_{s,n}(x)$ for the
P\"{o}schl--Teller potential ($eV_0 = 1.0$, $d = 3$, $m = 1$) and
$n = 0$ (blue), $1$ (orange), $2$ (green), $3$ (red). Even-$n$
states are symmetric and odd-$n$ states are antisymmetric about the
origin. All states decay rapidly for $|x| \gtrsim 15 \approx 5d$.}
\label{fig:pt_wf}
\end{figure}

The normalised wave functions $\psi_{s,n}(x)$ for $n=0,1,2,3$,
computed with $eV_0=1.0$, $d=3$, and $m=1$, are shown in
Fig.~\ref{fig:pt_wf}. The most prominent feature is the
definite-parity structure: states with even $n$ are symmetric about
the origin, whereas states with odd $n$ are antisymmetric. This is a
direct consequence of the even symmetry of $V_{\mathrm{PT}}(x)$.
All states are spatially confined within a region of order $d$; for
the parameters used here, the wave functions are already negligible
for $|x|\gtrsim 15 \approx 5d$. The number of nodes equals $n$, in
accordance with the oscillation theorem.

The two-component FV spinor is reconstructed from $\psi_{s,n}(x)$
through Eqs.~\eqref{eq:psi1_from_psid} and~\eqref{eq:psi2_from_psid},
\begin{align}
  \psi_{1,n}(x)
  &= \frac{1}{2}\left[
       1 + \frac{E}{m}
       + \frac{eV_0}{m\cosh^{2}(x/d)}
     \right]\psi_{s,n}(x),
  \label{eq:pt_psi1}\\[4pt]
  \psi_{2,n}(x)
  &= \frac{1}{2}\left[
       1 - \frac{E}{m}
       - \frac{eV_0}{m\cosh^{2}(x/d)}
     \right]\psi_{s,n}(x),
  \label{eq:pt_psi2}
\end{align}
and both components inherit the parity of $\psi_{s,n}$ because the
prefactors are even functions of $x$.

\begin{figure}[h!]
\centering
\includegraphics[width=0.90\linewidth]{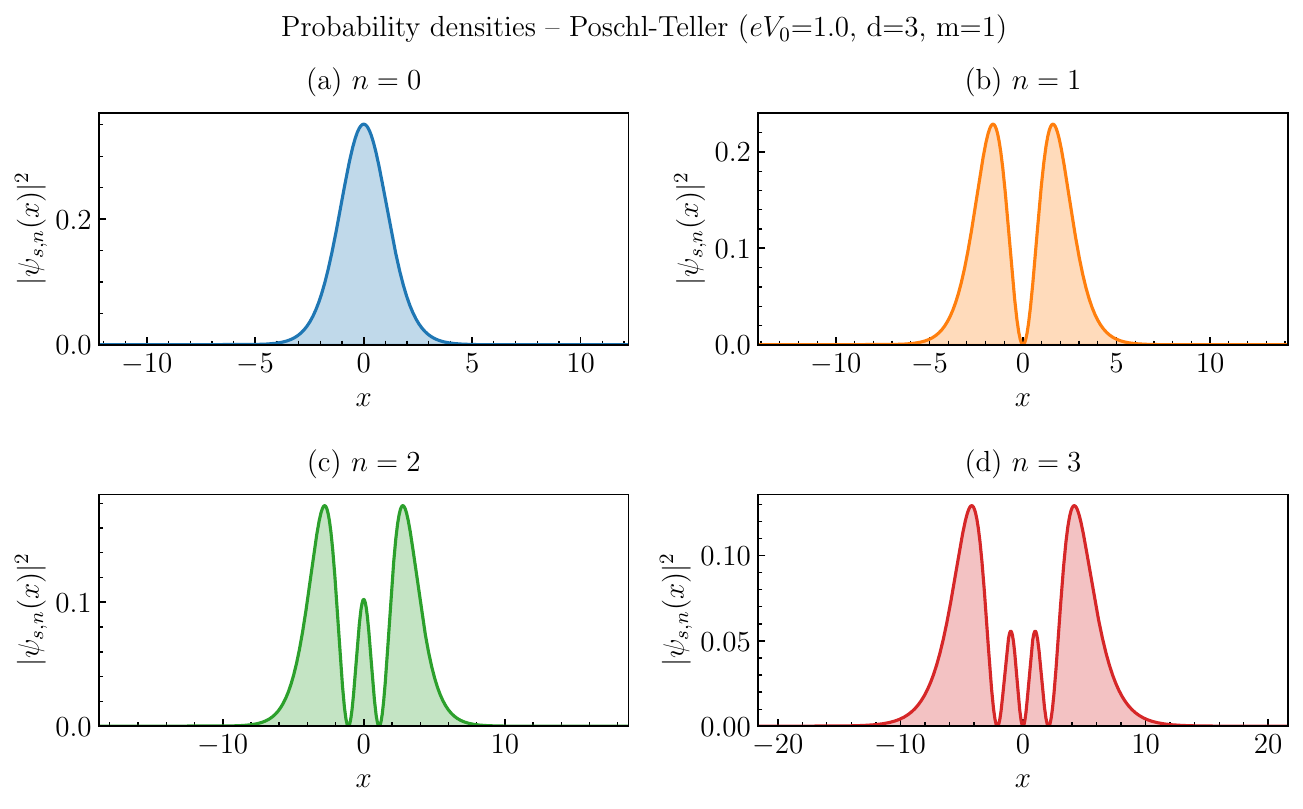}
\caption{Probability densities $|\psi_{s,n}(x)|^{2}$ for the
P\"{o}schl--Teller potential ($eV_0 = 1.0$, $d = 3$, $m = 1$).
(a)~$n = 0$: single peak at the origin. (b)~$n = 1$: two symmetric
peaks. (c)~$n = 2$: three peaks. (d)~$n = 3$: four peaks. All
distributions are confined within $|x| \lesssim 15$ and reflect the
even symmetry of the potential.}
\label{fig:pt_prob}
\end{figure}

The corresponding probability densities $|\psi_{s,n}(x)|^{2}$ are
shown in Fig.~\ref{fig:pt_prob}. The ground state [panel~(a)] has a
single peak centred at the origin. The first excited state
[panel~(b)] displays two symmetric peaks on either side of the
origin, as dictated by its odd-parity nodal structure. States (c) and
(d) show three and four peaks, respectively. The symmetric
distribution of peaks about $x=0$ is a direct signature of the even
symmetry of the PT potential.

Table~\ref{tab:pt_comparison} summarises the main structural
differences between the PT potential and the regularised Coulomb and
Cornell cases. The most significant distinction is the complete
regularity of the PT potential at the origin. A second key difference
is the finite number of bound states, which follows from the
short-range character of the well. As in the regularised Coulomb and
Cornell problems, parity remains a good quantum number on the full
line; what changes is the asymptotic behaviour of the interaction and,
consequently, the size of the discrete spectrum.

\begin{table}[h!]
\centering
\caption{Structural comparison of the regularised Coulomb, regularised
Cornell, and P\"{o}schl--Teller potentials within the FV formalism.}
\label{tab:pt_comparison}
\renewcommand{\arraystretch}{1.25}
\begin{tabular}{lccc}
\hline\hline
Property & Coulomb (cutoff) & Cornell (cutoff) & P\"{o}schl--Teller \\
\hline
Domain
  & $x \in \mathbb{R}$ & $x \in \mathbb{R}$ & $x \in \mathbb{R}$ \\
Behaviour at the origin
  & singular, regularised & singular, regularised & regular \\
Parity of eigenstates
  & definite & definite & definite \\
Number of bound states
  & infinite & finite (for fixed parameters) & finite \\
Asymptotic interaction
  & $1/|x|$ tail & linear confinement & short-range \\
Spectral determination
  & matching condition & matching condition & shooting method \\
\hline\hline
\end{tabular}
\end{table}

The two-component structure of the FV spinor for the
P\"{o}schl--Teller potential exhibits two qualitative features: a
definite parity inherited from the scalar wave function and a spatially
symmetric particle--antiparticle mixing profile. The local mixing
factor entering Eqs.~\eqref{eq:psi1_from_psid}
and~\eqref{eq:psi2_from_psid} is
\begin{equation}
  f(x) \equiv \frac{E - eV_{\rm PT}(x)}{m}
       = \frac{1}{m}\!\left(E + \frac{eV_0}{\cosh^2(x/d)}\right),
  \label{eq:pt_mixing}
\end{equation}
which is an even function of $x$ with a maximum
\begin{equation}
f(0)=\frac{E+eV_0}{m}
\end{equation}
at the origin and an asymptotic value
\begin{equation}
f(\infty)=\frac{E}{m}<1
\end{equation}
at large $|x|$. Since $f(x)$ is even, the prefactors
$[1\pm f(x)]$ are also even, and therefore $\psi_1$ and $\psi_2$
inherit the same parity as $\psi_s$.

\begin{figure}[h!]
\centering
\includegraphics[width=0.90\textwidth]{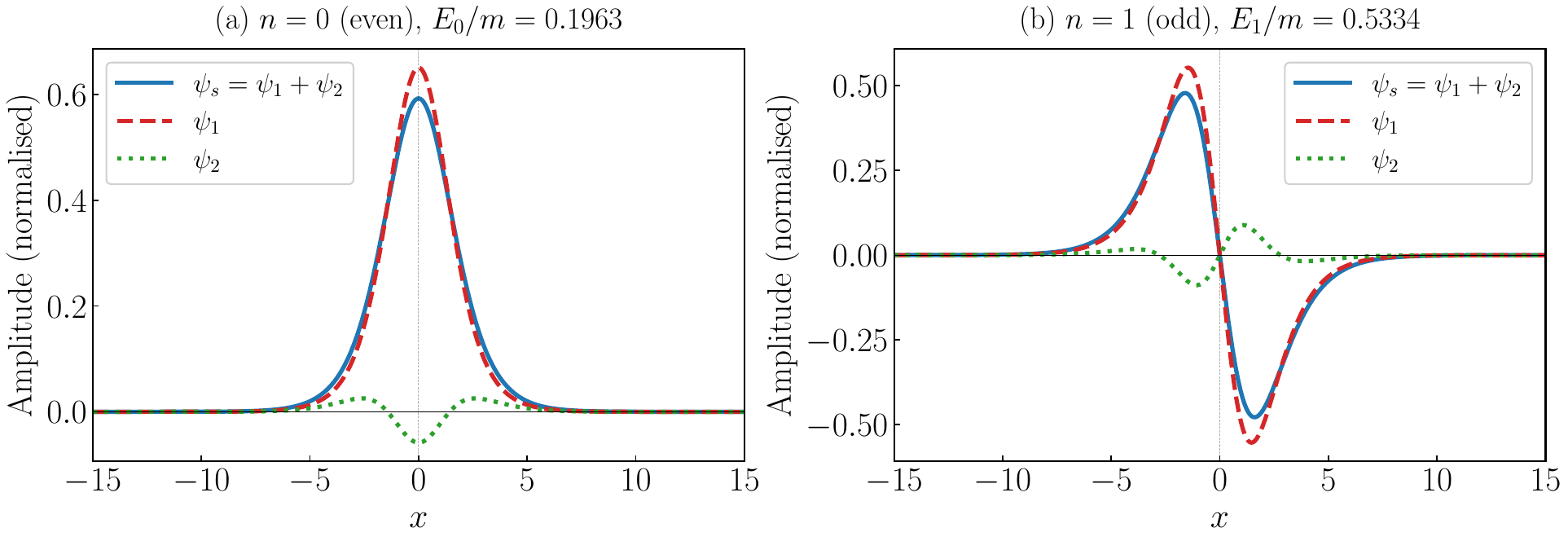}
\caption{FV spinor components $\psi_1$ (particle, red dashed),
$\psi_2$ (antiparticle, green dotted), and their sum
$\psi_s = \psi_1 + \psi_2$ (blue solid) for the
P\"{o}schl--Teller potential with $eV_0 = 1.0$, $d = 3$, and $m = 1$,
computed by the shooting method. Left panel: ground state $n = 0$
(even parity), $E_0/m = 0.1963$. Right panel: first excited state
$n = 1$ (odd parity), $E_1/m = 0.5334$. Both FV components share the
parity of $\psi_s$, since the mixing factor
$f(x) = (E + eV_0/\cosh^2(x/d))/m$ is even.}
\label{fig:pt_components}
\end{figure}

The FV components $\psi_1$ and $\psi_2$ for the two lowest states are
displayed in Fig.~\ref{fig:pt_components}, computed with
$eV_0=1.0$, $d=3$, and $m=1$. The ground state ($n=0$, even) has a
single bell-shaped $\psi_s$ centred at the origin, and both $\psi_1$
and $\psi_2$ are symmetric, as required by parity. The particle
component $\psi_1$ closely tracks $\psi_s$, while the antiparticle
component $\psi_2$, although smaller in absolute amplitude, is spread
symmetrically around the origin rather than being localised at a
singular core. The first excited state ($n=1$, odd) is antisymmetric:
$\psi_s$, $\psi_1$, and $\psi_2$ all vanish at the origin and form two
lobes of opposite sign. For the parameter set shown, the relative
weight of $\psi_2$ is larger for the deeply bound ground state than
for the less strongly bound excited state, reflecting the stronger
particle--antiparticle mixing induced by the smaller value of $E/m$.

\begin{figure}[h!]
\centering
\includegraphics[width=0.55\textwidth]{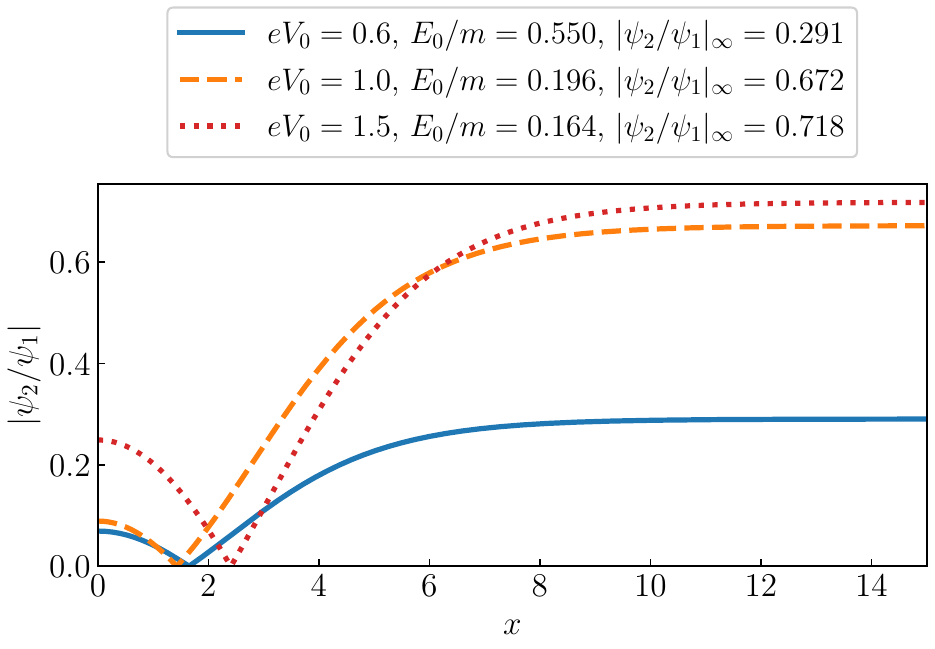}
\caption{Antiparticle-to-particle ratio $|\psi_2/\psi_1|
= |(1-f)/(1+f)|$ for the ground state ($n=0$, $d=3$, $m=1$) of the
P\"{o}schl--Teller potential and three well depths:
$eV_0 = 0.6$ (blue solid), $eV_0 = 1.0$ (orange dashed), and
$eV_0 = 1.5$ (red dotted). The ratio is smallest near the centre of
the well and approaches the asymptotic plateau
$|(1-E/m)/(1+E/m)|$ at large $|x|$. Deeper wells produce stronger
asymptotic particle--antiparticle mixing.}
\label{fig:pt_ratio}
\end{figure}

The dependence of the particle--antiparticle mixing on the well depth
is shown in Fig.~\ref{fig:pt_ratio}, which displays
$|\psi_2/\psi_1| = |(1-f)/(1+f)|$ for the ground state and three
values of $eV_0$. The ratio is smallest near the centre of the well,
where $f(x)$ is largest and particle dominance is strongest, and then
increases monotonically toward the asymptotic plateau
\begin{equation}
\left|\frac{1-E/m}{1+E/m}\right|.
\end{equation}
For deeper wells, the ground-state energy moves farther below the
continuum threshold, which increases the asymptotic plateau and hence
the overall relativistic mixing. Unlike the Coulomb and Cornell cases,
the profile is symmetric in $x$ and free of singular structure.

\begin{figure}[h!]
\centering
\includegraphics[width=0.90\textwidth]{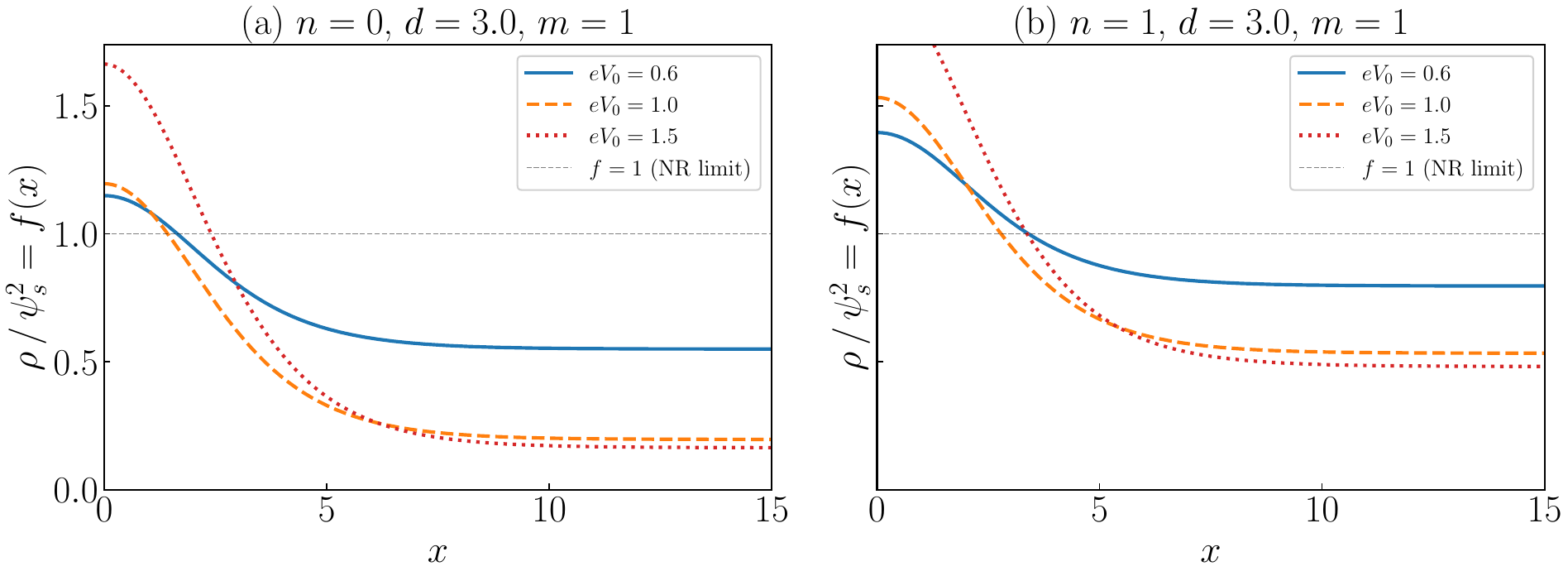}
\caption{Local mixing factor
$f(x) = (E + eV_0/\cosh^2(x/d))/m$,
equal to $\rho(x)/|\psi_s(x)|^2$, for the P\"{o}schl--Teller
potential with $d = 3$, $m = 1$, and $eV_0 = 0.6$ (blue solid),
$1.0$ (orange dashed), and $1.5$ (red dotted). Left panel: $n=0$.
Right panel: $n=1$. The dashed horizontal line at $f = 1$ marks the
crossover between $\rho > |\psi_s|^2$ and $\rho < |\psi_s|^2$.
Because $f(x)>0$ everywhere, the conserved charge density is always
positive.}
\label{fig:pt_density}
\end{figure}

The exact local relation
\begin{equation}
\frac{\rho(x)}{|\psi_s(x)|^2}=f(x)
\end{equation}
is displayed in Fig.~\ref{fig:pt_density} for $n=0$ and $n=1$ and
three values of $eV_0$. Since $f(x)>0$ throughout the domain, the
conserved FV charge density $\rho$ remains positive everywhere. Near
the origin, $f(0)=(E+eV_0)/m$ may exceed unity, so the conserved
density locally exceeds the naive density $|\psi_s|^2$. At large
$|x|$, however, $f(x)\to E/m<1$, and therefore
$\rho < |\psi_s|^2$ asymptotically. The crossover between these two
regimes occurs on the scale set by $d$.

\section{Woods--Saxon Potential}\label{s8}

The Woods--Saxon (WS) potential was originally introduced in nuclear
physics to describe the diffuse surface of the nuclear density
distribution~\cite{Woods1954}. In its one-dimensional form it reads
\begin{equation}
  V_{\mathrm{WS}}(x) = -\frac{V_0}{1 + e^{(x-R)/a}},
  \qquad V_0 > 0,
  \label{eq:ws_potential}
\end{equation}
where $V_0$ is the potential depth, $R$ is the midpoint of the
transition region, and $a>0$ is the diffuseness parameter controlling
the steepness of the surface. The potential decreases monotonically as
$x$ increases:
\begin{equation}
V_{\mathrm{WS}}(x)\to -V_0 \quad (x\to-\infty),
\qquad
V_{\mathrm{WS}}(x)\to 0 \quad (x\to+\infty).
\end{equation}
It is therefore neither even nor odd, in contrast with the
P\"{o}schl--Teller potential. As $a\to0$, the profile approaches a
sharp step; larger $a$ produces a smoother transition. This
flexibility has made the Woods--Saxon potential a standard tool in
nuclear structure calculations~\cite{Woods1954,Hosseinpour2015}.

Substituting $V=V_{\mathrm{WS}}(x)$ into the master
equation~\eqref{eq:KG_sumfun}, one obtains
\begin{equation}
  \frac{d^{2}\psi_{s}}{dx^{2}}
  + \left[\!\left(E + \frac{eV_0}{1+e^{(x-R)/a}}\right)^{\!2}
          - m^{2}\right]\psi_{s} = 0.
  \label{eq:ws_master}
\end{equation}
Expanding the square and writing
\begin{equation}
\kappa_+^2 \equiv m^{2}-E^{2},
\qquad
\kappa_-^2 \equiv m^{2}-(E+eV_0)^{2},
\end{equation}
the equation can be cast in the Schr\"odinger-like form
\begin{equation}
  \frac{d^{2}\psi_{s}}{dx^{2}}
  + \left[V_{\mathrm{eff}}(x)-\kappa_+^{2}\right]\psi_{s} = 0,
  \label{eq:ws_schrodinger_form}
\end{equation}
with
\begin{equation}
  V_{\mathrm{eff}}(x)
  = \frac{2eEV_0}{1+e^{(x-R)/a}}
    + \frac{(eV_0)^{2}}{\!\left(1+e^{(x-R)/a}\right)^{2}}.
  \label{eq:ws_Veff}
\end{equation}
Unlike the P\"{o}schl--Teller case, $V_{\mathrm{eff}}$ is not
symmetric: it interpolates smoothly between the constant value
$2eEV_0 + (eV_0)^2$ as $x\to-\infty$ and zero as $x\to+\infty$.

To expose the analytic structure of the equation, we introduce
\begin{equation}
u = \frac{1}{1+e^{(x-R)/a}},
\qquad u\in(0,1),
\end{equation}
so that
\begin{equation}
eV_{\mathrm{WS}}(x)=-eV_0\,u.
\end{equation}
Equation~\eqref{eq:ws_master} becomes
\begin{equation}
  u^{2}(1-u)^{2}\frac{d^{2}\psi_{s}}{du^{2}}
  - u(1-u)(1-2u)\frac{d\psi_{s}}{du}
  + a^{2}\!\left[-\kappa_+^{2} + 2eEV_0\,u + (eV_0)^{2}u^{2}\right]\psi_{s} = 0,
  \label{eq:ws_u_ode}
\end{equation}
which has regular singular points at $u=0$ and $u=1$. After removing
the asymptotic factors associated with the left and right decays, the
remaining equation belongs to the confluent Heun class~\cite{Ronveaux1995}. In contrast with the Kummer and Tricomi equations
encountered in the Coulomb and Cornell problems, the present equation
does not lead to a simple closed-form quantisation rule. For this
reason, the spectrum is determined numerically.

A bound state must decay on both sides of the well, so the conditions
\begin{equation}
\kappa_+^2>0,
\qquad
\kappa_-^2>0,
\label{eq:ws_bound_conditions}
\end{equation}
must both hold. In numerical practice, Eq.~\eqref{eq:ws_master} is
integrated from $x=-L$ to $x=+L$, with $L$ large enough that the
potential has reached its asymptotic limits at both boundaries. The
left asymptotic behaviour is
\begin{equation}
\psi_s(x)\propto e^{\kappa_- x},
\qquad x\to-\infty,
\end{equation}
whereas the right asymptotic behaviour is
\begin{equation}
\psi_s(x)\propto e^{-\kappa_+ x},
\qquad x\to+\infty.
\end{equation}
Accordingly, the shooting method is initialised at $x=-L$ with
\begin{equation}
\psi_s(-L)=e^{-\kappa_- L},
\qquad
\psi_s'(-L)=\kappa_- e^{-\kappa_- L},
\end{equation}
and the admissible energies are located as zeros of the mismatch at
$x=+L$. The wave functions are then normalised numerically over
$[-L,+L]$.

\begin{figure}[h!]
\centering
\includegraphics[width=0.90\linewidth]{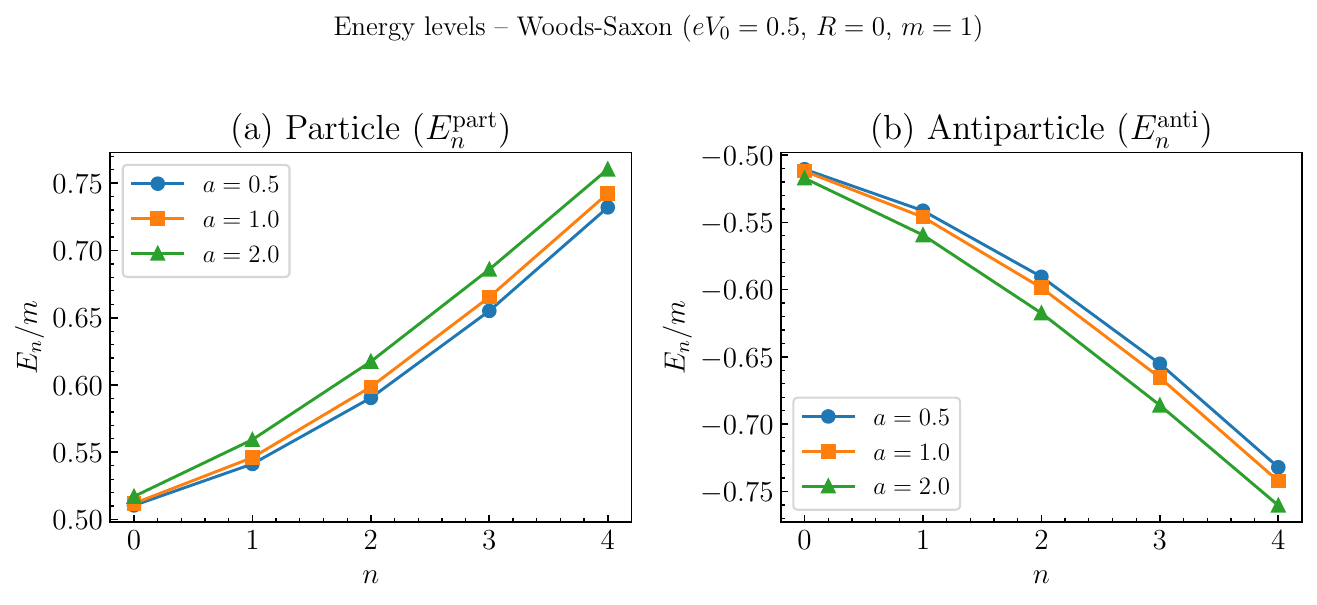}
\caption{First five energy levels $E_{n}/m$ as a function of the
quantum number $n$ for the Woods--Saxon potential ($m = 1$,
$eV_0 = 0.5$, $R = 0$) and three values of the diffuseness
parameter: $a = 0.5$ (circles), $a = 1.0$ (squares), and $a = 2.0$
(triangles). (a)~Particle branch $E_{n}^{(\mathrm{part})} > 0$.
(b)~Antiparticle branch $E_{n}^{(\mathrm{anti})} < 0$. Larger $a$
produces a slight upward shift of the levels, reflecting weaker
effective confinement in the diffuse-surface regime.}
\label{fig:ws_levels}
\end{figure}

The first five energy levels $E_n/m$ are displayed in
Fig.~\ref{fig:ws_levels} for $m=1$, $eV_0=0.5$, $R=0$, and three
values of the diffuseness parameter $a\in\{0.5,1.0,2.0\}$. Panel~(a)
shows the particle branch and panel~(b) the antiparticle branch. The
levels increase with $n$ and are more closely spaced than in the
Coulomb or Cornell cases, reflecting the finite depth and smoothness
of the WS profile. Increasing $a$ broadens the transition region and
slightly shifts the spectrum upward, since the effective confinement
becomes weaker as the surface becomes more diffuse.

\begin{figure}[h!]
\centering
\includegraphics[width=0.75\linewidth]{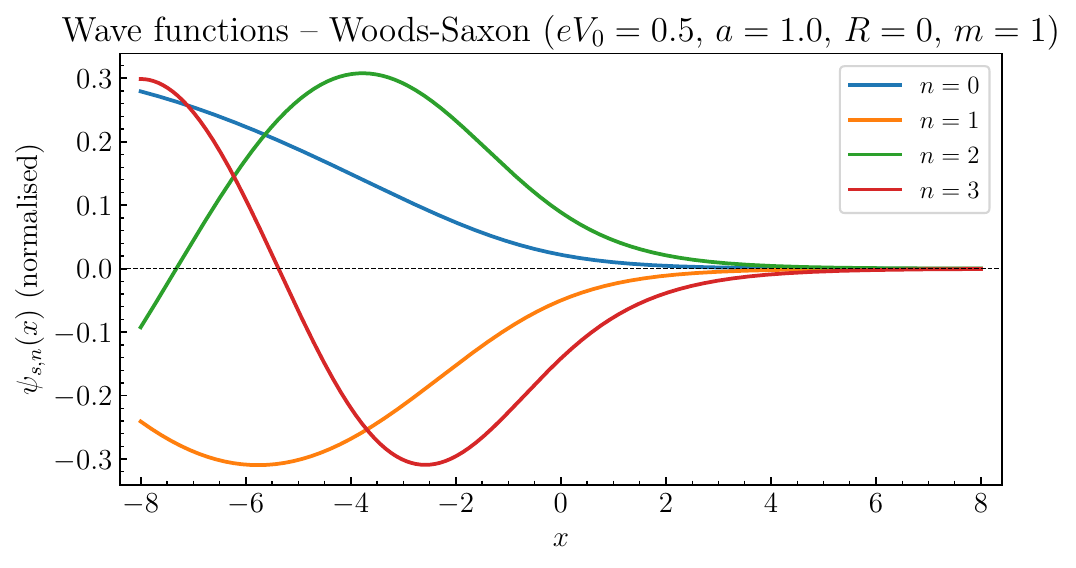}
\caption{Normalised wave functions $\psi_{s,n}(x)$ for the
Woods--Saxon potential ($eV_0 = 0.5$, $a = 1.0$, $R = 0$, $m = 1$)
and $n = 0$ (blue), $1$ (orange), $2$ (green), $3$ (red). All states
are localised predominantly in the deep-well region $x \lesssim 0$
and decay exponentially for $x \gg 0$. The absence of definite parity
is a direct consequence of the asymmetric profile of the WS
potential.}
\label{fig:ws_wf}
\end{figure}

The normalised wave functions $\psi_{s,n}(x)$ for $n=0,1,2,3$,
computed with $eV_0=0.5$, $a=1.0$, $R=0$, and $m=1$, are shown in
Fig.~\ref{fig:ws_wf}. The most immediately visible feature is the
absence of parity symmetry: since $V_{\mathrm{WS}}(x)$ is not even,
the eigenstates are neither symmetric nor antisymmetric about the
origin. All states are concentrated on the deep-well side,
$x\lesssim R=0$, and decay exponentially for $x\gg0$ with decay
constant $\kappa_+ = \sqrt{m^2-E_n^2}$. The number of nodes equals
$n$, consistent with the Sturm--Liouville oscillation theorem, and the
spatial extent of the wave function grows with $n$ as the energy
approaches the continuum threshold.

\begin{figure}[h!]
\centering
\includegraphics[width=0.90\linewidth]{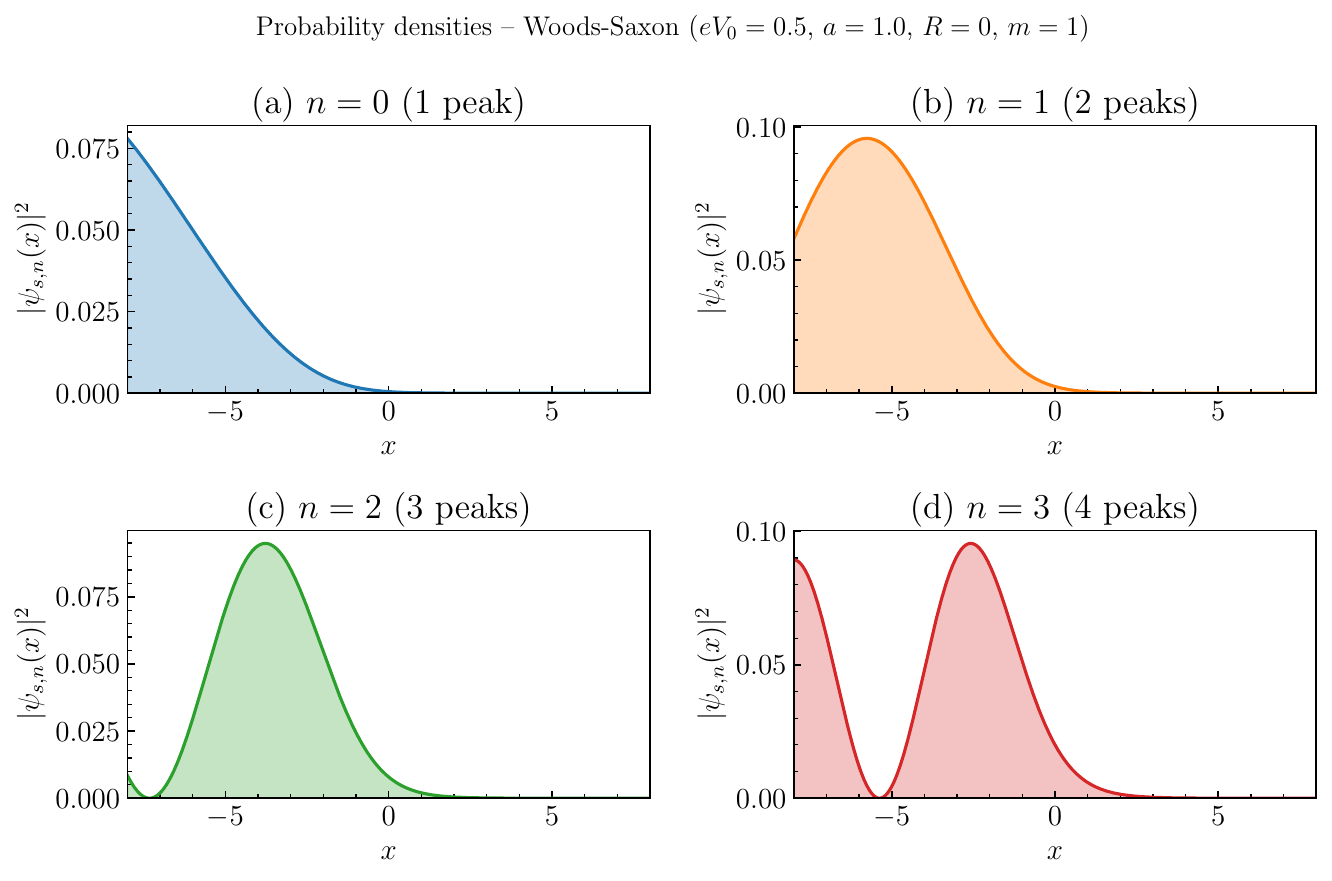}
\caption{Probability densities $|\psi_{s,n}(x)|^{2}$ for the
Woods--Saxon potential ($eV_0 = 0.5$, $a = 1.0$, $R = 0$, $m = 1$).
(a)~$n = 0$: single asymmetric peak. (b)~$n = 1$: two skewed peaks.
(c)~$n = 2$: three peaks. (d)~$n = 3$: four peaks. All distributions
are concentrated on the deep side of the well, a direct consequence
of the asymmetric WS profile.}
\label{fig:ws_prob}
\end{figure}

The corresponding probability densities $|\psi_{s,n}(x)|^{2}$ are
shown in Fig.~\ref{fig:ws_prob}. Their asymmetric structure is clear
in each panel: the dominant peaks are shifted toward the negative-$x$
side, where the potential is deeper. The number of peaks equals
$n+1$, again in agreement with the nodal structure of
Fig.~\ref{fig:ws_wf}. This left-skewed distribution is the direct
spatial signature of the asymmetric Woods--Saxon profile.

Table~\ref{tab:ws_comparison} summarises the main structural
differences among the four potentials treated in this work. The
Woods--Saxon case stands apart in two respects. First, it is the only
case here in which the bound-state equation naturally leads to the
confluent Heun class, so a direct numerical treatment is essential.
Second, because the potential is asymmetric, the eigenstates have no
definite parity and are spatially biased toward the deep side of the
well.

\begin{table}[h!]
\centering
\caption{Structural comparison of the potentials treated in this work
within the FV formalism.}
\label{tab:ws_comparison}
\renewcommand{\arraystretch}{1.25}
\begin{tabular}{lcccc}
\hline\hline
Property & Coulomb (cutoff) & Cornell (cutoff) & P\"{o}schl--Teller & Woods--Saxon \\
\hline
Parity of eigenstates
  & definite & definite & definite & absent \\
Number of bound states
  & infinite & finite (for fixed parameters) & finite & finite \\
Governing equation
  & Whittaker / matching & Tricomi / matching & numerical PT-type & confluent Heun / shooting \\
Closed-form spectrum
  & implicit & implicit & no simple closed form & no \\
\hline\hline
\end{tabular}
\end{table}

The two-component structure of the FV spinor for the Woods--Saxon
potential inherits the spatial asymmetry of the potential itself as
its most distinctive feature. The local mixing factor is
\begin{equation}
  f(x) \equiv \frac{E - eV_{\rm WS}(x)}{m}
       = \frac{1}{m}\!\left(
         E + \frac{eV_0}{1 + e^{(x-R)/a}}
       \right),
  \label{eq:ws_mixing}
\end{equation}
which is a strictly decreasing sigmoid function of $x$: it tends to
\begin{equation}
f(-\infty)=\frac{E+eV_0}{m}
\end{equation}
on the deep side and to
\begin{equation}
f(+\infty)=\frac{E}{m}
\end{equation}
on the shallow side. Unlike the Coulomb and Cornell potentials,
$f(x)$ has no singularity; unlike the P\"{o}schl--Teller potential, it
is not symmetric; and for the parameter range considered here it
remains positive throughout the domain.

\begin{figure}[h!]
\centering
\includegraphics[width=0.90\textwidth]{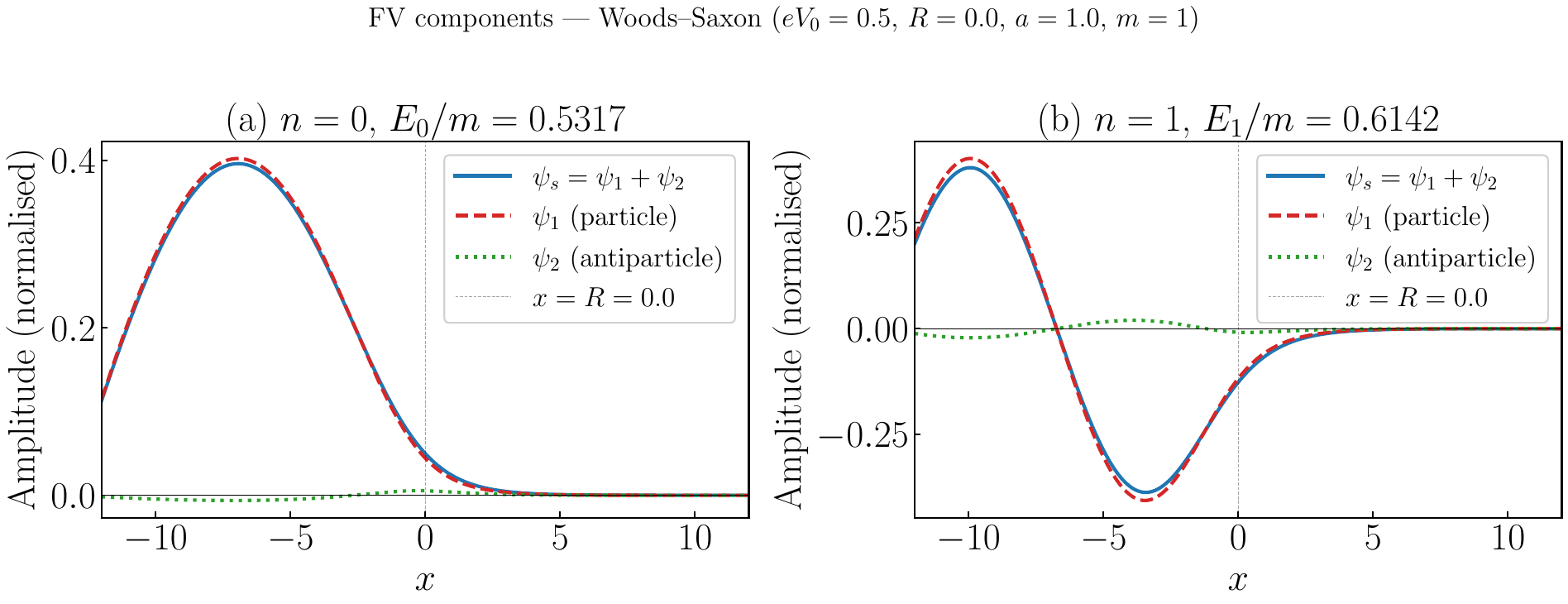}
\caption{FV spinor components $\psi_1$ (particle, red dashed),
$\psi_2$ (antiparticle, green dotted), and their sum
$\psi_s = \psi_1 + \psi_2$ (blue solid) for the Woods--Saxon
potential with $eV_0 = 0.5$, $R = 0$, $a = 1.0$, and $m = 1$.
The vertical dotted line marks $x = R = 0$.
(a) Ground state $n = 0$, $E_0/m = 0.5317$.
(b) First excited state $n = 1$, $E_1/m = 0.6142$.
The asymmetry of all three components directly reflects the asymmetric
profile of the WS potential. For the chosen parameters, the
antiparticle component remains small throughout the domain.}
\label{fig:ws_components}
\end{figure}

The FV components $\psi_s$, $\psi_1$, and $\psi_2$ computed for
$eV_0=0.5$, $R=0$, $a=1.0$, and $m=1$ are displayed in
Fig.~\ref{fig:ws_components} for $n=0$ and $n=1$. Figure \ref{fig:ws_components}(a) shows
the ground state ($E_0/m = 0.532$): both $\psi_s$ and $\psi_1$ are
single-peaked and concentrated on the deep side of the well. The
antiparticle component $\psi_2$ is very small throughout the domain,
because on the left one has
\begin{equation}
f(-\infty)=\frac{E_0+eV_0}{m}\approx 1.03,
\end{equation}
so the factor $(1-f)/2$ is nearly zero, while on the right the scalar wave function itself is already exponentially suppressed. Figure \ref{fig:ws_components}(b) shows the first excited state ($E_1/m = 0.614$): $\psi_s$ and
$\psi_1$ displays the one-node structure of the first excited state,
still concentrated on the deep side. The absence of parity symmetry is
manifest in both panels.

\begin{figure}[h!]
\centering
\includegraphics[width=0.55\textwidth]{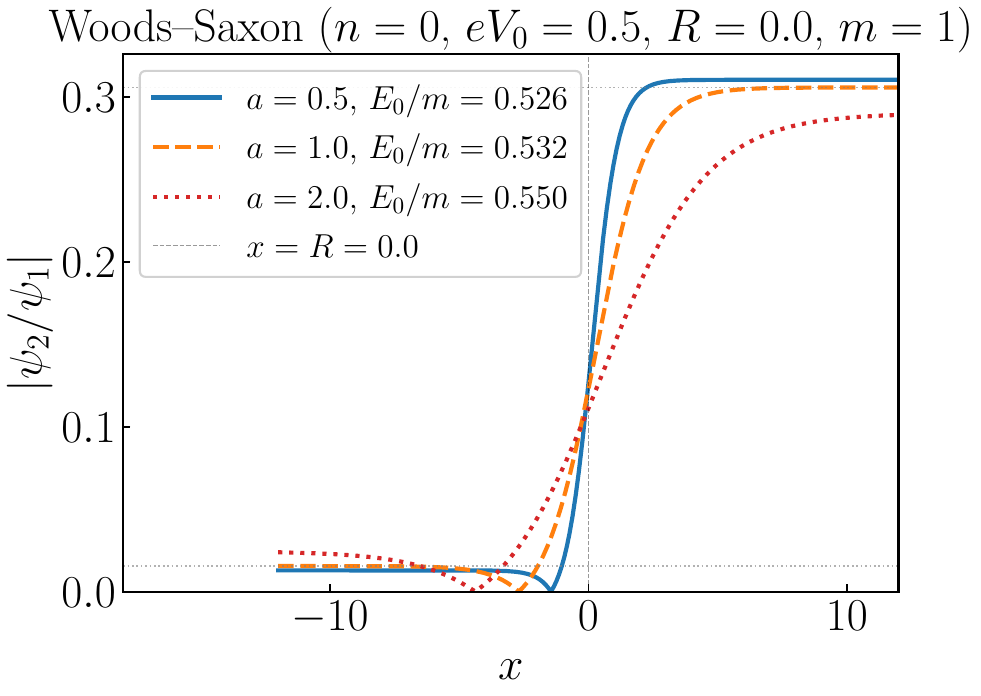}
\caption{Antiparticle-to-particle ratio $|\psi_2/\psi_1|
= |(1-f)/(1+f)|$ for the ground state ($n = 0$, $eV_0 = 0.5$,
$R = 0$, $m = 1$) of the Woods--Saxon potential and three values of
the diffuseness parameter: $a = 0.5$ (blue solid),
$a = 1.0$ (orange dashed), and $a = 2.0$ (red dotted).
The vertical dotted line marks $x = R = 0$.
The ratio exhibits the same sigmoidal spatial structure as the
potential itself: a small plateau on the deep side and a larger
plateau on the shallow side, connected across a transition region of
width controlled by $a$.}
\label{fig:ws_ratio}
\end{figure}

The spatial structure of the particle--antiparticle mixing is shown in
Fig.~\ref{fig:ws_ratio}, which plots
$|\psi_2/\psi_1| = |(1-f)/(1+f)|$ for the ground state and three
values of the diffuseness parameter $a$. The profile has the
characteristic sigmoid form of the WS potential itself. On the deep
side ($x\ll R$), the ratio approaches the small plateau associated
with $f(-\infty)\approx (E+eV_0)/m \approx 1$. As $x$ passes through
the transition region, $f$ decreases and the ratio rises toward the
shallow-side asymptotic plateau
\begin{equation}
\left|\frac{1-E/m}{1+E/m}\right|.
\end{equation}
The width of this transition is controlled directly by $a$: a smaller
$a$ produces a sharper rise, while a larger $a$ spreads the transition
over a broader spatial interval. This sigmoidal behaviour is unique
among the potentials studied in this work.

\begin{figure}[h!]
\centering
\includegraphics[width=0.90\textwidth]{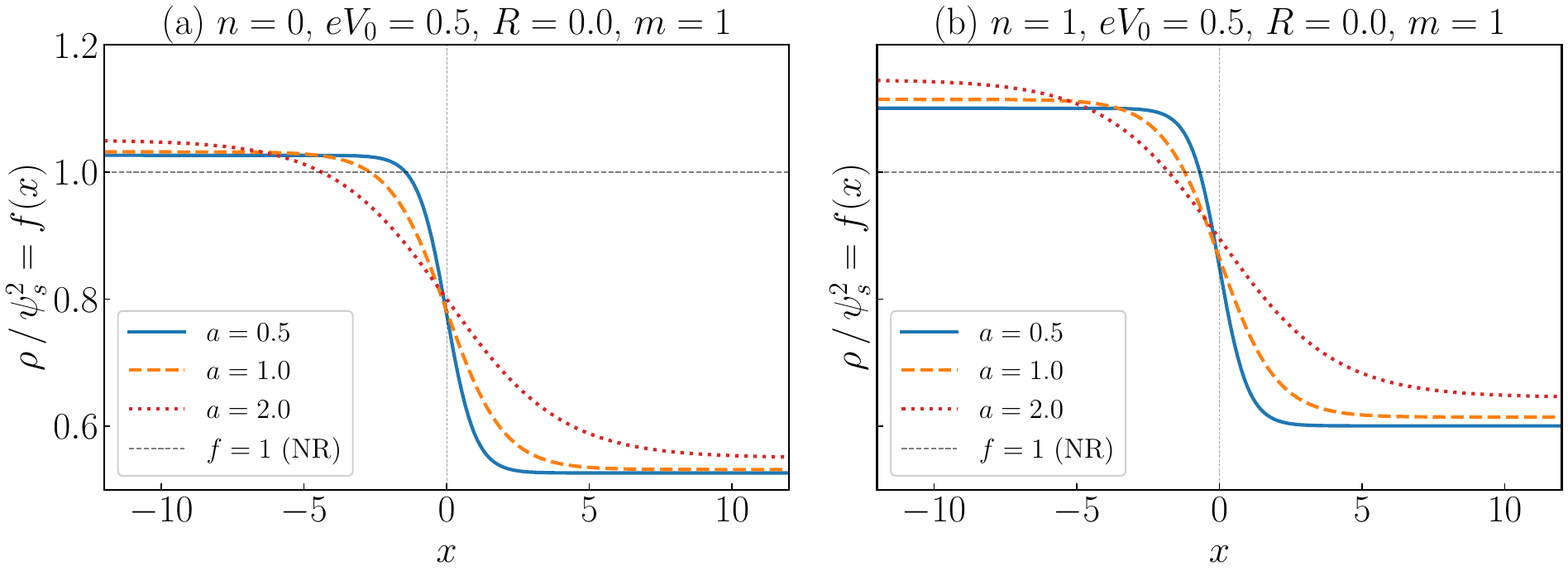}
\caption{Local mixing factor
$f(x) = (E + eV_0/(1 + e^{(x-R)/a}))/m$,
equal to the exact ratio $\rho(x)/|\psi_s(x)|^2$, for the
Woods--Saxon potential with $eV_0 = 0.5$, $R = 0$, $m = 1$, and
$a = 0.5$ (blue solid), $1.0$ (orange dashed), $2.0$ (red dotted).
The vertical dotted line marks $x = R = 0$.
(a) $n = 0$.
(b) $n = 1$.
In both panels $f(x)$ decreases smoothly from $(E+eV_0)/m$ on the
deep side to $E/m$ on the shallow side. The transition width is set
directly by the diffuseness parameter $a$.}
\label{fig:ws_density}
\end{figure}

Finally, the exact local ratio
\begin{equation}
\frac{\rho(x)}{|\psi_s(x)|^2}=f(x)
\end{equation}
is shown in Fig.~\ref{fig:ws_density} for $n=0$ and $n=1$ and three
values of $a$. In both panels, the profile is a sigmoid descending from
$(E+eV_0)/m$ on the deep side to $E/m<1$ on the shallow side. For the
parameters chosen here, $f(x)$ remains positive everywhere, so the conserved FV charge density never changes sign. The transition width is again controlled by $a$, which therefore acts as the spatial scale of the relativistic correction to the local charge density.

\section{Conclusion}\label{s_conclusion}

We have presented a unified analytical and numerical study of the one-dimensional Feshbach--Villars equation for spin-0 particles under five qualitatively distinct external potentials. In all cases, the FV formalism provides a transparent relativistic framework in which the dynamics is governed by the master equation
\begin{equation*}
  \psi_s''(x) + \bigl[(E-eV(x))^2-m^2\bigr]\psi_s(x)=0,
\end{equation*}
while the corresponding particle and antiparticle components are
reconstructed through
\begin{equation*}
  \psi_1 = \frac{1}{2}\left[1+\frac{E-eV(x)}{m}\right]\psi_s,
  \qquad
  \psi_2 = \frac{1}{2}\left[1-\frac{E-eV(x)}{m}\right]\psi_s.
\end{equation*}
This decomposition makes explicit how relativistic effects are encoded
locally through the mixing factor $(E-eV(x))/m$, and allows the charge
density and the internal FV spinor structure to be analysed on equal
footing with the spectrum and scalar wave function.

For the Coulomb problem, the one-dimensional singularity at the origin
was treated through a Loudon-type cutoff regularisation on the full
line. This procedure yields a mathematically controlled formulation in
which parity is well defined, odd and even states are treated on the
same footing, and the familiar odd--even near-degeneracy emerges as the
cutoff is reduced. The regularised formulation clarifies the role of
the deeply localised core state and shows that the physical finite-gap
branch is the one continuously connected to the ordinary bound-state
sector. In this sense, the singular problem is best understood as the
limit of a family of regular problems rather than as an isolated
starting point.

The power-exponential potential with $p=1$ places the system in a
complementary regime: the interaction is short-ranged, bounded, and
free of singularities, yet its relativistic FV spectrum differs
substantially from the Coulomb case. The analytic reduction to
Whittaker functions lead to a characteristic half-integer spectral
structure and to stationary states that are oscillatory rather than
exponentially localised. In the parameter scaling adopted here, these
states are naturally interpreted as intrinsically relativistic and do
not reduce to a conventional Schr\"odinger bound-state series in the
non-relativistic limit.

For the Cornell potential, the same cutoff strategy is used in the Coulomb
case provides a consistent full-line formulation of a singular-plus-
confining interaction. The spectrum is obtained from logarithmic-
derivative matching between the regularised inner solution and the
normalisable Tricomi-function exterior solution. The resulting
bound-state structure exhibits a clear even--odd pairing, with small
but finite splittings that vanish in the cutoff limit. The wave
functions display the characteristic interplay between short-distance
Coulomb attraction and long-distance linear confinement, while the FV
spinor reconstruction shows that the antiparticle component can be
significantly enhanced in the near-core region, even when the total state remains in the positive-energy gap sector.

The P\"oschl--Teller potential illustrates a different relativistic scenario: a smooth, even, short-range well on the full line, with definite-parity eigenstates and a finite number of bound states. In the FV formulation, the squared-potential contribution generates an additional $\mathrm{sech}^4(x/d)$ term, so the full relativistic problem is no longer identical to the textbook Schr\"odinger P\"oschl--Teller model. For the parameter range considered here, the bound-state spectrum is therefore most reliably obtained numerically by
shooting, while the parity structure, nodal ordering, FV component mixing, and charge-density behaviour remain completely transparent.

The Woods--Saxon potential stands apart through its spatial asymmetry.
Its eigenstates have no definite parity and are localised
preferentially on the deep side of the well, producing asymmetric wave
functions, asymmetric probability densities, and a sigmoidal spatial
profile for the particle--antiparticle mixing factor. At the level of
the differential equation, the Woods--Saxon case naturally leads to the
confluent Heun class rather than to the hypergeometric families
encountered in the other potentials. As a result, the numerical
shooting method is not merely convenient but essential for determining
the spectrum and the corresponding FV spinor structure.

Taken together, the five potentials analysed here reveal how the FV
formalism accommodates a broad range of relativistic scalar bound-state
problems within a single framework, while still preserving the
distinctive mathematical and physical signatures of each interaction.
The singular Coulomb and Cornell cases require regularisation and
matching; the P\"oschl--Teller and Woods--Saxon cases emphasise the role
of smooth finite-range geometry, and the power-exponential case highlights the possibility of intrinsically relativistic stationary states without a standard Schr\"odinger counterpart. Beyond the specific models studied here, the results provide a coherent set of analytical and numerical benchmarks for future investigations of one-dimensional relativistic bound states, including semiclassical, variational, and numerical approaches based on more general external fields.

\section*{Acknowledgments}

E. O. Silva acknowledges the support from Conselho Nacional de Desenvolvimento Cient\'{i}fico e Tecnol\'{o}gico (CNPq) (grants 306308/2022-3), Funda\c c\~ao de Amparo \`{a} Pesquisa e ao Desenvolvimento Cient\'{i}fico e Tecnol\'{o}gico do Maranh\~ao (FAPEMA) (grants UNIVERSAL-06395/22), and Coordena\c c\~ao de Aperfei\c coamento de Pessoal de N\'{i}vel Superior (CAPES) - Brazil (Code 001).

\section{Data Availability}

All numerical data supporting this study, including energies and wave functions, are available from the corresponding author upon request.

\section*{Conflict of Interests}

The authors declare no conflict of interest.


\begin{thebibliography}{46}%
	\makeatletter
	\providecommand \@ifxundefined [1]{%
		\@ifx{#1\undefined}
	}%
	\providecommand \@ifnum [1]{%
		\ifnum #1\expandafter \@firstoftwo
		\else \expandafter \@secondoftwo
		\fi
	}%
	\providecommand \@ifx [1]{%
		\ifx #1\expandafter \@firstoftwo
		\else \expandafter \@secondoftwo
		\fi
	}%
	\providecommand \natexlab [1]{#1}%
	\providecommand \enquote  [1]{``#1''}%
	\providecommand \bibnamefont  [1]{#1}%
	\providecommand \bibfnamefont [1]{#1}%
	\providecommand \citenamefont [1]{#1}%
	\providecommand \href@noop [0]{\@secondoftwo}%
	\providecommand \href [0]{\begingroup \@sanitize@url \@href}%
	\providecommand \@href[1]{\@@startlink{#1}\@@href}%
	\providecommand \@@href[1]{\endgroup#1\@@endlink}%
	\providecommand \@sanitize@url [0]{\catcode `\\12\catcode `\$12\catcode
		`\&12\catcode `\#12\catcode `\^12\catcode `\_12\catcode `\%12\relax}%
	\providecommand \@@startlink[1]{}%
	\providecommand \@@endlink[0]{}%
	\providecommand \url  [0]{\begingroup\@sanitize@url \@url }%
	\providecommand \@url [1]{\endgroup\@href {#1}{\urlprefix }}%
	\providecommand \urlprefix  [0]{URL }%
	\providecommand \Eprint [0]{\href }%
	\providecommand \doibase [0]{https://doi.org/}%
	\providecommand \selectlanguage [0]{\@gobble}%
	\providecommand \bibinfo  [0]{\@secondoftwo}%
	\providecommand \bibfield  [0]{\@secondoftwo}%
	\providecommand \translation [1]{[#1]}%
	\providecommand \BibitemOpen [0]{}%
	\providecommand \bibitemStop [0]{}%
	\providecommand \bibitemNoStop [0]{.\EOS\space}%
	\providecommand \EOS [0]{\spacefactor3000\relax}%
	\providecommand \BibitemShut  [1]{\csname bibitem#1\endcsname}%
	\let\auto@bib@innerbib\@empty
	\bibitem [{\citenamefont {Greiner}(2001)}]{B1}%
	\BibitemOpen
	\bibfield  {author} {\bibinfo {author} {\bibfnamefont {W.}~\bibnamefont
			{Greiner}},\ }\href {https://doi.org/10.1007/978-3-642-56826-8} {\emph
		{\bibinfo {title} {Quantum Mechanics: An Introduction}}},\ \bibinfo {edition}
	{5th}\ ed.\ (\bibinfo  {publisher} {Springer},\ \bibinfo {address} {Berlin},\
	\bibinfo {year} {2001})\BibitemShut {NoStop}%
	\bibitem [{\citenamefont {Schr{\"o}dinger}(1926)}]{B2}%
	\BibitemOpen
	\bibfield  {author} {\bibinfo {author} {\bibfnamefont {E.}~\bibnamefont
			{Schr{\"o}dinger}},\ }\href@noop {} {\bibfield  {journal} {\bibinfo
			{journal} {Ann. Phys.}\ }\textbf {\bibinfo {volume} {385}},\ \bibinfo {pages}
		{437} (\bibinfo {year} {1926})},\ \bibinfo {note} {no DOI assigned (pre-DOI
		era)}\BibitemShut {NoStop}%
	\bibitem [{\citenamefont {Fauseweh}(2024)}]{B3}%
	\BibitemOpen
	\bibfield  {author} {\bibinfo {author} {\bibfnamefont {B.}~\bibnamefont
			{Fauseweh}},\ }\href {https://doi.org/10.1038/s41467-024-46402-9} {\bibfield
		{journal} {\bibinfo  {journal} {Nat. Commun.}\ }\textbf {\bibinfo {volume}
			{15}},\ \bibinfo {pages} {2123} (\bibinfo {year} {2024})}\BibitemShut
	{NoStop}%
	\bibitem [{\citenamefont {Forbes}\ \emph {et~al.}(2024)\citenamefont {Forbes},
		\citenamefont {Youssef}, \citenamefont {Singh}, \citenamefont {Nape},\ and\
		\citenamefont {Ung}}]{B4}%
	\BibitemOpen
	\bibfield  {author} {\bibinfo {author} {\bibfnamefont {A.}~\bibnamefont
			{Forbes}}, \bibinfo {author} {\bibfnamefont {M.}~\bibnamefont {Youssef}},
		\bibinfo {author} {\bibfnamefont {S.}~\bibnamefont {Singh}}, \bibinfo
		{author} {\bibfnamefont {I.}~\bibnamefont {Nape}},\ and\ \bibinfo {author}
		{\bibfnamefont {B.}~\bibnamefont {Ung}},\ }\href
	{https://doi.org/10.1063/5.0185281} {\bibfield  {journal} {\bibinfo
			{journal} {Appl. Phys. Lett.}\ }\textbf {\bibinfo {volume} {124}},\ \bibinfo
		{pages} {110501} (\bibinfo {year} {2024})}\BibitemShut {NoStop}%
	\bibitem [{\citenamefont {Klein}\ and\ \citenamefont {Gordon}(1926)}]{k1}%
	\BibitemOpen
	\bibfield  {author} {\bibinfo {author} {\bibfnamefont {O.}~\bibnamefont
			{Klein}}\ and\ \bibinfo {author} {\bibfnamefont {W.}~\bibnamefont {Gordon}},\
	}\href@noop {} {\bibfield  {journal} {\bibinfo  {journal} {Z. Phys.}\
		}\textbf {\bibinfo {volume} {37}},\ \bibinfo {pages} {895} (\bibinfo {year}
		{1926})},\ \bibinfo {note} {no DOI assigned (pre-DOI era)}\BibitemShut
	{NoStop}%
	\bibitem [{\citenamefont {Dirac}(1928)}]{k2}%
	\BibitemOpen
	\bibfield  {author} {\bibinfo {author} {\bibfnamefont {P.~A.~M.}\
			\bibnamefont {Dirac}},\ }\href {https://doi.org/10.1098/rspa.1928.0023}
	{\bibfield  {journal} {\bibinfo  {journal} {Proc. R. Soc. Lond. A}\ }\textbf
		{\bibinfo {volume} {117}},\ \bibinfo {pages} {610} (\bibinfo {year}
		{1928})}\BibitemShut {NoStop}%
	\bibitem [{\citenamefont {Higgs}(1964)}]{k3}%
	\BibitemOpen
	\bibfield  {author} {\bibinfo {author} {\bibfnamefont {P.~W.}\ \bibnamefont
			{Higgs}},\ }\href {https://doi.org/10.1103/PhysRevLett.13.508} {\bibfield
		{journal} {\bibinfo  {journal} {Phys. Rev. Lett.}\ }\textbf {\bibinfo
			{volume} {13}},\ \bibinfo {pages} {508} (\bibinfo {year} {1964})}\BibitemShut
	{NoStop}%
	\bibitem [{\citenamefont {Englert}\ and\ \citenamefont {Brout}(1964)}]{k3-1}%
	\BibitemOpen
	\bibfield  {author} {\bibinfo {author} {\bibfnamefont {F.}~\bibnamefont
			{Englert}}\ and\ \bibinfo {author} {\bibfnamefont {R.}~\bibnamefont
			{Brout}},\ }\href {https://doi.org/10.1103/PhysRevLett.13.321} {\bibfield
		{journal} {\bibinfo  {journal} {Phys. Rev. Lett.}\ }\textbf {\bibinfo
			{volume} {13}},\ \bibinfo {pages} {321} (\bibinfo {year} {1964})}\BibitemShut
	{NoStop}%
	\bibitem [{\citenamefont {Aad}\ \emph {et~al.}(2012)\citenamefont {Aad} \emph
		{et~al.}}]{k3-2}%
	\BibitemOpen
	\bibfield  {author} {\bibinfo {author} {\bibfnamefont {G.}~\bibnamefont
			{Aad}} \emph {et~al.} (\bibinfo {collaboration} {ATLAS Collaboration}),\
	}\href {https://doi.org/10.1016/j.physletb.2012.08.020} {\bibfield  {journal}
		{\bibinfo  {journal} {Phys. Lett. B}\ }\textbf {\bibinfo {volume} {716}},\
		\bibinfo {pages} {1} (\bibinfo {year} {2012})}\BibitemShut {NoStop}%
	\bibitem [{\citenamefont {Chatrchyan}\ \emph {et~al.}(2012)\citenamefont
		{Chatrchyan} \emph {et~al.}}]{k3-3}%
	\BibitemOpen
	\bibfield  {author} {\bibinfo {author} {\bibfnamefont {S.}~\bibnamefont
			{Chatrchyan}} \emph {et~al.} (\bibinfo {collaboration} {CMS Collaboration}),\
	}\href {https://doi.org/10.1016/j.physletb.2012.08.021} {\bibfield  {journal}
		{\bibinfo  {journal} {Phys. Lett. B}\ }\textbf {\bibinfo {volume} {716}},\
		\bibinfo {pages} {30} (\bibinfo {year} {2012})}\BibitemShut {NoStop}%
	\bibitem [{\citenamefont {Battat}\ \emph {et~al.}(2016)\citenamefont {Battat}
		\emph {et~al.}}]{k4}%
	\BibitemOpen
	\bibfield  {author} {\bibinfo {author} {\bibfnamefont {J.~B.~R.}\
			\bibnamefont {Battat}} \emph {et~al.},\ }\href
	{https://doi.org/10.1016/j.physrep.2016.10.001} {\bibfield  {journal}
		{\bibinfo  {journal} {Phys. Rep.}\ }\textbf {\bibinfo {volume} {662}},\
		\bibinfo {pages} {1} (\bibinfo {year} {2016})}\BibitemShut {NoStop}%
	\bibitem [{\citenamefont {Peskin}\ and\ \citenamefont {Schroeder}(1995)}]{k5}%
	\BibitemOpen
	\bibfield  {author} {\bibinfo {author} {\bibfnamefont {M.~E.}\ \bibnamefont
			{Peskin}}\ and\ \bibinfo {author} {\bibfnamefont {D.~V.}\ \bibnamefont
			{Schroeder}},\ }\href {https://doi.org/10.1201/9780429503559} {\emph
		{\bibinfo {title} {An Introduction to Quantum Field Theory}}}\ (\bibinfo
	{publisher} {CRC Press},\ \bibinfo {address} {Boca Raton},\ \bibinfo {year}
	{1995})\BibitemShut {NoStop}%
	\bibitem [{\citenamefont {Asada}\ and\ \citenamefont {Futamase}(1997)}]{c1}%
	\BibitemOpen
	\bibfield  {author} {\bibinfo {author} {\bibfnamefont {H.}~\bibnamefont
			{Asada}}\ and\ \bibinfo {author} {\bibfnamefont {T.}~\bibnamefont
			{Futamase}},\ }\href {https://doi.org/10.1103/PhysRevD.56.R6062} {\bibfield
		{journal} {\bibinfo  {journal} {Phys. Rev. D}\ }\textbf {\bibinfo {volume}
			{56}},\ \bibinfo {pages} {R6062} (\bibinfo {year} {1997})}\BibitemShut
	{NoStop}%
	\bibitem [{\citenamefont {Critchfield}(1976)}]{c2}%
	\BibitemOpen
	\bibfield  {author} {\bibinfo {author} {\bibfnamefont {C.~L.}\ \bibnamefont
			{Critchfield}},\ }\href {https://doi.org/10.1063/1.522901} {\bibfield
		{journal} {\bibinfo  {journal} {J. Math. Phys.}\ }\textbf {\bibinfo {volume}
			{17}},\ \bibinfo {pages} {261} (\bibinfo {year} {1976})}\BibitemShut
	{NoStop}%
	\bibitem [{\citenamefont {Hosseinpour}\ and\ \citenamefont
		{Hassanabadi}(2015)}]{c3}%
	\BibitemOpen
	\bibfield  {author} {\bibinfo {author} {\bibfnamefont {M.}~\bibnamefont
			{Hosseinpour}}\ and\ \bibinfo {author} {\bibfnamefont {H.}~\bibnamefont
			{Hassanabadi}},\ }\href {https://doi.org/10.1142/S0217751X15501249}
	{\bibfield  {journal} {\bibinfo  {journal} {Int. J. Mod. Phys. A}\ }\textbf
		{\bibinfo {volume} {30}},\ \bibinfo {pages} {1550124} (\bibinfo {year}
		{2015})}\BibitemShut {NoStop}%
	\bibitem [{\citenamefont {Leite}\ \emph {et~al.}(2015)\citenamefont {Leite},
		\citenamefont {Belich},\ and\ \citenamefont {Bakke}}]{c4}%
	\BibitemOpen
	\bibfield  {author} {\bibinfo {author} {\bibfnamefont {E.~V.~B.}\
			\bibnamefont {Leite}}, \bibinfo {author} {\bibfnamefont {H.}~\bibnamefont
			{Belich}},\ and\ \bibinfo {author} {\bibfnamefont {K.}~\bibnamefont
			{Bakke}},\ }\href {https://doi.org/10.1155/2015/925846} {\bibfield  {journal}
		{\bibinfo  {journal} {Adv. High Energy Phys.}\ }\textbf {\bibinfo {volume}
			{2015}},\ \bibinfo {pages} {925846} (\bibinfo {year} {2015})}\BibitemShut
	{NoStop}%
	\bibitem [{\citenamefont {Bakke}\ and\ \citenamefont {Furtado}(2015)}]{c5}%
	\BibitemOpen
	\bibfield  {author} {\bibinfo {author} {\bibfnamefont {K.}~\bibnamefont
			{Bakke}}\ and\ \bibinfo {author} {\bibfnamefont {C.}~\bibnamefont
			{Furtado}},\ }\href {https://doi.org/10.1016/j.aop.2015.01.028} {\bibfield
		{journal} {\bibinfo  {journal} {Ann. Phys.}\ }\textbf {\bibinfo {volume}
			{355}},\ \bibinfo {pages} {48} (\bibinfo {year} {2015})}\BibitemShut
	{NoStop}%
	\bibitem [{\citenamefont {Santos}\ and\ \citenamefont {Barros}(2018)}]{c6}%
	\BibitemOpen
	\bibfield  {author} {\bibinfo {author} {\bibfnamefont {L.~C.~N.}\
			\bibnamefont {Santos}}\ and\ \bibinfo {author} {\bibfnamefont
			{J.}~\bibnamefont {Barros}, \bibfnamefont {Carlos~C.}},\ }\href
	{https://doi.org/10.1140/epjc/s10052-017-5476-3} {\bibfield  {journal}
		{\bibinfo  {journal} {Eur. Phys. J. C}\ }\textbf {\bibinfo {volume} {78}},\
		\bibinfo {pages} {13} (\bibinfo {year} {2018})}\BibitemShut {NoStop}%
	\bibitem [{\citenamefont {Hassanabadi}\ \emph {et~al.}(2012)\citenamefont
		{Hassanabadi}, \citenamefont {Molaee}, \citenamefont {Ghominejad},\ and\
		\citenamefont {Zarrinkamar}}]{c7}%
	\BibitemOpen
	\bibfield  {author} {\bibinfo {author} {\bibfnamefont {H.}~\bibnamefont
			{Hassanabadi}}, \bibinfo {author} {\bibfnamefont {Z.}~\bibnamefont {Molaee}},
		\bibinfo {author} {\bibfnamefont {M.}~\bibnamefont {Ghominejad}},\ and\
		\bibinfo {author} {\bibfnamefont {S.}~\bibnamefont {Zarrinkamar}},\ }\href
	{https://doi.org/10.1155/2012/489641} {\bibfield  {journal} {\bibinfo
			{journal} {Adv. High Energy Phys.}\ }\textbf {\bibinfo {volume} {2012}},\
		\bibinfo {pages} {489641} (\bibinfo {year} {2012})}\BibitemShut {NoStop}%
	\bibitem [{\citenamefont {Halliwell}(1987)}]{p2}%
	\BibitemOpen
	\bibfield  {author} {\bibinfo {author} {\bibfnamefont {J.~J.}\ \bibnamefont
			{Halliwell}},\ }\href {https://doi.org/10.1016/0370-2693(87)91011-2}
	{\bibfield  {journal} {\bibinfo  {journal} {Phys. Lett. B}\ }\textbf
		{\bibinfo {volume} {185}},\ \bibinfo {pages} {341} (\bibinfo {year}
		{1987})}\BibitemShut {NoStop}%
	\bibitem [{\citenamefont {Burd}\ and\ \citenamefont {Barrow}(1988)}]{p3}%
	\BibitemOpen
	\bibfield  {author} {\bibinfo {author} {\bibfnamefont {A.~B.}\ \bibnamefont
			{Burd}}\ and\ \bibinfo {author} {\bibfnamefont {J.~D.}\ \bibnamefont
			{Barrow}},\ }\href {https://doi.org/10.1016/0550-3213(88)90135-6} {\bibfield
		{journal} {\bibinfo  {journal} {Nucl. Phys. B}\ }\textbf {\bibinfo {volume}
			{308}},\ \bibinfo {pages} {929} (\bibinfo {year} {1988})}\BibitemShut
	{NoStop}%
	\bibitem [{\citenamefont {Bakke}(2024)}]{p1}%
	\BibitemOpen
	\bibfield  {author} {\bibinfo {author} {\bibfnamefont {K.}~\bibnamefont
			{Bakke}},\ }\href {https://doi.org/10.1209/0295-5075/ad3e72} {\bibfield
		{journal} {\bibinfo  {journal} {EPL}\ }\textbf {\bibinfo {volume} {146}},\
		\bibinfo {pages} {30004} (\bibinfo {year} {2024})}\BibitemShut {NoStop}%
	\bibitem [{\citenamefont {Ciurla}\ \emph {et~al.}(2002)\citenamefont {Ciurla},
		\citenamefont {Adamowski}, \citenamefont {Szafran},\ and\ \citenamefont
		{Bednarek}}]{p4}%
	\BibitemOpen
	\bibfield  {author} {\bibinfo {author} {\bibfnamefont {M.}~\bibnamefont
			{Ciurla}}, \bibinfo {author} {\bibfnamefont {J.}~\bibnamefont {Adamowski}},
		\bibinfo {author} {\bibfnamefont {B.}~\bibnamefont {Szafran}},\ and\ \bibinfo
		{author} {\bibfnamefont {S.}~\bibnamefont {Bednarek}},\ }\href
	{https://doi.org/10.1016/S1386-9477(02)00572-6} {\bibfield  {journal}
		{\bibinfo  {journal} {Physica E}\ }\textbf {\bibinfo {volume} {15}},\
		\bibinfo {pages} {261} (\bibinfo {year} {2002})}\BibitemShut {NoStop}%
	\bibitem [{\citenamefont {Feshbach}\ and\ \citenamefont
		{Villars}(1958{\natexlab{a}})}]{F1}%
	\BibitemOpen
	\bibfield  {author} {\bibinfo {author} {\bibfnamefont {H.}~\bibnamefont
			{Feshbach}}\ and\ \bibinfo {author} {\bibfnamefont {F.}~\bibnamefont
			{Villars}},\ }\href {https://doi.org/10.1103/RevModPhys.30.24} {\bibfield
		{journal} {\bibinfo  {journal} {Rev. Mod. Phys.}\ }\textbf {\bibinfo {volume}
			{30}},\ \bibinfo {pages} {24} (\bibinfo {year}
		{1958}{\natexlab{a}})}\BibitemShut {NoStop}%
	\bibitem [{\citenamefont {Bouzenada}\ and\ \citenamefont {Boumali}(2023)}]{F2}%
	\BibitemOpen
	\bibfield  {author} {\bibinfo {author} {\bibfnamefont {A.}~\bibnamefont
			{Bouzenada}}\ and\ \bibinfo {author} {\bibfnamefont {A.}~\bibnamefont
			{Boumali}},\ }\href {https://doi.org/10.1016/j.aop.2023.169302} {\bibfield
		{journal} {\bibinfo  {journal} {Ann. Phys.}\ }\textbf {\bibinfo {volume}
			{452}},\ \bibinfo {pages} {169302} (\bibinfo {year} {2023})}\BibitemShut
	{NoStop}%
	\bibitem [{\citenamefont {Bouzenada}\ \emph
		{et~al.}(2023{\natexlab{a}})\citenamefont {Bouzenada}, \citenamefont
		{Boumali}, \citenamefont {Vitoria}, \citenamefont {Ahmed},\ and\
		\citenamefont {Al-Raeei}}]{F3}%
	\BibitemOpen
	\bibfield  {author} {\bibinfo {author} {\bibfnamefont {A.}~\bibnamefont
			{Bouzenada}}, \bibinfo {author} {\bibfnamefont {A.}~\bibnamefont {Boumali}},
		\bibinfo {author} {\bibfnamefont {R.~L.~L.}\ \bibnamefont {Vitoria}},
		\bibinfo {author} {\bibfnamefont {F.}~\bibnamefont {Ahmed}},\ and\ \bibinfo
		{author} {\bibfnamefont {M.}~\bibnamefont {Al-Raeei}},\ }\href
	{https://doi.org/10.1016/j.nuclphysb.2023.116288} {\bibfield  {journal}
		{\bibinfo  {journal} {Nucl. Phys. B}\ }\textbf {\bibinfo {volume} {994}},\
		\bibinfo {pages} {116288} (\bibinfo {year} {2023}{\natexlab{a}})}\BibitemShut
	{NoStop}%
	\bibitem [{\citenamefont {Bouzenada}\ \emph
		{et~al.}(2023{\natexlab{b}})\citenamefont {Bouzenada}, \citenamefont
		{Boumali},\ and\ \citenamefont {Silva}}]{F4}%
	\BibitemOpen
	\bibfield  {author} {\bibinfo {author} {\bibfnamefont {A.}~\bibnamefont
			{Bouzenada}}, \bibinfo {author} {\bibfnamefont {A.}~\bibnamefont {Boumali}},\
		and\ \bibinfo {author} {\bibfnamefont {E.~O.}\ \bibnamefont {Silva}},\ }\href
	{https://doi.org/10.1016/j.aop.2023.169479} {\bibfield  {journal} {\bibinfo
			{journal} {Ann. Phys.}\ }\textbf {\bibinfo {volume} {458}},\ \bibinfo {pages}
		{169479} (\bibinfo {year} {2023}{\natexlab{b}})}\BibitemShut {NoStop}%
	\bibitem [{\citenamefont {Bouzenada}\ \emph
		{et~al.}(2023{\natexlab{c}})\citenamefont {Bouzenada}, \citenamefont
		{Boumali}, \citenamefont {Vitoria},\ and\ \citenamefont {Furtado}}]{F5}%
	\BibitemOpen
	\bibfield  {author} {\bibinfo {author} {\bibfnamefont {A.}~\bibnamefont
			{Bouzenada}}, \bibinfo {author} {\bibfnamefont {A.}~\bibnamefont {Boumali}},
		\bibinfo {author} {\bibfnamefont {R.~L.~L.}\ \bibnamefont {Vitoria}},\ and\
		\bibinfo {author} {\bibfnamefont {C.}~\bibnamefont {Furtado}},\ }\href@noop
	{} {\bibinfo {title} {{Feshbach}--{Villars} oscillator in a rotating frame in
			the cosmic string spacetime}} (\bibinfo {year} {2023}{\natexlab{c}}),\
	\Eprint {https://arxiv.org/abs/2311.02439} {arXiv:2311.02439 [hep-th]}
	\BibitemShut {NoStop}%
	\bibitem [{\citenamefont {Bouzenada}\ \emph
		{et~al.}(2023{\natexlab{d}})\citenamefont {Bouzenada}, \citenamefont
		{Boumali}, \citenamefont {Mustafa},\ and\ \citenamefont {Hassanabadi}}]{F6}%
	\BibitemOpen
	\bibfield  {author} {\bibinfo {author} {\bibfnamefont {A.}~\bibnamefont
			{Bouzenada}}, \bibinfo {author} {\bibfnamefont {A.}~\bibnamefont {Boumali}},
		\bibinfo {author} {\bibfnamefont {O.}~\bibnamefont {Mustafa}},\ and\ \bibinfo
		{author} {\bibfnamefont {H.}~\bibnamefont {Hassanabadi}},\ }\href@noop {}
	{\bibinfo {title} {{Feshbach}--{Villars} oscillator in the {G{\"o}del}-type
			spacetime}} (\bibinfo {year} {2023}{\natexlab{d}}),\ \Eprint
	{https://arxiv.org/abs/2304.12496} {arXiv:2304.12496 [hep-th]} \BibitemShut
	{NoStop}%
	\bibitem [{\citenamefont {Bouzenada}\ \emph
		{et~al.}(2023{\natexlab{e}})\citenamefont {Bouzenada}, \citenamefont
		{Boumali},\ and\ \citenamefont {Al-Raeei}}]{F7}%
	\BibitemOpen
	\bibfield  {author} {\bibinfo {author} {\bibfnamefont {A.}~\bibnamefont
			{Bouzenada}}, \bibinfo {author} {\bibfnamefont {A.}~\bibnamefont {Boumali}},\
		and\ \bibinfo {author} {\bibfnamefont {M.}~\bibnamefont {Al-Raeei}},\
	}\href@noop {} {\bibinfo {title} {Thermal properties of the
			{Feshbach}--{Villars} oscillator in a rotating frame}} (\bibinfo {year}
	{2023}{\natexlab{e}}),\ \Eprint {https://arxiv.org/abs/2302.13805}
	{arXiv:2302.13805 [quant-ph]} \BibitemShut {NoStop}%
	\bibitem [{\citenamefont {Wingard}\ \emph {et~al.}(2024)\citenamefont
		{Wingard}, \citenamefont {Garcia~Vallejo},\ and\ \citenamefont {Papp}}]{F8}%
	\BibitemOpen
	\bibfield  {author} {\bibinfo {author} {\bibfnamefont {D.}~\bibnamefont
			{Wingard}}, \bibinfo {author} {\bibfnamefont {A.}~\bibnamefont
			{Garcia~Vallejo}},\ and\ \bibinfo {author} {\bibfnamefont {Z.}~\bibnamefont
			{Papp}},\ }\href {https://doi.org/10.1007/s00601-024-01902-8} {\bibfield
		{journal} {\bibinfo  {journal} {Few-Body Syst.}\ }\textbf {\bibinfo {volume}
			{65}},\ \bibinfo {pages} {30} (\bibinfo {year} {2024})}\BibitemShut {NoStop}%
	\bibitem [{\citenamefont {Boumali}\ \emph {et~al.}(2024)\citenamefont
		{Boumali}, \citenamefont {Hamla},\ and\ \citenamefont {Chargui}}]{Hamla2024}%
	\BibitemOpen
	\bibfield  {author} {\bibinfo {author} {\bibfnamefont {A.}~\bibnamefont
			{Boumali}}, \bibinfo {author} {\bibfnamefont {A.}~\bibnamefont {Hamla}},\
		and\ \bibinfo {author} {\bibfnamefont {Y.}~\bibnamefont {Chargui}},\ }\href
	{https://doi.org/10.1007/s10773-024-05743-3} {\bibfield  {journal} {\bibinfo
			{journal} {Int. J. Theor. Phys.}\ }\textbf {\bibinfo {volume} {63}},\
		\bibinfo {pages} {200} (\bibinfo {year} {2024})}\BibitemShut {NoStop}%
	\bibitem [{\citenamefont {Silenko}(2022)}]{F9}%
	\BibitemOpen
	\bibfield  {author} {\bibinfo {author} {\bibfnamefont {A.~J.}\ \bibnamefont
			{Silenko}},\ }\href {https://doi.org/10.1103/PhysRevA.105.062211} {\bibfield
		{journal} {\bibinfo  {journal} {Phys. Rev. A}\ }\textbf {\bibinfo {volume}
			{105}},\ \bibinfo {pages} {062211} (\bibinfo {year} {2022})}\BibitemShut
	{NoStop}%
	\bibitem [{\citenamefont {Silenko}(2020)}]{F10}%
	\BibitemOpen
	\bibfield  {author} {\bibinfo {author} {\bibfnamefont {A.~J.}\ \bibnamefont
			{Silenko}},\ }\href {https://doi.org/10.1134/S1547477120020193} {\bibfield
		{journal} {\bibinfo  {journal} {Phys. Part. Nucl. Lett.}\ }\textbf {\bibinfo
			{volume} {17}},\ \bibinfo {pages} {116} (\bibinfo {year} {2020})}\BibitemShut
	{NoStop}%
	\bibitem [{\citenamefont {Silenko}(2013{\natexlab{a}})}]{F11}%
	\BibitemOpen
	\bibfield  {author} {\bibinfo {author} {\bibfnamefont {A.~J.}\ \bibnamefont
			{Silenko}},\ }\href {https://doi.org/10.1103/PhysRevD.88.045004} {\bibfield
		{journal} {\bibinfo  {journal} {Phys. Rev. D}\ }\textbf {\bibinfo {volume}
			{88}},\ \bibinfo {pages} {045004} (\bibinfo {year}
		{2013}{\natexlab{a}})}\BibitemShut {NoStop}%
	\bibitem [{\citenamefont {Silenko}(2013{\natexlab{b}})}]{Silenko2013}%
	\BibitemOpen
	\bibfield  {author} {\bibinfo {author} {\bibfnamefont {A.~J.}\ \bibnamefont
			{Silenko}},\ }\href {https://doi.org/10.1103/PhysRevD.88.045004} {\bibfield
		{journal} {\bibinfo  {journal} {Phys. Rev. D}\ }\textbf {\bibinfo {volume}
			{88}},\ \bibinfo {pages} {045004} (\bibinfo {year}
		{2013}{\natexlab{b}})}\BibitemShut {NoStop}%
	\bibitem [{\citenamefont {Feshbach}\ and\ \citenamefont
		{Villars}(1958{\natexlab{b}})}]{Feshbach1958}%
	\BibitemOpen
	\bibfield  {author} {\bibinfo {author} {\bibfnamefont {H.}~\bibnamefont
			{Feshbach}}\ and\ \bibinfo {author} {\bibfnamefont {F.}~\bibnamefont
			{Villars}},\ }\href {https://doi.org/10.1103/RevModPhys.30.24} {\bibfield
		{journal} {\bibinfo  {journal} {Rev. Mod. Phys.}\ }\textbf {\bibinfo {volume}
			{30}},\ \bibinfo {pages} {24} (\bibinfo {year}
		{1958}{\natexlab{b}})}\BibitemShut {NoStop}%
	\bibitem [{\citenamefont {Merad}\ \emph {et~al.}(2000)\citenamefont {Merad},
		\citenamefont {Chetouani},\ and\ \citenamefont {Bounames}}]{F12}%
	\BibitemOpen
	\bibfield  {author} {\bibinfo {author} {\bibfnamefont {M.}~\bibnamefont
			{Merad}}, \bibinfo {author} {\bibfnamefont {L.}~\bibnamefont {Chetouani}},\
		and\ \bibinfo {author} {\bibfnamefont {A.}~\bibnamefont {Bounames}},\ }\href
	{https://doi.org/10.1016/S0375-9601(00)00096-6} {\bibfield  {journal}
		{\bibinfo  {journal} {Phys. Lett. A}\ }\textbf {\bibinfo {volume} {267}},\
		\bibinfo {pages} {225} (\bibinfo {year} {2000})}\BibitemShut {NoStop}%
	\bibitem [{\citenamefont {Loudon}(1959)}]{Loudon1959}%
	\BibitemOpen
	\bibfield  {author} {\bibinfo {author} {\bibfnamefont {R.}~\bibnamefont
			{Loudon}},\ }\href {https://doi.org/10.1119/1.1934950} {\bibfield  {journal}
		{\bibinfo  {journal} {American Journal of Physics}\ }\textbf {\bibinfo
			{volume} {27}},\ \bibinfo {pages} {649} (\bibinfo {year} {1959})}\BibitemShut
	{NoStop}%
	\bibitem [{\citenamefont {Loudon}(2016)}]{Loudon2016}%
	\BibitemOpen
	\bibfield  {author} {\bibinfo {author} {\bibfnamefont {R.}~\bibnamefont
			{Loudon}},\ }\href {https://doi.org/10.1098/rspa.2015.0534} {\bibfield
		{journal} {\bibinfo  {journal} {Proceedings of the Royal Society A}\ }\textbf
		{\bibinfo {volume} {472}},\ \bibinfo {pages} {20150534} (\bibinfo {year}
		{2016})}\BibitemShut {NoStop}%
	\bibitem [{\citenamefont {Abramowitz}\ and\ \citenamefont
		{Stegun}(1972)}]{Abramowitz}%
	\BibitemOpen
	\bibfield  {author} {\bibinfo {author} {\bibfnamefont {M.}~\bibnamefont
			{Abramowitz}}\ and\ \bibinfo {author} {\bibfnamefont {I.~A.}\ \bibnamefont
			{Stegun}},\ }\href@noop {} {\emph {\bibinfo {title} {Handbook of Mathematical
				Functions with Formulas, Graphs, and Mathematical Tables}}}\ (\bibinfo
	{publisher} {Dover},\ \bibinfo {address} {New York},\ \bibinfo {year}
	{1972})\ \bibinfo {note} {9th Dover printing; originally published by the
		National Bureau of Standards (1964)}\BibitemShut {NoStop}%
	\bibitem [{\citenamefont {P{\"o}schl}\ and\ \citenamefont
		{Teller}(1933)}]{Poschl1933}%
	\BibitemOpen
	\bibfield  {author} {\bibinfo {author} {\bibfnamefont {G.}~\bibnamefont
			{P{\"o}schl}}\ and\ \bibinfo {author} {\bibfnamefont {E.}~\bibnamefont
			{Teller}},\ }\href {https://doi.org/10.1007/BF01331132} {\bibfield  {journal}
		{\bibinfo  {journal} {Zeitschrift f{\"u}r Physik}\ }\textbf {\bibinfo
			{volume} {83}},\ \bibinfo {pages} {143} (\bibinfo {year} {1933})}\BibitemShut
	{NoStop}%
	\bibitem [{\citenamefont {Cooper}\ \emph {et~al.}(1995)\citenamefont {Cooper},
		\citenamefont {Khare},\ and\ \citenamefont {Sukhatme}}]{Cooper1995}%
	\BibitemOpen
	\bibfield  {author} {\bibinfo {author} {\bibfnamefont {F.}~\bibnamefont
			{Cooper}}, \bibinfo {author} {\bibfnamefont {A.}~\bibnamefont {Khare}},\ and\
		\bibinfo {author} {\bibfnamefont {U.~P.}\ \bibnamefont {Sukhatme}},\ }\href
	{https://doi.org/10.1016/0370-1573(94)00080-M} {\bibfield  {journal}
		{\bibinfo  {journal} {Physics Reports}\ }\textbf {\bibinfo {volume} {251}},\
		\bibinfo {pages} {267} (\bibinfo {year} {1995})}\BibitemShut {NoStop}%
	\bibitem [{\citenamefont {Woods}\ and\ \citenamefont
		{Saxon}(1954)}]{Woods1954}%
	\BibitemOpen
	\bibfield  {author} {\bibinfo {author} {\bibfnamefont {R.~D.}\ \bibnamefont
			{Woods}}\ and\ \bibinfo {author} {\bibfnamefont {D.~S.}\ \bibnamefont
			{Saxon}},\ }\href {https://doi.org/10.1103/PhysRev.95.577} {\bibfield
		{journal} {\bibinfo  {journal} {Phys. Rev.}\ }\textbf {\bibinfo {volume}
			{95}},\ \bibinfo {pages} {577} (\bibinfo {year} {1954})}\BibitemShut
	{NoStop}%
	\bibitem [{\citenamefont {Hosseinpour}\ \emph {et~al.}(2015)\citenamefont
		{Hosseinpour}, \citenamefont {Hassanabadi},\ and\ \citenamefont
		{Salehi}}]{Hosseinpour2015}%
	\BibitemOpen
	\bibfield  {author} {\bibinfo {author} {\bibfnamefont {M.}~\bibnamefont
			{Hosseinpour}}, \bibinfo {author} {\bibfnamefont {H.}~\bibnamefont
			{Hassanabadi}},\ and\ \bibinfo {author} {\bibfnamefont {N.}~\bibnamefont
			{Salehi}},\ }\href {https://doi.org/10.1140/epjp/i2015-15236-8} {\bibfield
		{journal} {\bibinfo  {journal} {Eur. Phys. J. Plus}\ }\textbf {\bibinfo
			{volume} {130}},\ \bibinfo {pages} {236} (\bibinfo {year}
		{2015})}\BibitemShut {NoStop}%
	\bibitem [{\citenamefont {Ronveaux}(1995)}]{Ronveaux1995}%
	\BibitemOpen
	\bibinfo {editor} {\bibfnamefont {A.}~\bibnamefont {Ronveaux}},\ ed.,\
	\href@noop {} {\emph {\bibinfo {title} {Heun's Differential Equations}}}\
	(\bibinfo  {publisher} {Oxford University Press},\ \bibinfo {address}
	{Oxford},\ \bibinfo {year} {1995})\BibitemShut {NoStop}%
\end{thebibliography}
\end{document}